\documentclass[twocolumn,amsmath,amssymb,pra,longbibliography, superscriptaddress]{revtex4-2}
\usepackage{graphicx}
\usepackage{amsmath,amsfonts,amssymb,amsthm}
\usepackage{fancyhdr}
\usepackage{mathrsfs}
\usepackage{epstopdf}
\usepackage{setspace}
\usepackage{bm}
\usepackage[colorlinks]{hyperref}
\usepackage{xcolor}
\usepackage{dcolumn}
\usepackage{tikz}
\usetikzlibrary{quantikz}
\graphicspath{{figures/}}

\begin{document}
\title{Quantum algorithms for simulating systems coupled to bosonic modes using a hybrid resonator-qubit quantum computer}


\author{Juha Lepp\"akangas}
\affiliation{HQS Quantum Simulations GmbH, Rintheimer Str. 23, 76131 Karlsruhe, Germany}

\author{Pascal Stadler}
\affiliation{HQS Quantum Simulations GmbH, Rintheimer Str. 23, 76131 Karlsruhe, Germany}

\author{Dmitry Golubev}
\affiliation{HQS Quantum Simulations GmbH, Rintheimer Str. 23, 76131 Karlsruhe, Germany}

\author{Rolando Reiner}
\affiliation{HQS Quantum Simulations GmbH, Rintheimer Str. 23, 76131 Karlsruhe, Germany}

\author{Jan-Michael Reiner}
\affiliation{HQS Quantum Simulations GmbH, Rintheimer Str. 23, 76131 Karlsruhe, Germany}

\author{Sebastian Zanker}
\affiliation{HQS Quantum Simulations GmbH, Rintheimer Str. 23, 76131 Karlsruhe, Germany}

\author{Nicola Wurz}
\affiliation{IQM Quantum Computers, Georg-Brauchle-Ring 23-25, 80992 Munich, Germany}

\author{Michael Renger}
\affiliation{IQM Quantum Computers, Georg-Brauchle-Ring 23-25, 80992 Munich, Germany}

\author{Jeroen Verjauw}
\affiliation{IQM Quantum Computers, Georg-Brauchle-Ring 23-25, 80992 Munich, Germany}

\author{Daria Gusenkova}
\affiliation{IQM Quantum Computers, Georg-Brauchle-Ring 23-25, 80992 Munich, Germany}

\author{Stefan Pogorzalek}
\affiliation{IQM Quantum Computers, Georg-Brauchle-Ring 23-25, 80992 Munich, Germany}

\author{Florian Vigneau}
\affiliation{IQM Quantum Computers, Georg-Brauchle-Ring 23-25, 80992 Munich, Germany}

\author{Ping Yang}
\affiliation{IQM Quantum Computers, Georg-Brauchle-Ring 23-25, 80992 Munich, Germany}

\author{William Kindel}
\affiliation{IQM Quantum Computers, Georg-Brauchle-Ring 23-25, 80992 Munich, Germany}

\author{Hsiang-Sheng Ku}
\affiliation{IQM Quantum Computers, Georg-Brauchle-Ring 23-25, 80992 Munich, Germany}

\author{Frank Deppe}
\affiliation{IQM Quantum Computers, Georg-Brauchle-Ring 23-25, 80992 Munich, Germany}

\author{Michael Marthaler}
\affiliation{HQS Quantum Simulations GmbH, Rintheimer Str. 23, 76131 Karlsruhe, Germany}

%%%

\begin{abstract}
  Modeling composite systems of spins or electrons coupled to bosonic modes
  is of significant interest for many fields of applied quantum physics and chemistry.
  A quantum simulation can allow for the solution of quantum problems beyond
  classical numerical methods. However, implementing this on existing noisy quantum computers can be challenging due
  to the mapping between qubits and bosonic degrees of freedom, often requiring a large number of qubits or deep quantum circuits.
  In this work, we discuss quantum algorithms to solve composite systems by augmenting
  conventional superconducting qubits with microwave resonators used as computational elements. 
  This enables direct representation of bosonic modes by resonators.
  We derive efficient algorithms
  for typical models and propose a device connectivity that allows for feasible scaling of simulations with linear overhead.
  We also show how the dissipation of resonators can be a useful parameter for modeling continuous bosonic baths.
  Experimental results demonstrating these methods were obtained on the IQM Resonance cloud platform,
  based on high-fidelity gates and tunable couplers.
  These results present the first digital quantum simulation including a computational resonator on
  a commercial quantum platform.
\end{abstract}

\maketitle
%\tableofcontents

%---------------------------------------------------------------- Intro ----------------------------------------------------------------
%---------------------------------------------------------------- Intro ----------------------------------------------------------------
%---------------------------------------------------------------- Intro ----------------------------------------------------------------

\section{Introduction}
Quantum simulation is a key application of gate-based quantum computing~\cite{Fauseweh2024}.
An important target is the simulation of spins or electrons,
coupled to bosonic modes.
Such models are central for predicting properties in many fields of applied physics and chemistry, including
photovoltaics~\cite{Kippelen2009, Ostroverkhova2016}, biophysics~\cite{Ishizaki2012, Huelga_2013},
nanoelectronics~\cite{Ingold1992, SchoellerSchon1994, Leppakangas2023},
solid state physics~\cite{Bohm1953, Froehlich1954, Cooper1956, Holstein1959},
and quantum information devices~\cite{Shnirman2002}.
For near-term quantum computers, useful algorithms need to have low depth 
due to restrictions set by noise and gate errors~\cite{Barthi2022}.
Since modeling of bosonic modes using conventional all-qubit quantum computers needs non-trivial mappings
between bosons and qubits~\cite{Sawaya2020}, it can be challenging to perform useful quantum simulations
of bosonic systems on conventional quantum computers
before fully quantum error corrected devices are accessible.
This has motivated the search of algorithms that can run on
hybrid quantum devices~\cite{Mezzacapo2014, Langford2017, Wang2023, Katz2023, Dutta2024, Liu2024, Crane2024, Shapiro2024, Vu2025, EP4487263A1},
with programmable inherent bosonic modes.
A simple comparative analysis showing that such hybrid resonator-qubit architectures require 
fewer resources for simulating bosonic modes is shown in Table~\ref{table:coding_efficiency}. It should be noted that
even the polynomial overhead shown in Table~\ref{table:coding_efficiency} for binary mapping is clearly prohibitive when it
comes to implementations on NISQ devices. 

\begin{figure}[]
\centering
\includegraphics[width=\columnwidth]{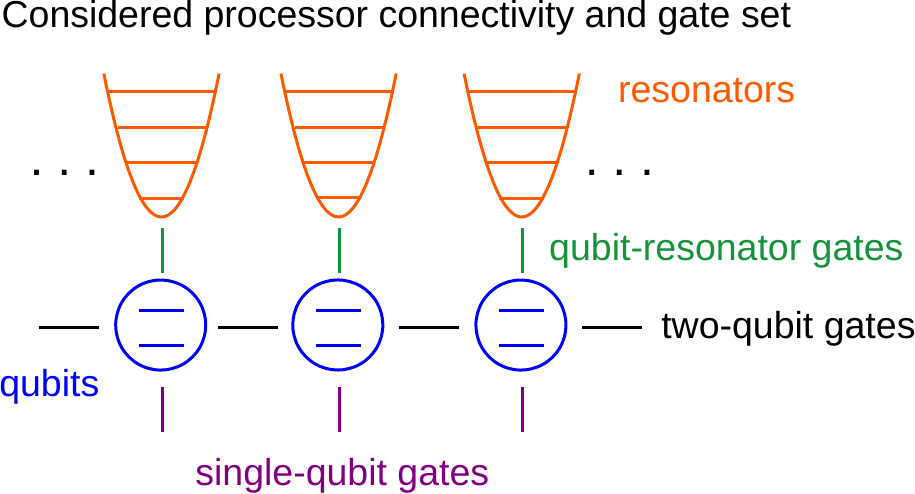}
\caption{%
An example of a connectivity which allows for low depth time propagation of most common types of system-boson models
(see Sec.~\ref{sec:models}).
Each qubit needs to couple to only one resonator and two other qubits.
In the quantum simulation, qubits represent either spin or electronic degrees of freedom,
while resonators represent bosonic modes.
}
\label{fig:connectivity}
\end{figure}

In this work, we develop quantum algorithms for a superconducting quantum computer that uses
microwave resonators as computational units~\cite{Mezzacapo2014, Langford2017, Liu2024, Crane2024, Shapiro2024, Vu2025}.
The approach combines conventional gate-based quantum computing with controlled interactions between qubits and microwave resonators.
This approach enables the representation of bosonic modes by resonators and the
use of the native Jaynes-Cummings interactions as a gate primitive.
Our paper adds to earlier works 
by introducing circuit decompositions and time propagation algorithms for most common system-boson models.
We also develop the idea of resonator noise utilization in digital hybrid quantum computing.
The paper also shows an explicit implementation of a time propagation of a spin-boson model on the IQM Resonance cloud. 

\begin{table*}
\begin{center}
\begin{tabular}{ c | c | c | c }
{\bf code} & \ {\bf included energy levels} \ & \ {\bf hardware components} \ & \ {\bf entangling gates} \ \\ [0.5ex]
\hline
resonator-qubit & arbitrary & $1$ qubit, $1$ resonator & $2$ \\
\hline
unary (all-qubit) & $d$ & $1 + d$ qubits & $O(d)$ \\
\hline
binary (all-qubit) & $d$ & $1 + \log_2(d)$ qubits & $O[d^2\log(d)]$ \\
\hline
\end{tabular}
\end{center}
\caption{Estimates for the qubit count and entangling-gate count when implementing
the spin-boson operator $\exp\left[\textrm{i}\phi\sigma_x(\hat b^\dagger + \hat b)\right]$
using resonator-qubit quantum circuits (Sec.~\ref{sec:qubit_resonator_circuits})
versus standard (all-qubit) unary and binary codings and including $d$ levels of the bosonic mode~\cite{Sawaya2020}.
The table highlights the key differences in the required
hardware resources and computational cost for each approach:
The resonator-qubit approach requires fewer hardware resources and shorter quantum algorithms, 
particularly for large $d$~values. It should be noted that the shown scalings of the all-qubit algorithms
are upper estimates and may be optimized, for example, by the use of additional (auxiliary) qubits~\cite{Liu2024}.
}
\label{table:coding_efficiency}
\end{table*}

We consider systems consisting of spins or electrons.
In the quantum simulation, the state of electrons
is described via occupations of fermionic orbitals,
which are either empty or filled (or certain superpositions of them).
The orbital occupations are then mapped directly to the state of qubits.
The same applies also for the state of modeled spins: it is mapped directly to the state of qubits.
This similarity holds even for the proposed time-propagation algorithms,
which are of the same form when implemented using the corresponding (spin or fermion) swap networks.

An important goal of this work is to develop feasible schemes for scaling resonator-qubit quantum simulations to larger systems.
For this purpose, we propose a straightforward architecture that is based on a simple unit cell with
nearest-neighbor qubits, each coupled to one resonator, see Fig.~\ref{fig:connectivity}.
We show that
for this device connectivity (which can of course be picked from a more general device connectivity)
and certain (common) system-boson models, we can write down time-evolution algorithms with low circuit depths.
These include, for example, linear and quadratic system-bath couplings in common spin-boson and electron-phonon models.
Our scheme does not require gates between the resonators, nor are direct
resonator drives (displacement gates) needed.

Another important goal of our work is to develop schemes where one uses
the noise in quantum computers as a resource~\cite{Lloyd1996, Tseng2010, Rost2020, Sun2021, Guimaraes2023, Leppakangas2023, Ma2024}.
In particular, we show that when using the proposed gate set and algorithms, 
resonator dissipation can be directly mapped to broadening of the bosonic modes in the simulated system.
This approach does not work for conventional all-qubit quantum simulations,
since noise of qubits representing bosons does not map~\cite{Fratus2022} to any established open quantum system physics.
This demonstrates the potential to leverage resonator noise as a resource rather than treating it as a limitation.
This mapping can also provide a clear intuition for
understanding the effects of resonator noise on the results of a resonator-qubit quantum simulation.

As first experimental realization, 
we present a quantum simulation of Jaynes-Cummings physics
on the IQM Resonance cloud platform. 
This example uses the high-fidelity resonator-qubit gates~\cite{Renger2025},
implemented using tunable couplers between a resonator and multiple qubits.
 A time propagation via Trotterization is presented, which solves the time evolution of the Jaynes-Cummings model,
utilizing higher excitation levels of the resonator.
While similar implementations of digital quantum simulations of spin-boson models have been presented earlier
in laboratory experiments~\cite{Langford2017},
our work presents the first demonstration on a commercial platform with cloud access.
This lays foundations for useful quantum simulations of bosonic systems on
publicly accessible hybrid quantum computers.

The paper is organized as follows. In Sec.~\ref{sec:what_to_solve}, we discuss the general form of
the models that can be solved efficiently using the presented resonator-qubit quantum algorithms.
We briefly review
several interesting classes of models, used in different fields of applied physics and chemistry.
We also go through general principles of gate-based quantum computing and digital quantum simulation.
In Sec.~\ref{sec:qubit_resonator_circuits}, we describe how
the earlier introduced different forms of spin-boson and electron-phonon couplings
can be decomposed using resonator-qubit gate primitives.
In Sec.~\ref{sec:swap_network}, we introduce efficient time propagation algorithms
based on a swap network optimized for typical system-boson models and restricted device connectivity.
In Sec.~\ref{sec:example}, we give an example that goes through all practical
steps needed for establishing a specific resonator-qubit quantum simulation. In
Sec.~\ref{sec:noise_mapping}, we analyze the effect of resonator dissipation and
introduce the mapping of resonator dissipation to broadening of the simulated bosonic modes.
In Sec.~\ref{sec:Experiment}, we present the experimental realization run on IQM Resonance.
Finally, a conclusion and outlook are given in Sec.~\ref{sec:conclusion}.

%---------------------------------------------------------------- What problem to solve and why ----------------------------------------------------------------
%---------------------------------------------------------------- What problem to solve and why ----------------------------------------------------------------
%---------------------------------------------------------------- What problem to solve and why ----------------------------------------------------------------

\section{What problem to solve and why}\label{sec:what_to_solve}
Our target models have the form of a system coupled to a bosonic bath. The system can be
interacting spins or electrons, whereas the bath can be phonons, photons, or generally any types of
bosonic excitations. Several interesting classes of models are discussed below.
The detailed quantum algorithms corresponding to these models are described in Secs.~\ref{sec:qubit_resonator_circuits}
and~\ref{sec:swap_network}.
Below, we will not discuss classical numerical methods that can be used to solve similar problems,
or the computational resources needed for doing that. For this, we refer the reader to recent literature, for example,
as presented in Refs.~\cite{Somoza2019, Tamascelli2019, Lambert2019, Vu2025, Tamascelli2018, Pleasance2020, Menczel2024}.

\subsection{Considered form of the Hamiltonian}
Although the quantum-computer gate set considered in this paper (Sec.~\ref{sec:qubit_resonator_circuits})
allows us to time propagate arbitrary
system-boson models~\cite{Liu2024}, we choose to consider only models that lead to short quantum
circuits.
These include composite systems of spins or electrons coupled to bosonic modes,
with the Hamiltonian
\begin{align}\label{eq:HamiltonianSimplified}
\hat H &= \hat H_\textrm{s} + \hat H_\textrm{b} + \hat H_\textrm{c} \, , 
\end{align}
where $\hat H_\textrm{s}$ describes the system, $\hat H_\textrm{b}$ the bosonic bath,
and $\hat H_\textrm{c}$ the coupling between them.
The bath Hamiltonian has the form (or is brought into the form)
of non-interacting bosons,
\begin{align}\label{eq:HamiltonianSimplified_bath}
\hat H_\textrm{b} &= \sum_k \omega_k \hat b^\dagger_k \hat b_k  \, .
\end{align}
Here, bosons of mode~$k$ are created and annihilated by the operators $\hat
b_k^\dagger$ and $\hat b_k$, correspondingly. In this paper we set $\hbar = 1$.
The considered system-bath coupling has the form
\begin{align}\label{eq:HamiltonianSimplified}
\hat H_\textrm{c} &= \sum_{jk}\left(v_{jk}\hat C_j\hat b_k^\dagger + v^*_{jk}\hat C_j^\dagger \hat b_k \right) \, ,
\end{align}
where $\hat C$ operates on the system only. For the most
efficient execution of the algorithm, $\hat H_\textrm{s}$ and $\hat C$ include products of up to two spin- or
fermion-operators, including fermion density-density operators.

\subsection{Use cases}\label{sec:models}
The variety of systems that can be modeled using the Hamiltonian introduced above is
vast and cannot be addressed here in all generality. Below we discuss some interesting use cases in
fields of photovoltaics, solid-state physics, quantum transport, quantum optics, and molecular chemistry.
A summary of the use cases and their key ingredients is given in Table~\ref{table:spin_boson_mapping}.
\begin{table*}
\begin{center}
\begin{tabular}{ c | c | c}
{\bf use cases} & \  {\bf represented by qubits} & \  {\bf represented by resonators}  \  \\ [0.5ex]
\hline
photovoltaics, quantum transport & excitons, quasiparticles, electrons & phonons, photons  \\ 
\hline
quantum optics & \   atoms, molecules  \  & photons \\ 
\hline
radical molecules (RPA) & important orbitals, active space & other orbitals, inactive space \\ 
\hline
\end{tabular}
\end{center}
\caption{%
Examples of models which may be solved efficiently using the resonator-qubit quantum algorithms and processor.
         Radical molecules~\cite{Shirazi2024} represent an example where the original fully electronic global system
         is mapped, under the random-phase approximation~(RPA), to a problem of interacting electrons coupled to fictitious bosonic modes.
         }\label{table:spin_boson_mapping}
\end{table*}

\subsubsection{Models for organic photovoltaics, solid-state physics, quantum transport}\label{sec:transport}
In research of materials for organic photovoltaics~\cite{Kippelen2009,
Ostroverkhova2016, Ishizaki2012, Huelga_2013, Campaioli2021},
a central quantity is the quasiparticle transport which is influenced by
coupling to vibrational modes (bosonic excitations) of the molecule.
The quasiparticles can be localized excitons (modeled as spins) or
dissociated electron-hole pairs. Spin-boson and electron-phonon models can be
used to make predictions of important properties such as charge transport and energy conversion.

Perhaps the simplest example model here is the Holstein model~\cite{Holstein1959} of exciton transport,
with Hamiltonian
\begin{align}\label{eq:general_type}
\hat H &= \sum_{i }\frac{\epsilon_{i}}{2}\sigma^i_z + \sum_{i<i'}(h_{ii'}\sigma^i_+ \sigma^{i'}_- + h^*_{ii'}\sigma^{i'}_+ \sigma^i_- )  \nonumber \\
&+ \sum_{j}\sum_{k}\sigma_z^j\left( v_{jk}\hat b_k^\dagger + v^*_{jk}\hat b_k \right) + \sum_{k} \omega_k \hat b^\dagger_k \hat b_k\, ,
\end{align}
where the first two terms on the right-hand side correspond to the system Hamiltonian $H_\textrm{s}$,
with exciton energies $\epsilon_i$ and hopping amplitudes $h_{ii'}$. The
coupling between the excitons and phonons is longitudinal, $\hat C_j = \hat C_j^\dagger =
\sigma_z^j$.
The Holstein model assumes that electron-hole bound states are local and the transport can
be treated to be neutral in electric charge.
A generalization of this is the Merrifield exciton model, where this condition is relaxed and electrons and holes are allowed
to be independent degrees of freedom, with strong mutual interaction, see for example Refs.~\cite{Binder2013, Campaioli2021}.

Similar models are also central for describing transport in solid state physics.
Phenomena such as formation of polarons~\cite{Froehlich1954} or Cooper pairs~\cite{Cooper1956}
are prominent examples.
The Hubbard-Holstein~\cite{Holstein1959} model describes interacting electrons on a discrete lattice with coupling to phonons.
The Hamiltonian has the form
\begin{align}\label{eq:general_type}
\hat H &= \sum_{i} \epsilon_{i} \hat n_i  + \sum_{i<i'} \epsilon_{ii'} \hat n_i \hat n_{i'}  + \sum_{i<i'}(t_{ii'} \hat c^\dagger_i \hat c_{i'} + t^*_{ii'} \hat c^\dagger_{i'} \hat c_i)  \nonumber \\
&+ \sum_{jk}\hat n_j \left( v_{jk}\hat b_k^\dagger + v^*_{jk}\hat b_k \right) + \sum_k \omega_k \hat b^\dagger_k \hat b_k\, ,
\end{align}
where $\hat c^{(\dagger)}$ is the electron annihilation (creation) operator and $\hat n_i = \hat
c_i^\dagger \hat c_i$ is the electron number operator. The electron-phonon coupling operator is
$\hat C_j = \hat n_i$.
The coupling to phonons is usually assumed to be local, i.e., $v_{jk}\neq 0$ only for $j=k$.
Such model is computationally more friendly to implement classically as well as quantum mechanically.
Letting $v_{jk}\neq 0$ also for $j\neq k$ generalizes the Hubbard-Holstein model to non-local couplings, for example,
accounting for a possibly wide spatial spread of individual bosonic modes.
Another generalization is the Fröhlich model~\cite{Froehlich1954}, which describes
electron-phonon coupling with the coupling Hamiltonian of the form
\begin{align}\label{eq:Froehlich}
\hat H_\textrm{c} &= \sum_{ijk}\left(v_{ijk}\hat c_i^\dagger \hat c_j \hat b_k^\dagger + \textrm{H.c.} \right)\, .
\end{align}
Here, the interaction comes with hopping (when $i\neq j$) between different sites/fermionic orbitals.
Furthermore, it should be mentioned that
such models can also describe transport in optoelectronic devices, where charge transport occurs
with emission/absorption of photons (instead of phonons). An example here is photon-assisted
single-charge tunneling~\cite{Ingold1992,Grimsmo2016, Westig2017, Grimm2019, Peugeot2021}, which
can be used to create non-classical states of light.

\subsubsection{Quantum optics and polariton chemistry}\label{sec:quantum_optics}
Quantum optics describe how light is emitted and absorbed by atoms and
molecules~\cite{WallsMilburn}. Topics such as collective emission (superradiance) and enhanced
absorption are interesting for the field of photovoltaics and quantum batteries.
Also, the possibility to engineer strong light-matter coupling has established the field of
polariton chemistry, where through the use of cavities one can inhibit, enhance, or steer certain
processes, by affecting the coupling to specific vibrational excitations~\cite{Schwarz2011}.
Here, the limit of ultra-strong coupling~\cite{Forn_Diaz_RevModPhys2019, Qin2024} can offer novel ways to control chemical bonds
and reactions~\cite{FriskKockum2019}.

A central model here is the Dicke model, which describes coupling of multiple spins with a cavity, or multiple cavities,
\begin{align}\label{eq:DickeModel}
\hat H = H_\textrm{s} + \sum_{ik}\sigma_x^i\left( v_{ik}\hat b_k^\dagger + v_{ik}^*\hat b_k \right) + \sum_k \omega_k \hat b^\dagger_k \hat b_k  \, .
\end{align}
Here, photons of cavity~$k$ are created and annihilated by the operators $\hat
b_k^\dagger$ and $\hat b_k$, respectively.
In the Dicke model, the coupling is transverse and thereby causes transitions between atomic or molecular energy levels,
described by the system Hamiltonian $\hat H_\textrm{s} = \sum_i\epsilon_i \sigma_z^i/2$.
The case of one atom and one bosonic mode is called the quantum Rabi (QR) model.
It should be mentioned that the system can also be generalized, for example, to Dicke-Ising model~\cite{Shapiro2024},
where $H_\textrm{s}$ can describe interactions between the atoms, for example,
through dipole-dipole coupling.

The Tavis-Cummings model describes the interaction of multiple spins with a radiation field
under the rotating-wave approximation, which conserves the number of excitations
\begin{align}\label{eq:TavisCummings}
\hat H = H_\textrm{s} + \sum_{ik}\left(v_{ik}\sigma_-^i \hat b_k^\dagger + v_{ik}^* \sigma_+^i \hat b_k \right) + \sum_k \omega_k \hat b^\dagger_k \hat b_k  \, .
\end{align}
The case of a single bosonic mode is called the Jaynes-Cummings (JC) model, which plays a
central role in our work, since it describes the natural interaction between a superconducting
transmon qubit~\cite{Koch2007, Blais2021_RevModPhys} and a microwave resonator. It will be the key tool for creating
qubit-resonator gates.

\subsubsection{Fully electronic global systems under random-phase approximation: Radical molecules}\label{sec:RadicalMolecules}
Similar models of electrons coupled to bosonic modes
can also be derived for fully electronic systems when the random-phase
approximation~\cite{Bohm1953} (RPA) is employed.
These include problems in quantum chemistry~\cite{Dunning1967, Shirazi2024}.
The RPA maps electron-electron interactions to interactions
between electrons and fictitious bosonic modes.

An example is the case of radical
molecules (radicals), where certain orbitals hold unpaired electrons, which makes them
chemically extremely reactive. Such molecules play central roles in biology and are of high pharmaceutical
interest.
A notable example is an enediyne compound, which is used in antitumor antibiotics due to its
exceptional reactions in the presence of DNA.
Two key orbitals are characterized by occupancies close to~1.
For such diradicals, the effect of the other orbitals may be included using the RPA~\cite{Shirazi2024},
corresponding to introducing coupling to a fictitious bosonic bath.

The Hamiltonian for a diradical molecule under RPA has the form
\begin{align}\label{eq:general_type}
\hat H &= \sum_{ij} t_{ij} \hat c_i^{\dagger} \hat c_j + \frac{1}{2}\sum_{ijj'i'} h_{ijj'i'} \hat c_{i}^{\dagger}\hat c_{j}^{\dagger} \hat c_{j'} \hat c_{i'} \nonumber \\
&+ \sum_{ijk}v_{ijk}\hat c_i^\dagger \hat c_j \left( \hat b_k^\dagger + \hat b_k \right) + \sum_k \omega_k \hat b^\dagger_k \hat b_k\, ,
\end{align}
where the second term on the right-hand side corresponds to the Coulomb interaction of electrons
in the key orbitals (active space) $i,j \in \{\vert
1\uparrow\rangle, \vert 1\downarrow\rangle, \vert 2\uparrow\rangle, \vert 2\downarrow\rangle\}$.
We see that the coupling between the core and
the fictitious bosonic modes is quadratic in the system operators and linear in the bosonic
operators. The form of the coupling is similar but slightly different from the Fröhlich model, Eq.~(\ref{eq:Froehlich}).
The details of the bosonic bath are determined by the excitation spectrum
of the original electronic bath (inactive space)~\cite{Shirazi2024}.

\subsection{Solving on quantum computers}
The above models can be solved using gate-based quantum computers. On the
quantum computer, states of the spins or fermionic orbitals can be represented directly by qubits. In
the case of conventional all-qubit quantum computers, the state of the bosonic modes needs to be
represented using special mappings, for example unary or binary coding~\cite{Sawaya2020}. However,
if also physical bosonic modes, such as microwave resonators, are used as computational elements, one
may represent the state of the bosonic mode directly by the physical modes~\cite{Mezzacapo2014, Langford2017, Liu2024, Vu2025},
as will be considered in the following.

\subsubsection{Optimal frame of time evolution}\label{sec:OptimalFrame}
Typically, microwave resonator frequencies in circuit-QED setups~\cite{Blais2021_RevModPhys}
are rarely tunable, meaning that direct resonator frequency shift gates, corresponding to bosonic-mode rotations,
cannot be implemented or the resulting gate is slow.
One possible implementation, using auxiliary qubits, is described in Sec.~\ref{sec:resonator_gates},
but it can lead to a significant increase in the complexity of the algorithm.
Instead, we want to time evolve models that have no free evolution of the bath, i.e., $\hat H_b=0$.
For the models presented above,
this is not usually the case initially, but we can achieve this form by transforming these models
to the rotating frame of the bosonic modes. Here the considered Hamiltonians have the form
\begin{align}\label{eq:HamiltonainInteractionPicture}
\hat H(t) = \hat H_\textrm{s} + \sum_{jk}\left(v_{jk}\hat C_j\hat b_k^\dagger e^{\textrm{i}\omega_k t} + v^*_{jk}\hat C_j^\dagger \hat b_k e^{-\textrm{i}\omega_k t} \right) \, .
\end{align}
We see that the modeled boson frequencies $\omega_k$ appear only as phases in the coupling terms.
Such phases can be implemented efficiently on standard gate-based quantum computers,
for example, using phased rotate-X (PRX) gates
as described in Sec.~\ref{sec:phases}.

\subsubsection{Time propagation}\label{sec:Trotterization}
Time evolution on quantum computer is based on discretizing the simulated time of the total
time-evolution operator $\hat U(t)$. For us, the Hamiltonian at each time step is different and our
starting point is the expansion
\begin{align}\label{eq:Trotter}
\hat U(t) &\approx e^{-\textrm{i} \tau \hat H((n-0.5)\tau) } \ldots e^{-\textrm{i} \tau \hat H((3/2)\tau)} e^{-\textrm{i} \tau \hat H((1/2)\tau)} \, ,
%\hat U(t) &\approx e^{-\textrm{i} \tau \hat H((n-1)\tau) } \ldots e^{-\textrm{i} \tau \hat H(\tau)} e^{-\textrm{i} \tau \hat H(0)} \, ,
\end{align}
where the total simulated time $t=n\tau$.
We do here time discretization according to midpoint values.
In the next step, we apply the Lie-Trotter (first-order) or Suzuki-Trotter (higher-order) formulas~\cite{Childs2021},
often called simply as orders of ``Trotterization'', to
the individual unitary operations, which in the case of the first-order formula means
approximating
\begin{align}\label{eq:Trotter_1st}
e^{-\textrm{i} \tau\hat H(m\tau)} &\approx \Pi_{i=1}^N e^{-\textrm{i} \tau \hat H_i(m\tau)}   \, ,
\end{align}
where $m+1/2$ is the number of the Trotter step and the partial Hamiltonians satisfy
\begin{align}
\hat H(m\tau)=\sum_{i=1}^N \hat H_i(m\tau) \, .
\end{align}
The motivation for introducing partial Hamiltonians is to be able to efficiently implement the
unitary operations~$\exp\left[-\textrm{i} \tau \hat H_i(m\tau)\right]$ on hardware.
Usually they correspond to individual single and two-body (or multi-body) terms.
In the second-order Trotterization, the time evolution is implemented as
\begin{align}\label{eq:Trotter_2nd}
e^{-\textrm{i} \tau\hat H(m\tau)} &\approx \Pi_{i=1}^N e^{-\textrm{i} \tau/2 \hat H_i(m\tau)} \Pi_{i=N}^1 e^{-\textrm{i} \tau/2 \hat H_i(m\tau)}   \, ,
\end{align}
which is preferred in our approach since the introduced resonator-qubit swap network (Sec.~\ref{sec:swap_network}) corresponds to this
order of Trotterization.

The Trotterization as well as the discretization of time (with time dependent Hamiltonian~$\hat H(t)$)
are approximations and come with errors,
whose forms are discussed in Appendix~\ref{sec:error_analysis}.

\subsubsection{Measurement of physical observables}
Central quantities to study in the considered models may be site-dependent excitation numbers,
transport, and/or various two- or multi-body correlation functions.
An example is the measurement of populations after an injection of an excitation at certain initial
site, the receptor. This is a typical procedure for models of
light-harvesting materials. The populations can be determined from averages of Pauli operators
describing the spins or fermionic orbitals, which can have the form
\begin{align}\label{eq:measure1}
P_i&(t) =  \frac{1}{2}\left[ 1 + \left\langle \sigma^i_z(t) \right\rangle \right] \, ,
\end{align}
The transport can also be estimated by
measuring correlation functions of the type
\begin{align}\label{eq:measure2}
& \left\langle \sigma^i_+(t) \sigma^j_-(t) \right\rangle \, .
\end{align}
Reliable estimations for such expectation values are obtained by averaging over thousands of
repetitions: averages have a sampling error
that scales like $1/\sqrt{N_\textrm{meas}}$, where $N_\textrm{meas}$ is the number of
measurements (shots).
It should be clear that also various
error mitigation methods can be introduced, for example, based on conservation of total excitation number.

For molecular systems the goal may be the measurement of properties such as
the singlet-triplet splitting, which can be inferred, for example, from the
oscillation period under simulated external drive. Other valuable information is
given by the one-particle (rdm1) and two-particles (rdm2) density matrices, $\langle\hat c_i^\dagger
\hat c_j \rangle$ and~$\langle\hat c_i^\dagger c_{i'}^\dagger \hat c_{j'} \hat c_{j} \rangle$,
which also correspond to measurement of Pauli-operator products of the
form~(\ref{eq:measure1}-\ref{eq:measure2}), with possible additional Pauli operators
to account for fermion-qubit mapping (e.g., Z operators if the Jordan-Wigner mapping was applied).
The obtained estimates for the rdms can be a useful
input for other material simulation methods, such as dynamical mean-field~\cite{Rozenberg1996} and
density-matrix embedding~\cite{Wouters2016, Sun_2020} theories.

In photonics, the properties of the light or correlations between the spins and light
may be at the center of analysis.
Expectation values such as
\begin{align}\label{eq:measure3}
& \left\langle \hat C_j(t) \hat b_k^\dagger(t) \right\rangle  
\end{align}
can be obtained by a simultaneous measurement of the qubit Pauli and resonator-quadrature
operators. Combined with the measurement of the boson numbers,
\begin{align}\label{eq:measure4}
& \left\langle \hat b_k^\dagger(t) \hat b_k(t) \right\rangle  \, ,
\end{align}
we can also estimate the total energy.
The measurement of arbitrary bosonic-mode properties, including full Wigner tomography,
can be done by combining resonator displacement pulses with entangling to additional probe qubits,
see for example Refs.~\cite{Hofheinz2009, Vlastakis2013, Langford2017, Kudra2022}.
In principle, these measurement algorithms can be implemented on hybrid quantum computers
using additional (auxiliary) qubits to effectively perform displacement gates (see Sec.~\ref{sec:resonator_gates})
as well as using them as the probe qubits.

%---------------------------------------------------------------- Resonator-qubit quantum circuits ----------------------------------------------------------------
%---------------------------------------------------------------- Resonator-qubit quantum circuits ----------------------------------------------------------------
%---------------------------------------------------------------- Resonator-qubit quantum circuits ----------------------------------------------------------------

\section{Resonator-qubit quantum circuits}\label{sec:qubit_resonator_circuits}
In this section, we define our elementary qubit-resonator gate and principles of decomposing
arbitrary system-boson coupling models.
A hybrid quantum algorithm for simulating the Rabi model, i.e., one spin coupled to one bosonic mode,
has been demonstrated experimentally in Ref.~\cite{Langford2017} also
using a superconducting qubit and a microwave resonator, proposed originally in Ref.~\cite{Mezzacapo2014}.
Our paper builds on these results, particularly by generalizing the algorithm to arbitrary system-boson models
assuming the considered device connectivity and gate set.
Also to note is the
recently proposed resonator-qubit algorithms for simulating the Dicke-Ising model in Ref.~\cite{Shapiro2024}
and the discussion of other hybrid oscillator-qubit quantum processors, corresponding algorithms,
and the universality of gate sets in Refs.~\cite{Liu2024, Crane2024}.

\subsection{Jaynes-Cummings gate as gate primitive}\label{sec:JaynesCummingsGate}
\begin{figure}[]
\centering
\includegraphics[width=\columnwidth]{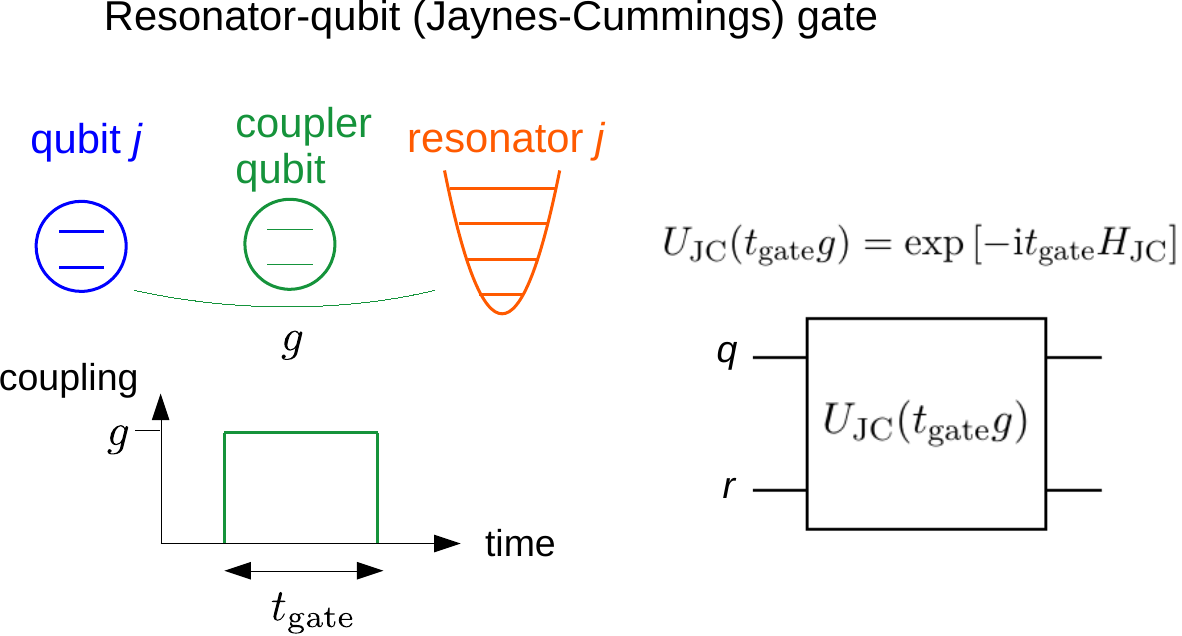}
\caption{%
Our elementary gate to decompose system-boson interactions is the Jaynes-Cummings (JC) gate. The JC
gate is based on the physical qubit-resonator interaction described by the JC Hamiltonian $\hat
H_\textrm{JC} = g(\sigma_- \hat b^\dagger + \sigma_+ \hat b)$, where $g$ is the
qubit-resonator coupling. 
Switching on this interaction for a time $t_\textrm{gate}$, e.g.,
using a tunable coupler~\cite{Renger2025} (Sec.~\ref{sec:Deneb_and_tunable_coupler}), we perform the gate $\hat U_\textrm{JC}(t_\textrm{gate}
g) \equiv \exp\left[-\textrm{i} t_\textrm{gate} \hat H_\textrm{JC}\right]$.
On the right is shown our circuit representation of this gate.
}
\label{fig:JC_gate}
\end{figure}

A native Jaynes-Cummings (JC) interaction between a qubit and a resonator is integrated as a gate primitive~\cite{Mezzacapo2014,
Langford2017}.
This is described by the JC Hamiltonian
\begin{align}\label{eq:JaynesCummings}
\hat H_\textrm{JC} &= g\left(\sigma_-  \hat b^\dagger + \sigma_+ \hat b\right) \, ,
\end{align}
where $g$ is the resonator-qubit coupling.
Switching on the interaction for a time
$t_\textrm{gate}$, e.g., using a tunable coupler, we perform the resonator-qubit gate
\begin{align}
\hat U_\textrm{JC}(t_\textrm{gate} g) &\equiv \exp\left[-\textrm{i} t_\textrm{gate} \hat H_\textrm{JC}\right] \, .
\end{align}
Our circuit representation of this gate is shown in Fig.~\ref{fig:JC_gate}, where ``$q$'' refers to
the qubit and ``$r$'' to the resonator. We often drop the argument $t_\textrm{gate} g$, for
simplicity. Together with a universal qubit gate set, we can efficiently
decompose all the discussed interactions between spins (or electrons) and bosonic modes (Sec.~\ref{sec:models}). According to
Ref.~\cite{Liu2024}, such gate set is even universal in the total system-boson Hilbert space.

\begin{figure}[]
\centering
\includegraphics[width=0.8\columnwidth]{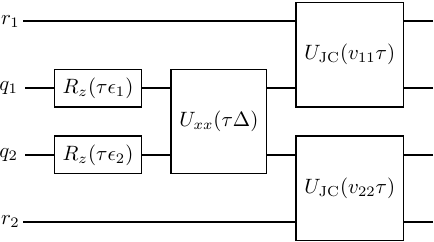}
\caption{%
  An example circuit implementing interactions between two spins and two bosonic modes. Here
  the modeled spin splittings $\epsilon_i\sigma_z^i/2$ are implemented via single-qubit Z-rotations
  and the spin-spin coupling $\Delta\sigma_x^1\sigma_x^2$ via two-qubit gate $\hat U_{xx}$. The
  spin-boson couplings $v_{ii}(\sigma_-^i b_i^\dagger + \sigma_+^i b_i)$ are implemented by the
  Jaynes-Cummings (JC) gates. }
\label{fig:example_QRA}
\end{figure}

We note that in our definition of the JC gate, the qubit and the resonator are assumed to be
on-resonance during the execution of the gate. An alternative definition would be to allow for a
finite off-resonance in order to simultaneously implement qubit Z-rotations~\cite{Mezzacapo2014,
Langford2017}, which can sometimes be advantageous but leads to a more complicated calibration process.
In this paper, the unitary operations describing spin splittings or electron orbital energies
will then be applied separately by single-qubit Z-rotations.
An example circuit of one Trotter step modeling interactions
between two spins and two bosonic modes is shown in Fig.~\ref{fig:example_QRA}.

\subsection{Decompositions of linear couplings (coupling types~$\sigma \hat b$, $\hat n \hat b$)}\label{sec:Decomposing}
The JC gate natively implements a coupling of the form $\sigma_- \hat b^\dagger +
\sigma_+ \hat b$. If the system-boson coupling in the simulated  model is not of this form, it needs
to be decomposed into two or more JC gates. Here, we consider the case of linear
couplings, by which we mean that the coupling operator~$\hat C$ is a linear operator in terms of
Pauli operators or electron-number operators.

\subsubsection{Quantum Rabi gate}\label{sec:Decomposing_QR}
\begin{figure}[]
\centering
\includegraphics[width=0.9\columnwidth]{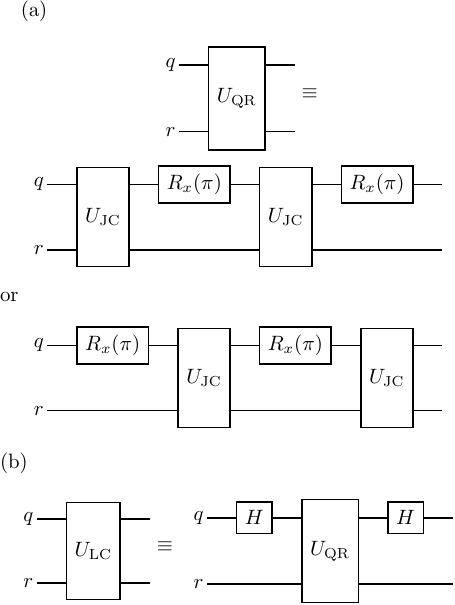}
\caption{Example circuits implementing (a) 
the quantum-Rabi gate $\hat U_\textrm{QR}(t_\textrm{gate} g) = \exp\left[-\textrm{i}t_\textrm{gate} g\sigma_x(\hat b^\dagger + \hat b) \right]$,
and (b) the longitudinal-coupling gate  $U_\textrm{LC}(t_\textrm{gate} g) = \exp\left[-\textrm{i}t_\textrm{gate} g \sigma_z(\hat b^\dagger + \hat b) \right]$.
In (a) the order between the JC gates and X-flips can be changed.
In particular, in the second-order Trotterization of time propagation, both orderings will appear (see Fig.~\ref{fig:example_fig2}).
}
\label{fig:example_QR}
\end{figure}

One possible form of the model coupling is $v\sigma_x(\hat b^\dagger + \hat b)$, which appears, for
example, in quantum optics and is there called the quantum Rabi (QR) model
(Sec.~\ref{sec:quantum_optics}). To implement this, we need to add the counter-rotating terms to the
JC Hamiltonian, i.e., we need to add
\begin{align}
\hat H_\textrm{AJC} = g\left(\sigma_+ \hat b^\dagger + \sigma_- \hat b\right)\, .
\end{align}
Since $\hat H_\textrm{AJC} = \sigma_x \hat H_\textrm{JC} \sigma_x$, the needed computational
processes is implemented 
by surrounding the JC gate by $R_x(\pi)$ (i.e., X) gates~\cite{Mezzacapo2014,Langford2017,Shapiro2024}.
For the implementation of the full QR gate,
\begin{align}
\hat U_\textrm{QR}(t_\textrm{gate} g) =\exp\left[-\textrm{i}t_\textrm{gate} g\sigma_x(\hat b^\dagger + \hat b) \right] \, ,
\end{align}
we can then use the computation sequence that sequentially applies
interactions $\hat H_\textrm{JC}$ and $\sigma_x \hat H_\textrm{JC} \sigma_x$, as shown in Fig.~\ref{fig:example_QR}(a).
Here the order of implementing the two contributions is not fixed.
Indeed, in the second-order Trotterization of time propagation, both orderings will appear.
Such constructions are not exact and introduce a Trotter error, which is discussed more detailed in
Appendix~\ref{sec:error_analysis}.

\subsubsection{Longitudinal coupling gate}\label{sec:longitudinal_coupling}
We can now continue with this approach and start creating also other forms of linear couplings. In
the Holstein models of the exciton and quasiparticle transport (Sec.~\ref{sec:transport}), the
couplings have the form $\sigma_z(\hat b^\dagger + \hat b)$ or the fermionic equivalent $\hat
n (\hat b^\dagger + \hat b)$. Since electronic number operators map to spin-Z operators
(via the Jordan-Wigner transformation), both of these couplings correspond to gate operations of type
\begin{align}
U_\textrm{LC}(t_\textrm{gate} g)=\exp\left[-\textrm{i}t_\textrm{gate} g \sigma_z(\hat b^\dagger + \hat b) \right] \, ,
\end{align}
One possible computational sequence to create this is to surround the QR gate
by Hadamard gates, effectively changing the qubit-X operator to Z-operator.
This decomposition is shown in Fig.~\ref{fig:example_QR}(b).
The same effect can be achieved by surrounding the QR gate by $\pi/2$ rotate-Y gates.
It is clear that with a similar
construction, a coupling to an arbitrary-direction spin-operator can be constructed.

\subsection{Decompositions of quadratic couplings (coupling types~$\sigma_i \sigma_j \hat b$,~$\hat c_i^\dagger \hat c_j \hat b$)}\label{sec:Decomposing2}
JC gates together with a complete set of qubit gates lets us decompose also other forms
system-boson interactions. In Sec.~\ref{sec:models}, we showed classes of models which included
quadratic coupling operators, by which we mean operators~$\hat C$ that include pairs of spins or electronic degrees of freedom.

\begin{figure}[]
\centering
\includegraphics[width=0.55\columnwidth]{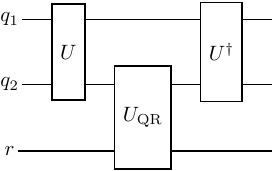}
\caption{A formal circuit implementing quadratic system-boson interactions of the form $\hat S (\hat b^\dagger + \hat b)$.
Here we use a unitary operation $\hat U$ which satisfies $\hat S = \hat U^\dagger \sigma_x \hat U$.
The X-gates of the QR gate operate on the lower qubit ($q_2$).
}
\label{fig:example_general}
\end{figure}

\begin{figure*}
\centering
\includegraphics[width=1.3\columnwidth]{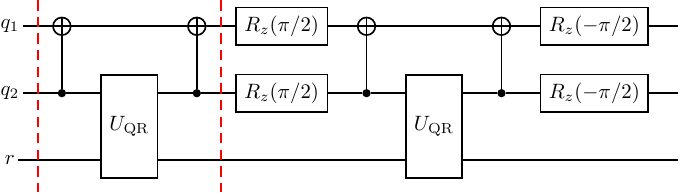}
\caption{Example of circuit creating system-boson interactions of a reactive molecule with
  electron-boson coupling of type $H = \hat c_1^\dagger \hat c_2 (\hat b^\dagger + \hat b) +
  \textrm{H. c.}$, see Sec.~\ref{sec:RadicalMolecules}.
  Similar circuits can also describe interaction between charge transport and emission/absorption of light~\cite{Crane2024}.
  When replacing the quantum-Rabi (QR) gates by Jaynes-Cummings gates, we instead
  obtain the (Fröhlich) polaron model with electron-phonon coupling of type $\hat H = \hat c_1^\dagger
  \hat c_2 \hat b^\dagger + \hat c_2^\dagger \hat c_1 \hat b$. }\label{fig:radical_decompostions}
\end{figure*}

We start by formally showing how to construct any coupling of the form $\hat S (\hat b^\dagger + \hat b)$, where
$\hat S$ 
is a two-spin operator, or generally a multi-spin operator. Assuming we have found a
decomposition of operator $\hat U$, which satisfies $\hat S = \hat U^\dagger \sigma_x \hat U$,
the formal construction of interaction $\hat S (\hat b^\dagger + \hat b)$
looks like the one shown in Fig.~\ref{fig:example_general}.
A simple example is the implementation of the model coupling $H_\textrm{c} = \sigma^1_x \sigma^2_x
(\hat b^\dagger + \hat b)$, which is generated by the first part of the circuit shown in
Fig.~\ref{fig:radical_decompostions} (marked by the dashed lines). By surrounding this by Hadamard gates to both qubits changes the coupling to
the form $H_\textrm{c} = \sigma^1_z \sigma^2_z (\hat b^\dagger + \hat b)$.
A slightly more complex example is the implementation of the interaction $H = \left(\sigma^1_+
  \sigma^2_- + \sigma^1_- \sigma^2_+\right) (\hat b^\dagger + \hat b)$, which can be done by the
same construction as of interaction XX+YY between two spins, which corresponds to the full circuit
shown in Fig.~\ref{fig:radical_decompostions}. This corresponds to models of radical molecules
(Sec.~\ref{sec:RadicalMolecules}).
This decomposition has also been introduced in Ref.~\cite{Crane2024} in the context of
charge transport with emission/absorption of light.

Another possible coupling has the form $\left(\hat C \hat b_k^\dagger + \hat C^\dagger \hat b
\right)$, where $\hat C \neq \hat C^\dagger$. An example is the Fröhlich Hamiltonian describing
electron-phonon interaction, Eq.~(\ref{eq:Froehlich}). The formal approach introduced here remains valid,
assuming that there exists a unitary transformation $\hat U$ for which $\hat C = \hat U^\dagger
\sigma_- \hat U $. The only difference will be that the final circuit will be constructed around the
JC gates instead of QR gates. A possible way to search for the correct transformation is by first
studying the case where we replace the bosonic operators by complex numbers, i.e., inserting~$\hat
a=a$, in which case the JC gate corresponds to a (small-angle) X-rotation. In the next step, one
writes down a transformation that implements the resulting two-spin interaction. For example, for
$a\hat C + a^*\hat C^\dagger = a\sigma_-^1 \sigma_+^2 + \textrm{H.c.}$, we implement the full
spin-spin interaction by some construction of XX+YY interaction. Lastly, one goes back and replaces
X-rotations depending on the amplitudes~$a$ by the corresponding JC gates. The result corresponds to
the full circuit in Fig.~\ref{fig:radical_decompostions}, with replacing the QR gates by the JC
gates.

\subsection{Phases of coupling operators}\label{sec:phases}
As described in Sec.~\ref{sec:OptimalFrame}, we time evolve the model in the rotating frame of
bosonic modes, where the system-boson couplings have linearly increasing phases as a function of the simulated time, see
Eq.~(\ref{eq:HamiltonainInteractionPicture}). This means that for each Trotter step, we need to
implement slightly different phases of couplings. This can be done by inserting qubit
Z-rotations to both sides of the JC gate~\cite{Langford2017}, as described below.
Furthermore, rotate-Z gates can be implemented efficiently on standard gate-based quantum computers,
for example, by using a decomposition into two phased rotate-X (PRX) gates~\cite{McKay_2017},
$R_z(\varphi) = R_{\varphi/2}(\pi) R_{0}(\pi)$
[where PRX $R_\varphi(\theta)=R_z(\varphi)R_x(\theta)R_z(-\varphi)$
and we have neglected a global phase].

\begin{figure}[]
\centering
\includegraphics[width=0.85\columnwidth]{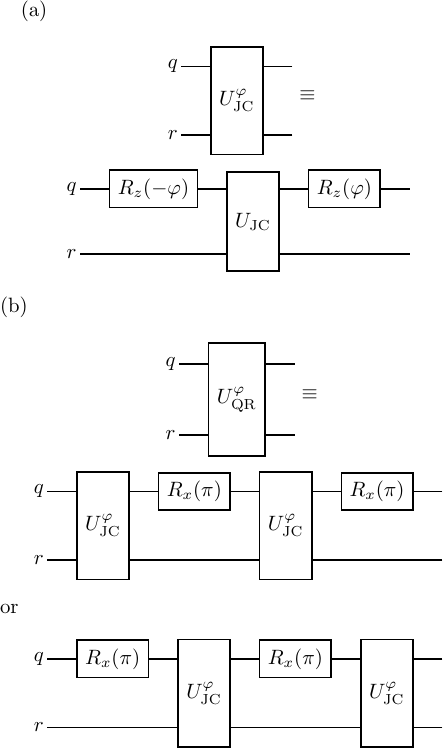}
\caption{(a) Definition of the phased Jaynes-Cummings gate~$\hat U^\varphi_\textrm{JC}$, which adds an arbitrary phase~$\varphi$ to the JC coupling amplitude.
(b) Definition of the phased quantum-Rabi (JC) gate~$\hat U^\varphi_\textrm{QR}$.
These decompositions are used to time evolve the corresponding models in the rotating frame of the bosonic modes (Sec.~\ref{sec:OptimalFrame}).
}
\label{fig:JC_with_phase}
\end{figure}

To see how this works, consider the case of time propagating the JC model, with linearly increasing phases
\begin{align}
  H_\textrm{JC}(\tau) &= g\sigma_-  b^\dagger e^{\textrm{i}\omega m\tau} + g\sigma_+ b e^{-\textrm{i}\omega m\tau} \equiv H_\textrm{JC}^{\omega m\tau}   \, ,
\end{align}
where~$\omega$ is the bosonic mode frequency and $m+1$ is the current Trotter step. Since
\begin{align}
H_\textrm{JC}^\varphi = e^{-\textrm{i}\varphi\sigma_z/2} H_\textrm{JC} e^{\textrm{i}\varphi\sigma_z/2} = R_z(\varphi) H_\textrm{JC} R_z(-\varphi) \, ,
\end{align}
the phases can be implemented by additional Z-rotations of angles~$\varphi = \pm \omega m\tau$. We
now mark such construction in the quantum circuits with the gate $\hat U^\varphi_\textrm{JC}$, as
shown in Fig.~\ref{fig:JC_with_phase}(a).
This construction can also be used generally. For example, in the final QR model time propagation algorithm,
we use circuits shown in Fig.~\ref{fig:JC_with_phase}(b).

\subsection{Pure resonator gates using auxiliary qubits}\label{sec:resonator_gates}
It is possible to implement arbitrary (pure) resonator gates using the
considered elementary gate set, i.e., the JC gate and a complete set of qubit gates~\cite{Liu2024}.
This can be done by using certain qubits as auxiliary (aux) qubits, i.e., not computational qubits.
Below we discuss three different types of such approach.

The simplest example is the creation of the bosonic-mode displacement gate
\begin{align}
\hat U &=\exp\left[-\textrm{i}t_\textrm{gate}g (\hat b^\dagger + \hat b)\right] \, . \nonumber
\end{align}
This can be realized
with the help of the LC gate~$\exp\left[-\textrm{i}t_\textrm{gate}g \sigma_z^\textrm{aux}(\hat b^\dagger + \hat b)\right]$
(Sec.~\ref{sec:longitudinal_coupling}), performed
with the auxiliary qubit initialized to the state~$\langle \sigma_z^\textrm{aux}\rangle=1$.

Another example is the implementation of the bosonic-mode rotation
\begin{align}
\hat U &=\exp\left[-\textrm{i}\phi \hat b^\dagger \hat b\right] \, . \nonumber
\end{align}
Ideally, this gate is implemented natively via dispersive interaction between an auxiliary qubit and resonator, i.e.,
by executing the JC gate off-resonance between the auxiliary qubit and the resonator.
If this is not supported, the dispersive interaction can also be simulated using JC gates and rotate-Z gates,
e.g., by Trotterizing the operation
$\exp\left[-\textrm{i} \tau_\textrm{aux} \hat H \right]$,
where $\hat H= \epsilon\sigma_z^\textrm{aux}/2 + g\sigma_-^\textrm{aux} \hat b^\dagger + g\sigma_+^\textrm{aux} \hat b$.
In the dispersive limit, $\epsilon\gg g$,
this corresponds to time propagating the resonator
according to the Hamiltonian $(g^2/\epsilon) \sigma_z^\textrm{aux}\hat b^\dagger \hat b$.
We then choose the ``auxiliary'' time $\tau_\textrm{aux} = \phi \epsilon / g^2 $ (and initialize
the auxiliary qubit to state~$\langle \sigma_z^\textrm{aux}\rangle=1$).

The third example is the implementation of higher-order bosonic terms, such as boson-boson interaction
\begin{align}
\hat U &=\exp\left[-\textrm{i}\phi (\hat b^\dagger \hat b)^2\right] \, . \nonumber
\end{align}
One possible way to implement this is to use the construction~\cite{Liu2024}
$\exp\left(\hat A\right)\exp\left(\hat B\right)\exp\left(-\hat A\right)\exp\left(-\hat B\right)\approx \exp\left([\hat A,\hat B]\right)$.
We have neglected here contributions of higher orders in $\hat A$ and $\hat B$.
If we choose $\hat A=\textrm{i}\sqrt{\phi/2}\sigma_x^\textrm{aux}\hat b^\dagger \hat b$ and $\hat B=\textrm{i}\sqrt{\phi/2}\sigma_y^\textrm{aux}\hat b^\dagger \hat b$,
we get the operation $\exp\left([\hat A,\hat B]\right)=\exp\left(-\textrm{i}\phi\sigma_z^\textrm{aux} (\hat b^\dagger \hat b)^2\right)$.
We can then implement the desired gate using the aux-qubit initialization $\langle \sigma_z^\textrm{aux}\rangle=1$.

%------------------------------------------------- Time propagation algorithms ----------------------------------------------------
%------------------------------------------------- Time propagation algorithms ----------------------------------------------------
%------------------------------------------------- Time propagation algorithms ----------------------------------------------------

\section{Time propagation algorithms}\label{sec:swap_network}
We will discuss a quantum algorithm for time propagation using resonator-qubit gates,
based on a hybrid swap network for common system-boson models.

We primarily discuss time propagation in systems where spins couple to bosonic modes. However, our results and
conclusions also apply to fermionic couplings, with the only difference being the replacement of standard SWAP
gates with fermionic SWAP (f-SWAP) gates~\cite{Kivlichan2018}. We will discuss
specific ways to reduce gate depth depending on the form of the Hamiltonian, but also quite general ways to optimize 
gate counts for swap networks exist in the literature~\cite{Hagge2020}.
Swap algorithms for hybrid resonator-qubit quantum computers have also been discussed in Refs.~\cite{Crane2024, Shapiro2024}.

The physical layout of the proposed device (Fig.~\ref{fig:connectivity}) is designed such that it supports an efficient
execution of swap network.
It should be noted that such network could also be picked from a more general
device connectivity.
Also, a more general device connectivity
may be more efficient for time propagating models with similar connectivity.

\subsection{Linear swap network}\label{sec:swap_linear}
\begin{figure}
\includegraphics[width=\columnwidth]{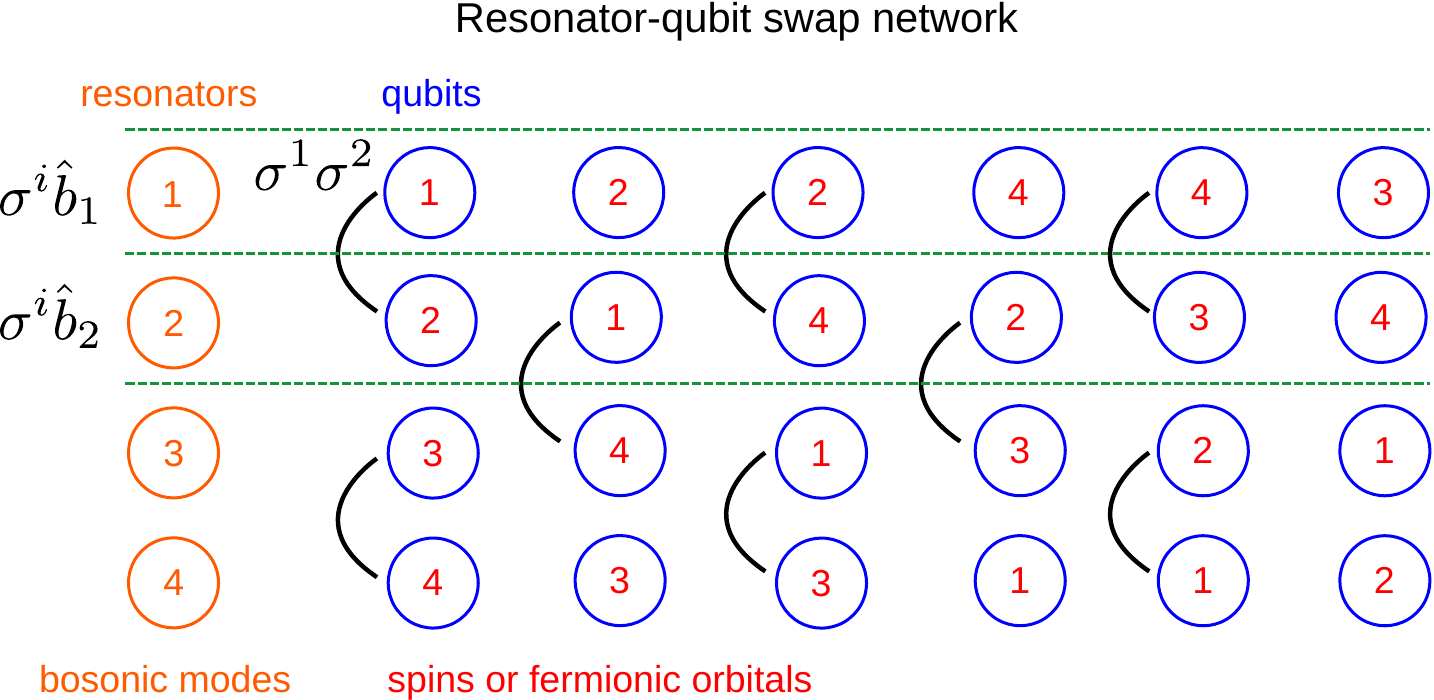}
\caption{Resonator-qubit swap network for considered system-boson models with linear couplings. In
  the algorithm, all pairs of spins (or fermionic orbitals) will be adjacent at least once, where the
  interactions between these spins can be applied. Additionally, every spin will be near each
  bosonic mode (blue horizontal slots) at least once, where corresponding system-boson interactions
  of the form $\sigma^i \hat b_k$ can be applied. }
\label{fig:figure_swap_network_linear}
\end{figure}
The swap network
can be used to efficiently implement all couplings in the Hamiltonian using nearest-neighbor
hardware connectivity. In practice, it implements the second-order Trotter formula introduced
in Sec.~\ref{sec:Trotterization}. It time evolves the global system over one Trotter time step~$\tau$.

The linear swap network we introduce first is similar to the common swap algorithm known in the
literature, see for example Ref.~\cite{Kivlichan2018}.
This algorithm is visualized in Fig.~\ref{fig:figure_swap_network_linear}.
Here, one swaps neighboring spins
(i.e.~swaps states of the neighboring qubits) in alternating layers of even and odd spins pairs. During
the swap network, all spins neighbor each other at least once within $N$ layers of swaps, where $N$
is the number of qubits. For us this is also the number of bosonic modes, $N=N_\textrm{b}$.
Arbitrary two-body interactions can be implemented at layers where the corresponding spins are
nearest neighbors. They can also be integrated into the SWAP gate itself~\cite{Kivlichan2018}.
Furthermore, within $2N_\textrm{b}$ layers of swaps, each spin neighbors each resonator at least
once, where the corresponding system-boson interactions can be applied. In this way, one can implement
all model interactions that include maximally two parties within a (Trotter) circuit depth that is
linear in~$N_\textrm{b}$.

Above we assume that the number of simulated spins equals the number of resonators $N_\textrm{b}$
(and qubits in the layout of Fig.~\ref{fig:connectivity}). However, it should be also noted that the
number of modeled spins~$n_\textrm{s}$ (or fermionic orbitals) can also be smaller than the number of
bosonic modes~$N_\textrm{b}$. In such simulations, one uses the $N_\textrm{b}-n_\textrm{s}$ ``free''
qubits to bus spins near each bosonic mode~\cite{Leppakangas2023}. The depth of the algorithm stays the same. Important to
note is also that since the bosonic modes are non-interacting, no gates between the
resonators are needed.

\subsection{Quadratic swap network}\label{sec:swap_quadratic}
\begin{figure}
\includegraphics[width=0.8\columnwidth]{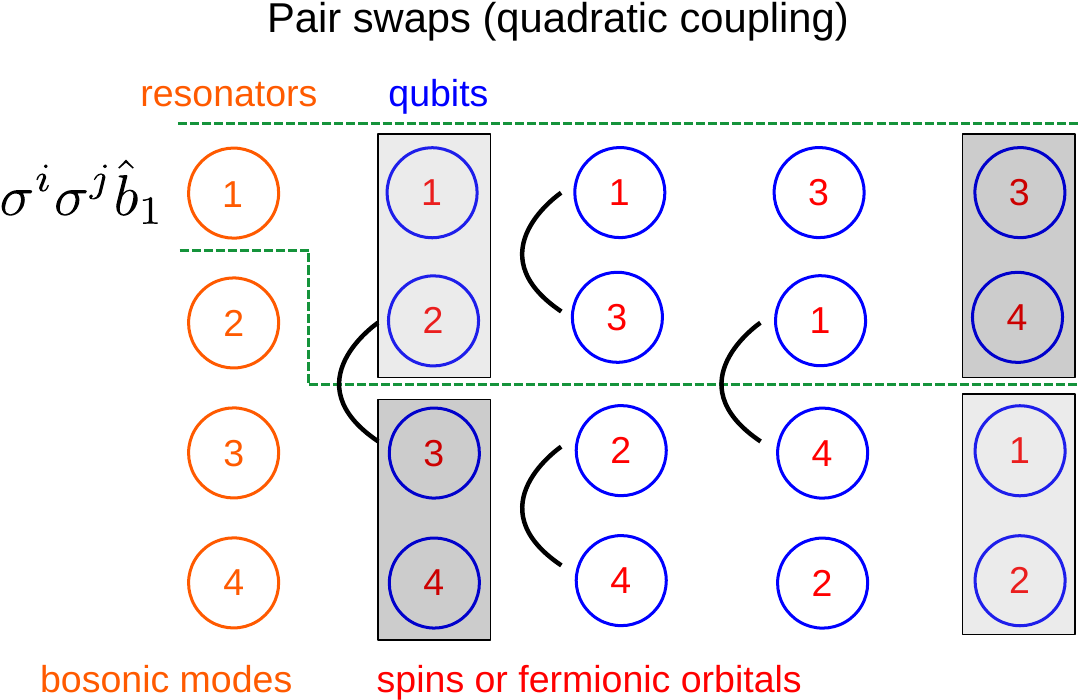}
\caption{For system-boson models with quadratic coupling operators, such as
  terms~$\sigma^i\sigma^j\hat b_k$, the resonator-qubit swap network is generalized so that spins
  are also transported in pairs. Each pair swap can be realized by four swap gates. When a pair is
  adjacent to a bosonic mode, either with the left or the right qubit, the corresponding coupling
  can be implemented. An example of a quantum circuit implementing a quadratic coupling is shown
  in Fig.~\ref{fig:radical_decompostions}. }
\label{fig:figure_swap_network_pair_swap}
\end{figure}

The quadratic network is an extension of the linear one, allowing for implementing
model interactions that include two arbitrary system operators coupling to arbitrary
bosonic modes. In particular, the fermion-boson interaction is always of this form, unless the two
fermionic operators describe the same fermionic orbitals, e.g., for $\hat c^\dagger_i \hat c_i (\hat
b_k + \hat b_k^\dagger)$, in which case it maps to single-spin Z-operator and linear network. The
naming ``quadratic'' refers at the same time to the degree of the system coupling as well as to the
scaling of the circuit depth of this network.

We assume now that the model has the most general case of connectivity, i.e., the model has
interactions of the form $\sigma^i \sigma^j \hat b $ for arbitrary
indices~$i,j,k$. For this case we first introduce the concept of an ``external'' swap network, which is
equivalent to the one described in Sec.~\ref{sec:swap_linear}, i.e., alternating layers of even and
odd swaps between the qubits. As earlier, all linear interactions can be implemented within this
network. The external network is run for $2N_\textrm{b}$ layers, so that every spin is adjacent to every
bosonic mode at least once through the iteration.

During the external swap network, after every odd layer and implementation of the corresponding
linear interactions, we run an “even-pair swap network” and an “odd-pair swap network”. These nested
pair-swap networks are needed to get each pair of spins in contact with every bosonic mode. The
even-pair swap network uses pair swap as described in Fig.~\ref{fig:figure_swap_network_pair_swap},
to swap around even pairs of spins~$(01), (23), (45),\ldots$, so that each pair gets coupled with
every bosonic mode at least once during the even-pair swap network, either with the left or the
right spin. After every layer of the even-pair swaps, the appropriate interactions are implemented.
The odd-pair swap network works similarly, only it moves around odd pairs, i.e., pairs~$(12), (34)$,
and so on. The external swap network is needed to reshuffle the pairs of fermionic modes.

Note that for models with fermionic orbitals instead of spins, we use the f-SWAP gates and implement
the interactions always when orbitals are adjacent to each other. In this case, no
additional parity terms, such as Jordan-Wigner transformations appear: the mapping from electron
hoppings to quadratic qubit interactions, such as $\sigma_+^i\sigma_-^j \hat b_k$, is direct.
It should also be noted that the sketched algorithm is for the most general case,
with maximal model connectivity, but can be simplified for models with restricted connectivity.

A summary of the scaling of the Trotter circuit depths and total gate counts for different model classes
is given in Table~\ref{table:spin_boson_circuitDepthsAndGateCount}.
\begin{table*}
\begin{center}
\begin{tabular}{ c | c | c | c }
{\bf Model class} & \  {\bf Use cases} \ & \ {\bf Number of couplings}  \ & \  {\bf Circuit depth}   \  \\ [0.5ex]
\hline
n. n., $\sigma^i \hat b_i$ & exciton transport & $N_\textrm{b}$ & $1$ \\ 
\hline
a.t.a., $\sigma^i \hat b_k$ & exciton transport, quantum optics & $N_\textrm{b}^2$ & $N_\textrm{b}$ \\ 
\hline
quadratic, n.n., $\sigma^i \sigma^{i+1} \hat b_{i}$ & electron-phonon coupling & $N_\textrm{b}$ & $1$ \\ 
\hline
quadratic. a.t.a., $\sigma^i \sigma^j \hat b_k$ & electron-phonon, radical molecules & $N_\textrm{b}^3$ & $N_\textrm{b}^2$ \\ 
\hline
1~spin, $N_\textrm{b}$~bosonic modes, $\sigma^0 \hat b_k$ & spin-boson model & $N_\textrm{b}$  & $N_\textrm{b}$\\ 
\hline
\end{tabular}
\end{center}
\caption{%
Scaling of the circuit depths as a function of the bosonic-mode number~$N_\textrm{b}$
for different models discussed in Sec.~\ref{sec:models}.
A nearest-neighbor model connectivity is marked as ``n.n.'' and all-to-all as ``a.t.a.''.
We assume that the number of spins $n_\textrm{s}=N_\textrm{b}$, except for the quadratic n.n.~$n_\textrm{s}=N_\textrm{b}+1$
and for the spin-boson model $n_\textrm{s}=1$.
}\label{table:spin_boson_circuitDepthsAndGateCount}
\end{table*}

%---------------------------------------------------------------- Example ----------------------------------------------------------------
%---------------------------------------------------------------- Example ----------------------------------------------------------------
%---------------------------------------------------------------- Example ----------------------------------------------------------------

\section{Example}\label{sec:example}
In this section we go through all practical steps of establishing a resonator-qubit quantum simulation
of a given system-boson model. We discuss steps related to definition of the model, calibration of
gates, time propagation algorithm, and optimization of the gate sequences.

\subsection{Model definition}
As a concrete example we choose the Dicke model including two bosonic modes and two spins
with the Hamiltonian
\begin{align}\label{eq:example_Hamiltonian}
\hat H &= \sum_{i=1}^2\frac{\epsilon_i}{2}\sigma_z^i +\epsilon_{12}\sigma^1_z\sigma^2_z \nonumber \\
  &+ \sum_{i=1}^2\sum_{k=1}^2v_{ik}\sigma_x^i\left( \hat b_k^\dagger + \hat b_k \right) + \sum_{k=1}^2 \omega_k \hat b^\dagger_k \hat b_k  \, .
\end{align}
We assume that all couplings $v_{ik}$ are finite.
We note that the choice of the inner-system Hamiltonian~$\hat H_\textrm{s}$ is here rather arbitrary
and does not affect the form of the final algorithm.

In the quantum simulation we time propagate the quantum state in the
rotating frame of the bosonic modes, where the same Hamiltonian has the form
\begin{align}\label{eq:example_Hamiltonian_rf}
\hat H(t) &= \sum_{i=1}^2\frac{\epsilon_i}{2}\sigma_z^i +\epsilon_{12}\sigma^1_z\sigma^2_z \nonumber\\
    &+ \sum_{i=1}^2\sum_{k=1}^2v_{ik}\sigma_x^i\left(  \hat b_k^\dagger e^{\textrm{i}\omega_k t} +  \hat b_ke^{-\textrm{i}\omega_k t} \right) \, .
\end{align}
We see that the bosonic-mode frequencies $\omega_k$ appear only as phases of the coupling terms.

\subsection{Decomposition of model interactions}
The decomposition of each spin-boson interaction operator follows the description given for the QR
gate in Sec.~\ref{sec:Decomposing_QR}. Here, each QR gate is decomposed into two JC gates and two
X-gates.

\subsection{Calibration of resonator-qubit gates}
The simulated time step $\tau$ is chosen such that the error in time propagation stays small.
This generally needs a separate analysis of Trotter error~\cite{Childs2021},
with guidelines given in Appendix~\ref{sec:error_analysis}.
However, qualitatively, we can state that we must have $\tau v_{ik}\ll 1$.
After fixing the time step $\tau$ to a certain wished value,
the following calibration of the resonator-qubit gates can be performed.

We consider a processor with two qubits and two resonators with the nearest-neighbor
connectivity as shown in Fig.~\ref{fig:connectivity}. The calibration corresponds to fixing the
physical resonator-qubit interaction times to $t_\textrm{gate}^k=\tau v_{ik}/g_k$
(Sec.~\ref{sec:JaynesCummingsGate}), where $k$ is the resonator index. The calibration then looks like
the following:
\begin{enumerate}
    \item The first resonator-qubit pair is calibrated to two possible interaction times: $\tau v_{11} /g_1$ and $\tau v_{21} /g_1$
    \item The second resonator-qubit pair is calibrated to two possible interaction times: $\tau v_{12} /g_2$ and $\tau v_{22} /g_2$
\end{enumerate}
Of course, a set of single-qubit and two-qubit gates need to be calibrated also. In particular, we
assume that single-qubit X-gates can be implemented with arbitrary phase~\cite{McKay_2017} (phased rotate-X)
and that we have certain entangling two-qubit gate (such as CNOT) that
can be used to decompose the qubit-qubit ZZ-interaction and the SWAP gate.

\subsection{Time evolution algorithm}
\begin{figure*}
\centering
\includegraphics[width=1.65\columnwidth]{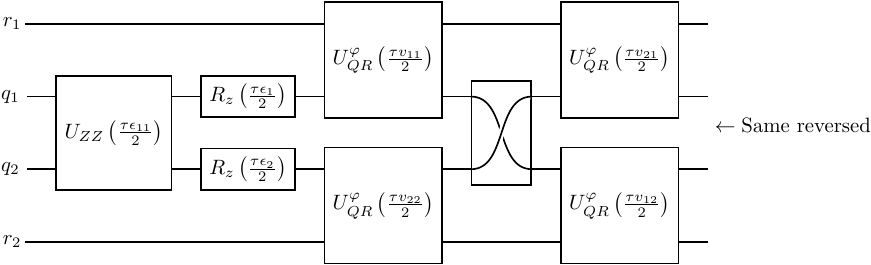}
\caption{%
  A second-order Trotter circuit for the considered Dicke model including two spins and two bosonic
  modes.
The crossing lines mark the swap gates.
We note that the form of the swap algorithm stays similar for arbitrary many spins (qubits) and bosonic modes (resonators).
}
\label{fig:example_fig1}
\end{figure*}

For the time propagation we use the second-order
Trotterization implemented by the resonator-qubit swap network.
This algorithm is visualized in Fig.~\ref{fig:example_fig1}. 
\begin{figure*}
\centering
\includegraphics[width=1.75\columnwidth]{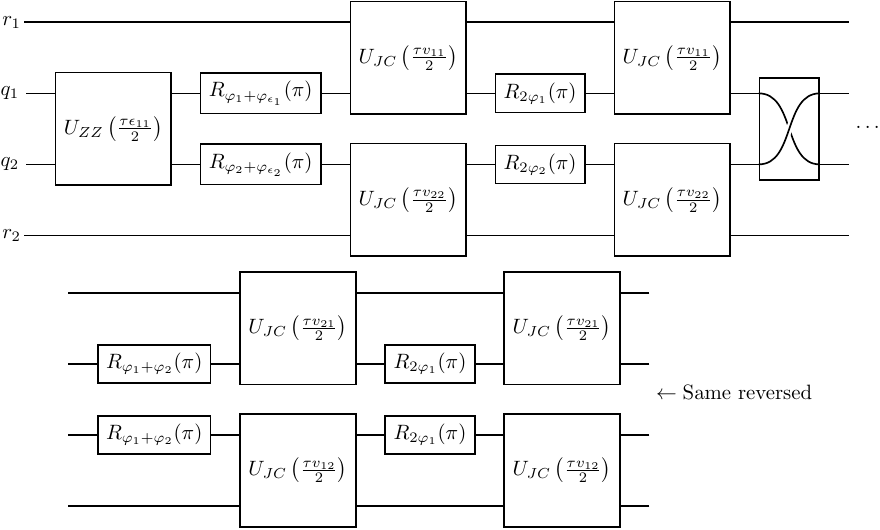}
\caption{Circuit that is equivalent to Fig.~\ref{fig:example_fig1}. Here we show explicitly the
  decompositions of the phased Rabi gates~$\hat U_{QR}^\varphi$ by Jaynes-Cummings
  gates~$\hat U_{JC}$ and the use of virtual gates [i.e., phases~$\varphi$ of the gates
  $R_{\varphi}(\pi)$] to implement the spin-splittings and bosonic-mode frequencies. Here $\varphi_i
  = -\frac{m\tau\omega_i}{2}$, $\varphi_{\epsilon_i} = -\frac{\tau\epsilon_i}{4}$, and $m+1/2$ is the
  number of the Trotter step. We note that the form of the swap algorithm stays similar for arbitrary many spins and
  bosonic modes.}
\label{fig:example_fig2}
\end{figure*}
In particular, when placing the JC gates of the QR gates
in the
``forward-part'' of the circuit, we tactically insert it in the order $H_\textrm{JC}X\, H_\textrm{JC}
X$.
And when coming backwards, we insert it in the order $X H_\textrm{JC}X\,
H_\textrm{JC}$.
This ordering lets us combine all single-qubit Z-rotations with the X-gates, by including them as
phases~$\varphi$ of the PRX gates
\begin{align}\label{eq:phased_rotate_X}
R_{\varphi}(\pi) &= R_z(\varphi) R_x(\pi) R_z(-\varphi) \, .
\end{align}
The resulting optimized circuit is shown explicitly in Fig.~\ref{fig:example_fig2}, where
$\varphi_\epsilon = -\frac{\tau\epsilon}{4}$, $\varphi_i = -\frac{m\tau\omega_i}{2}$, and $m+1/2$ is the
number of the Trotter step.
Furthermore, we emphasize
that this trick of combining Z-rotations and X-gates in the swap network works also for
arbitrary many resonator-qubit pairs, providing a very efficient algorithm for time propagating models
with arbitrary many spins and bosonic modes. Numerical simulations on running this circuit on hardware are
shown in Fig.~\ref{fig:noise_mapping}.

%-------------------------------------------- Effect of resonator dissipation ----------------------------------------------------
%-------------------------------------------- Effect of resonator dissipation ----------------------------------------------------
%-------------------------------------------- Effect of resonator dissipation ----------------------------------------------------

\section{Effect of resonator dissipation}\label{sec:noise_mapping}
The dominant source of errors in the considered hybrid quantum computer is most probably resonator damping and dephasing (dissipation).
Below, we show that these noise channels map to broadening of the bosonic modes in the simulated system.
This result not only allows a clear understanding of the effect of the resonator noise on the results of the quantum simulation,
but also demonstrates the potential to leverage resonator noise as a resource for
simulating continuous baths.

\subsection{Gate noise model}
We start by introducing a notation which allows to easily interpret incoherent errors as non-unitary operations.
We now assume that resonator noise is predominantly damping and dephasing and that qubit noise can be neglected
(although similar noise mapping as introduced below can be established also between qubits and spins~\cite{Fratus2022}).
This means that there is
an incoherent error of qubit-resonator gates which is due to resonator dissipation during
the application of the JC gate.
This is described by the Lindblad master-equation
\begin{align}\label{eq:physical_noise}
\dot {\hat \rho}_\textrm{physical} &= \textrm{i}[\hat \rho_\textrm{physical}, \hat H_\textrm{JC}]+ {\cal L}_N[\hat \rho_\textrm{physical}] \, ,
\end{align}
where the noise Lindbladian~${\cal L}_N$ has the form
\begin{align}\label{eq:noise_Linbbladian}
{\cal L}_N[\rho] &= \sum_k{\cal L}_{N_k}[\rho]=  \sum_{k}\gamma_k\left( \hat b_k \hat \rho \hat b_k^\dagger - \frac{1}{2}\left\{\hat b_k^\dagger \hat b_k , \hat\rho \right\} \right) \nonumber \\
&+ \sum_{k}2\Gamma_k\left( \hat n_k \hat \rho \hat n_k - \frac{1}{2}\left\{\hat n_k^2 , \hat\rho \right\} \right) \, ,
\end{align}
and~$\hat\rho$ is the density matrix of the full system-boson pair and $\hat n_k = \hat b_k^\dagger
\hat b_k$.
The resonator~$k$ damping corresponds to the
term proportional the damping rate~$\gamma_k$ and similarly for the dephasing with
rate~$\Gamma_k$. The gate is applied for time~$t_\textrm{gate}$, as described in Sec.~\ref{sec:JaynesCummingsGate}.
Under the assumption that the gate performs a small-angle unitary rotation ($t_\textrm{gate}g\ll 1$),
the effect of the noise can be modeled as the ideal gate followed a non-unitary operation~\cite{Fratus2023}
\begin{align}\label{eq:gate_noise_model}
\hat U  &\rightarrow  e^{t_\textrm{gate}\mathcal{L}_N} {\cal U} \, ,
\end{align}
where on the right hand side the effect of the gate is described by the corresponding
superoperator~${\cal U}$.
Error during idling will be modeled similarly, but is neglected in the discussion, for simplicity.

\subsection{Noise mapping}
Eq.~(\ref{eq:gate_noise_model}) describes the effect of a noisy JC gate in the quantum
circuit. We are interested in how such noise appears in the simulated system.
This is defined by all the noise collected during one Trotter circuit, and by the coherent gates between
them~\cite{Fratus2022}.
This effective noise is here a simple sum of all
individual Lindbladians describing noisy gates in one Trotter circuit
\begin{align}\label{eq:noise_mapping_result}
\tau \mathcal{L}_{\text{eff}} = \sum_{k}D_kt_\textrm{gate}{\mathcal{L}}_{N_{k}} \, ,
\end{align}
where~$D_k$ is the number of JC gates performed on resonator~$k$.
For simplicity, we assume that $t_\textrm{gate}$ is the same for all JC gates.
This result follows from the fact that our circuit does not perform any large-angle operations directly
on resonators, i.e., since $t_\textrm{gate}g \ll 1$. (The possible X-gates operate on qubits.)
Corrections to this result are higher orders in~$\tau\propto t_\textrm{gate}g$~\cite{Leppakangas2023}.
The time evolution of the simulated system then follows the master equation
\begin{align}\label{eq:simulated_noise}
\dot {\hat \rho}_\textrm{simulation} &= \textrm{i}[\hat \rho_\textrm{simulation}, \hat H]+ {\cal L}_{\text{eff}}[\hat \rho_\textrm{simulation}] \, .
\end{align}

\begin{figure}[]
\centering
\includegraphics[width=\columnwidth]{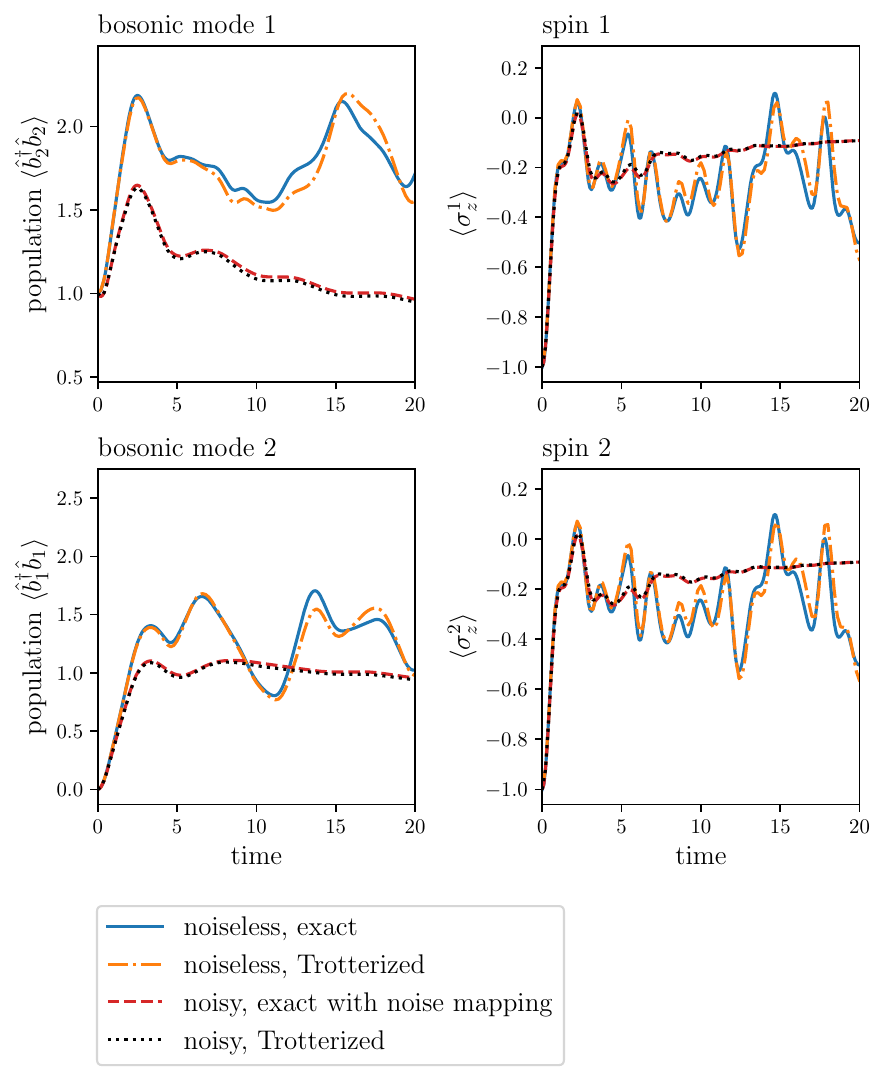}
\caption{Numerical simulation of Trotterized time evolution of the Dicke model with two spins and
two bosonic modes, as defined in Eq.~(\ref{eq:example_Hamiltonian_rf}).
As initial state, we set the first bosonic mode to its first excited state and
others to ground. The used time evolution algorithm is shown in Fig.~\ref{fig:example_fig2}. We
compare the Trotterized time evolution to the continuous (non-Trotterized) numerical solution of
the original problem, obtained by including $d=8$~levels for each bosonic mode. This solution is here
reproduced with a small Trotter error, which can be reduced by choosing a smaller Trotter time
step~$\tau$. Most of the Trotter error arises from high populations of the bosonic mode
(Appendix~\ref{sec:error_analysis}).
In the presence of gate noise,
we compare the Trotterized time evolution to the numerical solution of a master equation with
added noise Lindbladian~${\cal L}_\textrm{eff}$, as defined in
Eq.~(\ref{eq:noise_mapping_result}). The results of the noisy Trotterization matches to 
this model. In the numerical simulations, we use the Trotter time step $\tau = 0.2$ and model
parameters $\epsilon_1=\epsilon_2=0.5$, $\epsilon_{12}=1$, $\omega_1=\omega_2 = 1.0$,
$v_{11}=v_{22}=-v_{12}=-v_{21}= 0.5$. When considering the effect of noise, we include resonator
damping during the application of the Jaynes-Cummings gates with rates
$t_\textrm{gate}\gamma_{1/2} = 0.005$. }
\label{fig:noise_mapping}
\end{figure}

If~$D_k$ is the same for all resonators ($D_k =D$), Eq.~(\ref{eq:noise_mapping_result})
yields the effective noise rates in the simulated system
\begin{align}
\gamma_{\textrm{eff},k}  &=  \frac{Dt_\textrm{gate}}{\tau} \gamma_k  \label{eq:gate_noise_model_damping} \\
\Gamma_{\textrm{eff},k}  &=  \frac{Dt_\textrm{gate}}{\tau} \Gamma_k  \, . \label{eq:gate_noise_model_dephasing}
\end{align}
We can also write down a simple formula describing the
broadening of bosonic modes in the simulated system
\begin{align}
\gamma_{\textrm{eff},k}^* &=  \frac{D t_\textrm{gate}}{\tau}\gamma^*_k \, , \label{eq:model_broadening}
\end{align}
where $\gamma^*_k = \gamma_k/2+\Gamma_k$ is the linewidth of the resonator~$k$.
The parameter~$D$ is usually given by the circuit depth (Table~\ref{table:spin_boson_circuitDepthsAndGateCount})
multiplied by the number of JC gates needed to implement each specific coupling term in the model (Secs.~\ref{sec:Decomposing}-\ref{sec:Decomposing2}). 
In Fig.~\ref{fig:noise_mapping}, we verify this relation numerically
by reproducing the result of gate-based time propagation in the presence of resonator noise
by the solution of the effective Lindbladian, Eq.~(\ref{eq:simulated_noise}).

The found relations
let us map the resonator dissipation to simulated bosonic mode dissipation.
Furthermore,
if the bosonic modes describe a bosonic bath of a system-bath model, which couple to the system via the operator
\begin{align}
\hat X=\sum_{k} v_{k}(\hat b_{k}^\dagger + \hat b_{k}) \, ,
\end{align}
then the bath spectral function in the simulated system has the form~\cite{Leppakangas2023}
\begin{align}\label{eq:broadened_mode}
S(\omega) &= \int_{-\infty}^\infty dt e^{\textrm{i}\omega t} \left\langle \hat X(t) \hat X(0) \right\rangle_0 \nonumber \\
&= 2\pi\sum_k v_k^2 \frac{2\gamma_{\textrm{eff},k}^*}{(\gamma_{\textrm{eff},k}^*)^2 + (\omega-\omega_{k})^2} \, ,
\end{align}
where the trace is calculated according to the free evolution of the bath (where $\hat H_\textrm{c}=0$).
In particular, the peak widths are defined by the chosen Trotter time step~$\tau$, see
Eqs.~(\ref{eq:gate_noise_model_damping}-\ref{eq:model_broadening}). This result means that
it is possible to tailor various spectral functions by optimizing the (digitally implemented) simulation
parameters~$v_k,\omega_k,\tau$~\cite{Leppakangas2023}.

\subsection{Estimation of broadening for different models}
It is now straightforward
to make estimates for the bosonic-mode broadening for different model classes
and resonator qualities. As a practical example, consider a
Hamiltonian with parameters of the order~$1$ and time propagation with Trotter
time step $\tau=0.2$.
Assuming JC gate infidelity $\sim t_\textrm{gate}\gamma^*=10^{-3}$,
the broadening of each bosonic mode for Dicke-type models with all-to-all couplings and
$N_\textrm{b}=n_\textrm{s}=10$~modes is
according to Eq.~(\ref{eq:model_broadening})
$\gamma_\textrm{eff}^* = 2 \times 10 \times 10^{-3}/0.2=0.1$,
where the multiplication by two comes from the needed two JC gates per coupling term.
This is to be compared with the size of Hamiltonian parameters (which was set to~$1$).
Furthermore, since the circuit depth scales here
linearly with~$N_\textrm{b}$, increasing the simulation size to
$N_\textrm{b}=100$~bosonic modes, the same broadening is achieved by improving the resonator quality such that
$t_\textrm{gate}\gamma^*=10^{-4}$. On the other hand, a nearest-neighbor
spin-boson interaction model has a constant depth and, for example, with 100~bosonic modes gives broadening $0.1$
already for $t_\textrm{gate}\gamma^*=10^{-2}$.

We note that by increasing~$\tau$, we can decrease the broadening, but this will increase the Trotter error.
In practice, one needs to find an optimal tradeoff between mode broadening and Trotter error. %~\cite{Leppakangas2023}.

Since the above considered resonator gate errors are in the limit of feasible values
for state-of-the-art technology, it is rather clear that, in practice,
the quantum simulation will always come with a noticeable
broadening of bosonic modes. The conclusion here is then that the most promising application of the
device is the modeling of systems coupled to broadened bosonic modes or a continuous bath.

%---------------------------------------------------------------- Demonstration on IQM resonance ----------------------------------------------------------------
%---------------------------------------------------------------- Demonstration on IQM resonance ----------------------------------------------------------------
%---------------------------------------------------------------- Demonstration on IQM resonance ----------------------------------------------------------------

\section{Demonstration on IQM resonance}\label{sec:Experiment}
\begin{figure}[]
\centering
\includegraphics[width=0.9\columnwidth]{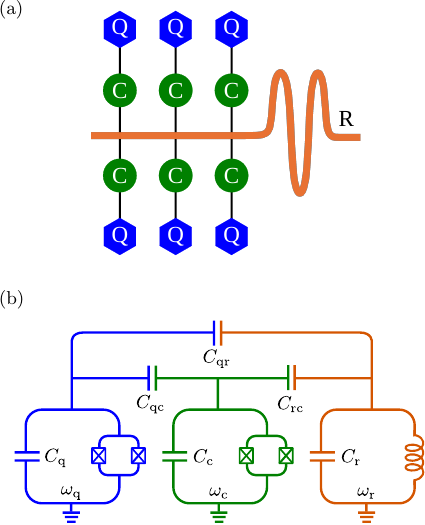}
\caption{%
(a) The quantum processor~\cite{Renger2025} (\textit{Deneb})which was used for this paper
consists of a transmission-line resonator coupled to six superconducting qubits
via tunable couplers.
(b) A circuit schematic of a computational qubit (left) coupled to the resonator (right) in parallel
via frequency-tunable coupler-qubit (middle) and a direct qubit-resonator capacitor (top).
The interference between these two paths defines a tunable qubit-resonator coupling.
}
\label{fig:setup}
\end{figure}

Here, we present the central technical details of the hybrid quantum processor~\cite{Renger2025} we used 
and the quantum simulations we performed via the IQM Resonance cloud. The quantum processor is called Deneb.
This demonstration includes also calibration of the JC gates on the algorithmic level.
The results highlight the feasibility to integrate computational resonators
when using commercial quantum-computing platforms and to perform system-boson quantum simulations using the algorithms we presented.

\subsection{Quantum processor with tunable qubit-resonator coupler}\label{sec:Deneb_and_tunable_coupler}
The unit cell of the quantum processor we used
is a qubit coupled to a microwave resonator. The qubit is a
superconducting transmon~\cite{Koch2007,Blais2021_RevModPhys} which is an anharmonic nonlinear LC-resonator.
In simple terms, the qubit is a capacitor shunted by a DC-SQUID, which provides both the nonlinearity necessary
for the circuit to be approximated as a qubit, and the DC-SQUID imparts frequency tunability on the
qubit. The phased rotate-X (PRX) gate can be implemented by Rf drives on the qubit.
The qubit frequency can be tuned by the magnetic flux through the qubits DC-SQUID loop,
which is needed for the implementation of the JC gates.

The quantum processor consists of six transmon qubits
coupled to one common resonator, Fig.~\ref{fig:setup}(a).
The resonator is a transmission-line resonator which can be modeled as a linear LC circuit.
The simplest way to connect a qubit to the resonator is to directly couple these two elements with
a coupling capacitor (or coupling inductor). If this direct coupling would be
implemented, the physical JC Hamiltonian would be directly implemented. However, here want to use
a more flexible and programmable processor.

Therefore, the coupling between each qubit and the resonator is controlled by embedding a tunable coupler qubit between
the qubit and resonator~\cite{Yan2018, Marxer2023}. 
A circuit schematic is shown in Fig.~\ref{fig:setup}(b).
By tuning the frequency of this coupler, we tune the coupling strength between
the qubit and resonator, allowing us to define our qubit-resonator gates.
In simple terms, this circuit provides two coupling paths: (i) a direct path and (ii) coupler mediated path;
the interference between these two paths defines the coupling strength. Tuning the frequency of the
coupler tunes this interference. In principle, this allows fast tuning between fully
extinguishing coupling, and having the coupling on, to implement the qubit-resonator JC gate as
described in Sec.~\ref{sec:JaynesCummingsGate}. A more detailed description of the device
and its benchmarking is given in Refs.~\cite{Renger2025, Algaba2022}.

\subsection{Time propagation of a spin-boson model}\label{sec:demonstration_Deneb}
We time evolve a Jaynes-Cummings model on the Deneb quantum processor 
using access via IQM Resonance cloud.
As described above, the gate set of Deneb includes the JC gate
and the PRX gate.
The JC gate
is calibrated such that it performs a population exchange between the states $\vert 1_q, 0_r\rangle $
and $\vert 0_q, 1_r\rangle $, where $q$ refers to the qubit and $r$ to the resonator.
The JC gate comes with a fixed unknown phase (rotate-Z) on the qubit. However,
we will discuss a calibration method on the algorithmic level below, which will allow us to remove this effect.

The Jaynes-Cummings model we consider describes interaction between one spin and one bosonic mode
and has the Hamiltonian
\begin{align}\label{eq:JaynesCummings}
\hat H = \frac{\delta\epsilon}{2}\sigma_z + v\sigma_-\hat b^\dagger + v\sigma_+\hat b  \, .
\end{align}
The Hamiltonian is in the combined rotating frame of the bosonic mode and the spin,
defined by the frequency of the bosonic mode~$\omega_0$.
(For this model transforming also the spin to the rotating frame is beneficial,
since then time-dependent phases disappear.)
Here, the spin spitting~$\delta\epsilon$ is the difference of the energies in the original frame,
$\delta\epsilon = \epsilon-\omega_0$.

\begin{figure}[]
\centering
\includegraphics[width=0.9\columnwidth]{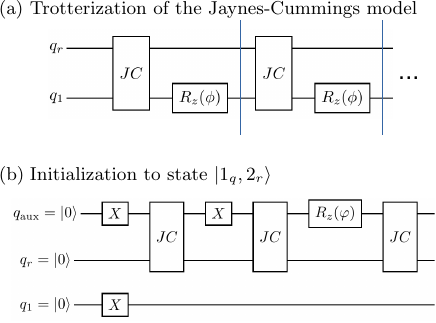}
\caption{%
Trotterization and initialization of digitized Rabi oscillations on the third excitation manifold of the resonator-qubit pair.
The shown gate ``$JC$'' corresponds to a Jaynes-Cummings gate calibrated to perform a population exchange in the first excitation manifold.
(a) The Trotter circuit consists of single Jaynes-Cummings gate followed by a rotate-Z
gate with angle~$\phi$. This added phase term is to
compensate for the phase coming with the used Jaynes-Cummings gate on an algorithmic level, and
to implement energy-level difference between the modeled spin and bosonic mode.
(b) Before Trotterization, we initialize the qubit-resonator pair to the state~$\vert 1_q, 2_r\rangle$ with the help of an auxiliary qubit.
}
\label{fig:demonstration_initialization}
\end{figure}

In the first-order Trotterization of the time evolution, the
unitary operation describing time evolution over the time step~$\tau$
becomes
\begin{align}\label{eq:JaynesCummings}
\exp\left(-\textrm{i} \tau \hat H\right) &\rightarrow R_z(\tau\delta\epsilon) U_\textrm{JC}(\tau v) \, .
\end{align}
The Trotterization algorithm is shown in Fig.~\ref{fig:demonstration_initialization}(a).
Each rotate-Z gate is created by two PRX gates
using the decomposition $R_z(\varphi) = R_{\varphi/2}(\pi) R_{0}(\pi)$.
The phase~$\varphi$ also includes a
compensation for the (above-mentioned) additional Z-rotation coming with the JC gate.
The origin of this is a frame-tracking correction of iSWAP-type gates~\cite{Arute2020, Huang2023}.
This correction is then implemented on the algorithmic level.

\begin{figure*}[]
\centering
\includegraphics[width=2\columnwidth]{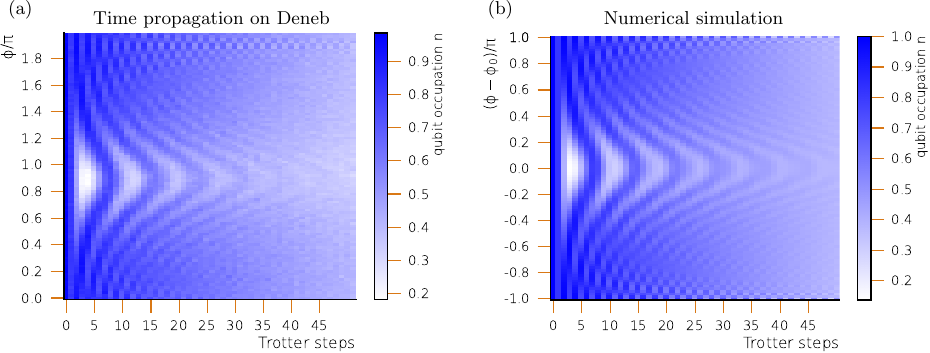}
\caption{%
Demonstration of digitized Rabi oscillations on the third excitation manifold of the resonator-qubit pair.
(a) Results from the IQM Resonance quantum cloud.
This simulation reproduces the well known reduction
of the oscillation period when the bosonic mode and the spin are off-resonance  (see for example Ref.~\cite{Braumueller2017}).
For the (central) phase $\phi_0/\pi\approx 0.9$ we cancel out the fixed-phase gate that comes together with the implementation of the JC gate.
Phases different from $\phi_0$ correspond to modeling a finite energy-level difference between the spin and the bosonic mode.
(b) Numerical simulation of the Trotterized time evolution on the third excitation manifold
with resonator damping of strength $\gamma t_\textrm{gates}= 0.03$,
where $t_\textrm{gates}$ is the total physical time needed to run one Trotter step.
}
\label{fig:demonstration3}
\end{figure*}

For the time propagation, the angle of the JC gate~$\tau v$ needs to be small,
ideally much smaller than 1. However, as mentioned earlier,
the JC gate for our present application was calibrated to perform a population exchange in the first excitation manifold,
corresponding to angle~$\pi/2$ and it was key goal of this demonstration to only use gates and features that are publicly available
via standard cloud access. 
To establish an effective small-angle JC gate,
we time evolve the JC model
in the third excitation manifold, i.e., we time evolve Rabi oscillations between the states
\begin{align}
\vert 1_q, 2_r\rangle \leftrightarrow \vert 0_q, 3_r\rangle \, .
\end{align}
Here the application of the used JC gate creates
effectively a small-angle rotation: since the physical coupling between the resonator states $\vert
2_r\rangle$ and $\vert 3_r\rangle$ is $\sqrt{3}$ times stronger than between the resonator states $\vert
0_r\rangle$ and $\vert 1_r\rangle$, the effective rotation angle within the same interaction time
$t_\textrm{gate}$ is
\begin{align}
& \sqrt{3}\times \frac{\pi}{2} \approx \pi - 0.42 \, ,
\end{align}
where the phase~$\pi$ introduces an additional minus sign to the state and can be neglected here.
We can then use the under-rotation~$0.42$ as an effective small-angle gate in the
Trotterization of the JC interaction dynamics.
A similar effective small-angle JC gate can be constructed also
in the second excitation manifold,
see Appendix~\ref{Appendix:Additional_Deneb_simulations}.

The initialization of the qubit-resonator pair to the state~$\vert
1_q, 2_r\rangle$ is done with the help of an auxiliary qubit, by the circuit shown in
Fig.~\ref{fig:demonstration_initialization}(b). This circuit involves two JC gates with auxiliary
qubit rotate-Z of angle
$\varphi$ in between, whose value is optimized such that the measured ground state population of the
auxiliary qubit is maximized (here close to~$0.9$).
In this case, the resonator is with high probability
initialized to state~$\vert 2_r\rangle$.
Later, we also post-select the time-propagation results so that only measurements where the
auxiliary qubit was found to be in the ground state are accepted.

The result of the quantum simulation is shown in
Fig.~\ref{fig:demonstration3}(a), where we run the simulation for different values of $\phi$. We
identify the ``central'' value~$\phi_0/\pi\approx 0.9$ as the value for a perfectly calibrated JC gate.
A deviation from the central phase,~$\delta\phi=\phi-\phi_0$,
corresponds to an additional $R_z(\delta\phi)$, which in turn corresponds to the
energy-level difference between the spin and the bosonic mode via the relation $\delta\phi=\tau\delta\epsilon$.
The results correspond to averaging over~2000 measurement shots.
This simulation reproduces the well known reduction
of the oscillation period when the bosonic mode and the spin are off-resonance  (see for example Ref.~\cite{Braumueller2017}).
The numerical simulation shown in Fig.~\ref{fig:demonstration3}(b)
corresponds to an accurate model of the hybrid quantum computation with resonator damping of
strength $\gamma t_\textrm{gates}= 0.03$,
where $t_\textrm{gates}$ is the total physical time needed to run one Trotter step,
consistent with the Deneb gate times and resonator T1 (Appendix~\ref{app:calibration}).
We estimate that this is the main source of decoherence (decay of Rabi oscillations) in the demonstration.
The Trotter error in the time-propagation is seen mostly as the ``disturbance'' near $\vert\delta\phi\vert \lesssim \pi$
(Appendix~\ref{sec:error_analysis}).

The almost perfect agreement between the experiment and theory
shows that it is feasible to accurately time propagate bosonic-mode dynamics
using resonators as computational elements via cloud access.
Further improvements on this platform
will be achieved by the inclusion of a variable-angle JC~gate to the Deneb gate-set,
realized via a pulse level access~\cite{IQM_Pulla}.
This will also allow for time propagating other types of spin-boson models
(Sec.~\ref{sec:models}) and performing noise mapping (Sec.~\ref{sec:noise_mapping}),
i.e., utilizing resonator dissipation for time evolving a system coupled to a continuous bosonic bath.

%----------------------------------------------------- Conclusions -------------------------------------------------
%----------------------------------------------------- Conclusions -------------------------------------------------
%----------------------------------------------------- Conclusions -------------------------------------------------

\section{Conclusion and outlook}\label{sec:conclusion}
In conclusion,
we discussed quantum algorithms to model composite systems of spins or electrons coupled to bosonic modes.
The key idea was to improve simulation feasibility
by representing bosonic modes directly by microwave resonators and composing system-boson
interactions using native qubit-resonator interactions.
We introduced a minimal design of the quantum processor connectivity and tailored quantum algorithms that can
efficiently time propagate many classes of useful system-boson models.
It should be kept in mind that a more general processor connectivity could allow for more
efficient quantum algorithms to be developed for model classes that are similar to the device
connectivity.

An important finding of our work is that for simulations with a large number of bosonic modes (and
resonators), the effect of resonator dissipation will be non-negligible even for most efficient
algorithms. Further, we also find that the effect of resonator dissipation can be mapped to
broadening of bosonic modes in the simulated system.
Putting these two results together, we find that modeling of a continuous bosonic bath is the most
realistic use case for the considered quantum processor in near
future.
Moreover, we believe that modeling of bosonic systems without any broadening is realistic only for
fully error-corrected conventional (all-qubit) quantum computers, when available in the future. And
even then,
it is fair to expect that the hybrid resonator-qubit processor described here would have its
advantages in the modeling of continuous bosonic baths, due to its straightforward implementation of
bath-mode broadening via inherent dissipation, which for error-corrected quantum computers would
need creation of non-unitary gates, e.g., via the use of auxiliary qubits and quantum feedback control.

Finally, another possible and very intriguing use case for the resonator-qubit quantum
computer could be the study of a quantum advantage in simulation of bosonic
systems, i.e., performing computations that are not possible to be modeled using classical
computers. In particular, large systems of spins and electrons coupled to multiple bosonic modes in
the limit of ultra-strong coupling~\cite{FriskKockum2019} can be expected to be optimal for this purpose.

\section*{Acknowledgments}
This work was supported by the German Federal Ministry of Education and Research,
through projects Q-Exa (13N16065) as well as QSolid (13N16155),
and by the European Union’s Horizon program number 101046968 (BRISQ).

%----------------------------------- Appendix A ----------------------------------------
%----------------------------------- Appendix A ----------------------------------------
%----------------------------------- Appendix A ----------------------------------------

\appendix
\newpage

\section{Demonstration of Rabi oscillations in the second excitation manifold}\label{Appendix:Additional_Deneb_simulations}
\begin{figure*}[]
\centering
\includegraphics[width=2\columnwidth]{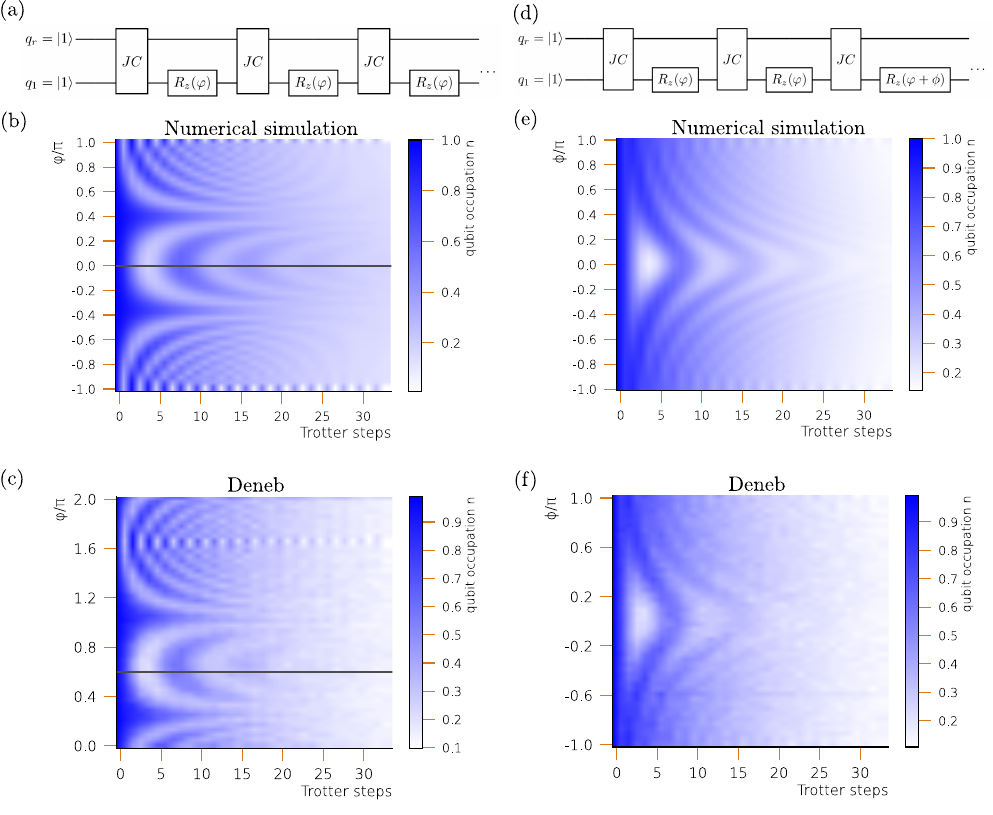}
\caption{%
(a) Trotter circuit for the calibration (determination of phase corrections) of the JC gates.
The qubit-resonator pair is always initialized to state $\vert 1_q, 1_r\rangle$.
(b, c) Determination of phase corrections.
The correct phase correction is indicated by the black line.
(d) Trotterization of time propagation using three JC gates as one effective small-angle JC gate.
After every third JC gate an additional phase~$\phi$ will be inserted, modeling the energy-level difference
between the spin and bosonic mode, $\phi=\tau\delta\epsilon$.
(e, f) Theoretical and experimental results for the digitized Rabi oscillations with different values
of the phase~$\phi$, i.e., energy-level difference between the modeled spin and bosonic mode.
The resonator damping rate used in the numerical simulation is the same as in Fig.~\ref{fig:demonstration3}:
due to three JC gates and phase corrections per Trotter step we have here $\gamma t_\textrm{gates}=3\times 0.03$.
}
\label{fig:demonstration1}
\end{figure*}

In the second demonstration, we time evolve Rabi oscillations in the second excitation manifold.
Here we first start the time evolution from the resonator-qubit state $\vert 1_q, 1_r\rangle$ and
time propagate Rabi oscillations between the states
\begin{align}
\vert 1_q, 1_r\rangle \leftrightarrow \vert 0_q, 2_r\rangle \, .
\end{align}
As earlier, for this to be possible, the applied gate needs to correspond to a small-angle rotation
in this manifold. Since the physical coupling between the resonator states $\vert 1_r\rangle$ and
$\vert 2_r\rangle$ is $\sqrt{2}$ times stronger than between the states $\vert 0_r\rangle$ and $\vert
1_r\rangle$, the effectively generated angle from~$n$ consecutive JC gates is
\begin{align}
n=1:\,\,\,\,\,\,\,  &  \sqrt{2}\pi/2 \approx \pi - 0.92 \nonumber \\
n=2:\,\,\,\,\,\,\,  & 2\sqrt{2}\pi/2 \approx \pi + 1.30 \nonumber \\
n=3:\,\,\,\,\,\,\,  & 3\sqrt{2}\pi/2 \approx 2\pi + 0.38 \nonumber 
\end{align}
We then find that for $n=3$ consecutive JC gates we establish a small-angle over-rotation 0.38 (the
phase~$2\pi$ can be neglected). The time evolution can then be build from Trotter
circuits consisting of three JC gates separated by frame tracking corrections with phase~$\varphi$.
The correct value of~$\varphi$ is found by changing
its value from 0 to~$2\pi$ and making comparison to theory. We do this using the circuit shown in Fig.~\ref{fig:demonstration1}(a).
The theoretical result is shown in Fig.~\ref{fig:demonstration1}(b), which is then compared to the
experimental result shown in Fig.~\ref{fig:demonstration1}(c).
After finding the correct value of~$\varphi$, we fix the
phase correction~$\varphi$ to this value and add an extra phase~$\phi$ to every
third JC gate, to simulate the effect of finite energy-level difference between the spin and bosonic
mode, as shown in Fig.~\ref{fig:demonstration1}(d).
The theoretical and experimental results for the time propagation are shown in Figs.~\ref{fig:demonstration1}(e) and~\ref{fig:demonstration1}(f).
Similarly to the demonstration shown in Fig.~\ref{fig:demonstration3},
we again reproduce the increase of the oscillation period for finite energy-level difference~$\delta\epsilon$, i.e.,
for finite added phase~$\phi=\tau\delta\epsilon$.

%----------------------------------- Appendix B ----------------------------------------
%----------------------------------- Appendix B ----------------------------------------
%----------------------------------- Appendix B ----------------------------------------

\section{Calibration data}\label{app:calibration}
A thorough benchmarking of Deneb is presented in Ref.~\cite{Renger2025}.
The calibration data collected during circuit execution is shown in Tables~\ref{tab:qubit_resonator1}
and~\ref{tab:qubit_resonator2}.
The single-qubit gate times are $40$~ns and the JC (MOVE) gate times vary between $80-96$~ns~\cite{Renger2025}.
\begin{table}[h]
\centering
\begin{tabular}{|l|c|c|c|c|c|}
\hline
 & $T_1$ & $T_2$ (Ramsey) & PRX err. & Readout err. & JC err. \\
\hline
Q1 & 27 $\mu\text{s}$ & 40 $\mu\text{s}$ & 0.0008 & 0.01 & 0.0075 \\
\hline
Q4 & 41 $\mu\text{s}$ & 26 $\mu\text{s}$ & 0.0006 & 0.01 & 0.0076 \\
\hline
R & 5.6 $\mu\text{s}$ & 11 $\mu\text{s}$ & - & -  & - \\
\hline
\end{tabular}
\caption{Benchmark data for the qubits and resonator used in the demonstration presented in Fig.~\ref{fig:demonstration3},
collected on January 22, 2025, during circuit execution. The qubit Q1 was used as the auxiliary qubit
and Q4 as the computational qubit.}
\label{tab:qubit_resonator1}
\end{table}
\begin{table}[h]
\centering
\begin{tabular}{|l|c|c|c|c|c|}
\hline
 & $T_1$ & $T_2$ (Ramsey) & PRX err. & Readout err. & JC err. \\
\hline
Q3 & 39 $\mu\text{s}$ & 26 $\mu\text{s}$ & 0.0004 & 0.01 & 0.0085 \\
\hline
R & 5.5 $\mu\text{s}$ & 11 $\mu\text{s}$ & - & -  & - \\
\hline
\end{tabular}
\caption{Benchmark data for the qubit and resonator used in the demonstration presented in Fig.~\ref{fig:demonstration1},
collected on January 27, 2025, during circuit execution.}
\label{tab:qubit_resonator2}
\end{table}

%----------------------------------- Appendix C ----------------------------------------
%----------------------------------- Appendix C ----------------------------------------
%----------------------------------- Appendix C ----------------------------------------

\section{Advantage in simulation of bosonic modes}
In its given task, the qubit-resonator quantum simulator can be drastically more efficient than conventional all-qubit quantum computers.
We can make a simple (upper) estimate for
the hardware-component count and entangling-gate count when implementing the spin-boson operator
\begin{align}
\exp\left[\textrm{i}\phi\sigma_{x}(\hat b^\dagger + \hat b)\right] \, ,
\end{align}
using resonator-qubit algorithms and standard unary and binary codings of all-qubit quantum
computers. As described in Sec.~\ref{sec:Decomposing}, for resonator-qubit algorithms this is an
easy task: the operation needs only two entangling (JC) gates for modeling any number of bosons.
However, for all-qubit quantum computers the bosonic operator $\hat b^\dagger + \hat
b$ needs to be mapped to a sum of Pauli strings of qubits encoding the Hilbert space of the bosonic
mode.
It is known that one needs $2(p-1)$ Control-Z (or Control-X) gates to create a Pauli string of length $p$,
assuming all-to-all connectivity.
We note also that the full operator, $\sigma_x(\hat b^\dagger + \hat b)$, includes also the
spin-qubit which needs to entangle with all the qubits encoding the bosonic mode, leading increase of the Pauli strings by one.
To estimate the total cost of implementing the operator $\hat b^\dagger + \hat b$, one needs
to evaluate the sum of gates implementing all different Pauli strings. To get a finite number, one needs to
reduce the Hilbert space of bosonic modes to a finite number of energy levels. We consider now
including $d$~levels of the bosonic mode. The results are given in
Table~\ref{table:coding_efficiency} and are based on following the analysis made in
Ref.~\cite{Sawaya2020}. We see that for large $d$, either the qubit count (unary coding) or the
entangling-gate count (binary coding) becomes large. On the other hand, in the resonator-qubit
quantum simulation, the same operation needs only two entangling gates, for arbitrary~$d$. Since the
fabrication of resonators is not more complicated than that of qubits, it is clear that the
resonator-qubit quantum computer is always significantly more efficient than conventional quantum
computer when~$d$ is large.
It should be noted that the scalings of the all-qubit algorithms
are upper estimates and may be optimized, for example, by the use of additional (auxiliary) qubits~\cite{Liu2024}.

%----------------------------------- Appendix D ----------------------------------------
%----------------------------------- Appendix D ----------------------------------------
%----------------------------------- Appendix D ----------------------------------------

\section{Error sources and error mitigation}\label{sec:error_analysis}
Here we discuss the error in time propagation due to Trotterization.
Particularly interesting is the behavior of the error as a function of the boson number.
We also discuss the expected forms of coherent JC-gate errors and possible ways to mitigate them.

\subsection{Trotter error}
The Trotter error~\cite{Childs2021} emerges when separating the individual partial Hamiltonians $\hat H_i$ in
$\exp\left(-\textrm{i}\tau\hat H\right)= \exp\left(-\textrm{i}\tau\sum_i\hat H_i\right)$
to consecutive operations $\Pi_i\exp\left(-\textrm{i}\tau\hat H_i\right)$, 
see Eqs.~(\ref{eq:Trotter_1st}) and~(\ref{eq:Trotter_2nd}).
The error per time step $\tau$ may be described by an additional error term~$\hat\alpha$ in the
effective time-propagation operator~$\exp\left[-\textrm{i}\tau\hat H + \hat\alpha \right]$.
For the first-order formula we have the leading-order error term
\begin{align}\label{eq:Trotter_error1}
\hat\alpha_1 &= \left(-\textrm{i}\tau\right)^2 \frac{1}{2} \sum_{i_1<i_2}\left[H_{i_2}, H_{i_1}\right]   \, ,
\end{align}
where we have applied the Baker-Campbell-Hausdorff (BCH) formula when deriving this.
The total error accumulated over the whole time propagation is generally hard to predict,
but can be bound by the spectral norm
multiplied by the number of Trotter steps~\cite{Childs2021}. This may however overestimate the
actual error in the time evolution. Alternatively, the error may also be interpreted as the
term~$\propto\hat\alpha/\tau$ in the time-propagated Hamiltonian.

For the second-order formula, $\hat\alpha$ does not generally have such simple general expression,
so usually one rather
directly gives an upper estimate for the noise scaling~\cite{Childs2021}
\begin{align}\label{eq:Trotter_error2}
\vert\vert \hat\alpha_2 \vert\vert &< \tau^3 \sum_{i,j,k}\vert\vert\left[H_{i},\left[H_j, H_k\right]\right]\vert\vert \, .
\end{align}
(Also a more tight bound can also be formulated for the second order~\cite{Childs2021,Schubert2023}.)
For simple circuits, $\hat \alpha_2$ can however be calculated straightforwardly analytically,
as done below.

\subsubsection{Quantum Rabi (QR) gate}\label{sec:Trotter_error_formulas}
Using this approach, we can qualitatively study
the form of the error in decomposing the quantum Rabi (QR) gate.
For simplicity, we first mark $\hat A = \hat H_\textrm{JC}$ and $\hat B= \hat H_\textrm{AJC}$.
In the first-order Trotterization, such as 
$\hat U_\textrm{QR}=e^{-\textrm{i}\tau\hat A}e^{-\textrm{i}\tau\hat B}$,
the leading-order error is then described by the commutator
\begin{align}
\hat\alpha_1 &= -\frac{1}{2}\tau^2[\hat A, \hat B] \nonumber \\
           &= -\frac{1}{2}(\tau v)^2\left[(\sigma_-\hat b^\dagger + \sigma_+\hat b), (\sigma_+\hat b^\dagger + \sigma_-\hat b) \right] \nonumber \\
&= \frac{1}{2}(\tau v)^2\sigma_z \left[ \hat b^{\dagger^2} - \hat b^2  \right] \, .
\end{align}
We notice that the error operator has the fom of a conditional squeezing.
The magnitude of the boson operators can be estimated to scale like~$\vert\vert \hat b^2 \vert\vert \sim N_\textrm{bosons}$.
This leads to the qualitative estimate that the Trotter error in this construction scales like~$\sim (v \tau\sqrt{N_\textrm{bosons}})^2$.
We have then dependency on the number of bosons.

In the second-order Trotterization, the Quantum-Rabi gate can be constructed like
$\hat U_\textrm{QR}=e^{-\textrm{i}\tau\hat A/2}e^{-\textrm{i}\tau\hat B/2}e^{-\textrm{i}\tau\hat B/2}e^{-\textrm{i}\tau\hat A/2}$.
A direct use of the BCH formula gives that the leading-order error is described by the operator
\begin{align}
\hat\alpha_2 &=\textrm{i}\left(\frac{\tau}{2}\right)^3\left(-\frac{1}{2}[\hat A + \hat B, [\hat A, \hat B]] + \frac{1}{6}[\hat A - \hat B, [\hat A, \hat B]] \right) \nonumber \\
& \approx \textrm{i}\left(\frac{\tau}{2}\right)^3\left(-\frac{1}{2}[\hat A + \hat B, [\hat A, \hat B]]\right) \nonumber \\
&= \textrm{i}\left(\frac{\tau v}{2}\right)^3 \left[\frac{1}{2}\sigma_x(\hat b^\dagger + \hat b), \sigma_z \left( \hat b^{\dagger^2} - \hat b^2 \right)\right] \, .
\end{align}
We can then estimate that the Trotter error scales in such construction like~$\sim (v \tau\sqrt{N_\textrm{bosons}})^3$.
Again notable is the dependency on the boson number.
In practice, this implies that the Trotter time step may need to be reduced when the number of bosons increases,
possibly with dependency $\tau\propto 1/\sqrt{N_\textrm{bosons}}$.

\subsubsection{Longitudinal coupling (LC) gate}
This gate is constructed from the quantum-Rabi gate by additional Hadamard gates,
as shown in Fig.~\ref{fig:example_QR}(b).
Since there is no additional error in this
construction in comparison to the QR gate, the Trotter error analysis is similar to the above case.
The only difference will be that the Hadamard gates will effectively transform~$\sigma_x$ operators to~$\sigma_z$
operators, and vice versa.

\subsection{Trotter error 2: Error in discretizing the time-dependent Hamiltonian}
The discretization in Eq.~(\ref{eq:Trotter}) is not exact since the Hamiltonian is time dependent.
Such an error is a part of the (total) Trotterization error.
Indeed, when we are referring to an observed Trotter error in numerical simulations, both error mechanisms contribute.
This discretization error is not present in common quantum algorithms, based on time-independent Hamiltonians.
The optimization of the time propagation algorithm and its error for time-dependent Hamiltonians is generally less understood,
but has been studied for example in Ref.~\cite{Ikeda2023}.

However, for the specific algorithm types considered in this paper~\cite{Mezzacapo2014, Langford2017, Shapiro2024}
we can describe this error by above formulas.
This is since for a time dependency originating in transformation to the rotating frame
(and using midpoints time discretization),
the final simulation algorithm is equivalent to removing the time-dependent phases and performing
Trotterization with bosonic-mode rotation gates~$\exp(-\textrm{i}\hat H_\textrm{b}\tau/2)$,
placed at the beginning and at the end of each second-order Trotterization sequence.
This means that the error is described by
commutators with $H_\textrm{b}$.
In particular, commuting $H_\textrm{b}$ with coupling $\hat H_\textrm{c}$
gives a term of form $\sim\omega_k\hat H_\textrm{c}$ (with interchanged $\hat b_k$ and $\hat b_k^\dagger$)
and thereby a contribution to the total error whose magnitude scales like~$v\tau\sqrt{N_\textrm{bosons}}\times \omega_k \tau$.
Another commutation with $H_\textrm{b}$ then gives a contribution proportional to ~$v\tau\sqrt{N_\textrm{bosons}} \times (\omega_k \tau)^2$,
and similarly for other possible terms in Eq.~(\ref{eq:Trotter_error2}).
This result implies that the Trotter time step may need to be reduced when the bosonic-mode frequencies are increased,
possibly with dependency $\tau\propto 1/\omega_k$.

\subsection{Trotter error in the main-text simulations}
\begin{figure*}[]
\centering
\includegraphics[width=1.8\columnwidth]{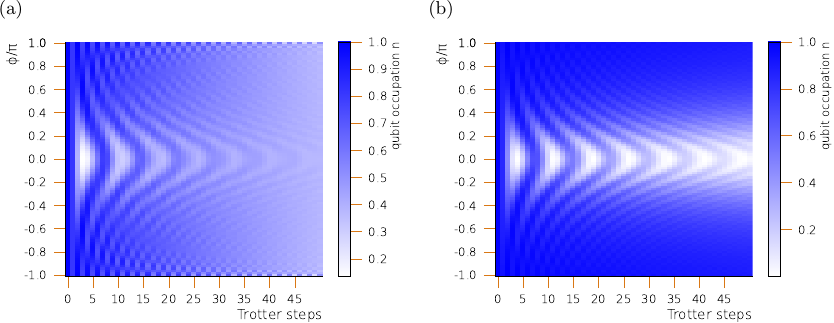}
\caption{%
(a) Numerical simulation of the demonstration in the third excitation manifold.
This corresponds to Trotterized time evolution of the Jaynes-Cummings model.
(b) Corresponding numerically exact solution of the Jaynes-Cummings model in the first excitation manifold,
when plotted with the same time discretization~$\tau$.
The general difference in the contrast is due to different relaxation dynamics towards the ground state,
whereas the oscillation pattern of (a) towards the edges $\vert\phi/\pi\vert = 1$ is distorted due to the Trotter error.
}
\label{fig:trotter_error_demonstration3}
\end{figure*}

We now analyze the Trotter error in the simulation shown in Fig.~\ref{fig:demonstration3}.
To do this
we compare here the numerically exact master equation solution for Rabi oscillations in the first excitation manifold
to the numerical simulation of the Trotterized time evolution in the third excitation manifold using the available JC gates.
The comparison is made in this way, since correct relaxation dynamics
from higher manifolds cannot be reproduced using the available JC gates
(since they can be used as a small-angle gates only in the third excitation manifold, Sec.~\ref{sec:demonstration_Deneb}).
This comparison is given in Fig.~\ref{fig:trotter_error_demonstration3}.
We plot the data in the same time steps~$\tau$ and phase steps~$\phi$,
the latter being related to the simulated energy level difference~$\delta\epsilon$ by $\phi=\tau\delta\epsilon$.
The difference in the pattern is due to the Trotterization error,
whereas difference in the contrast is due to different relaxation dynamics towards the ground state.

\begin{figure*}[]
\centering
\includegraphics[width=1.9\columnwidth]{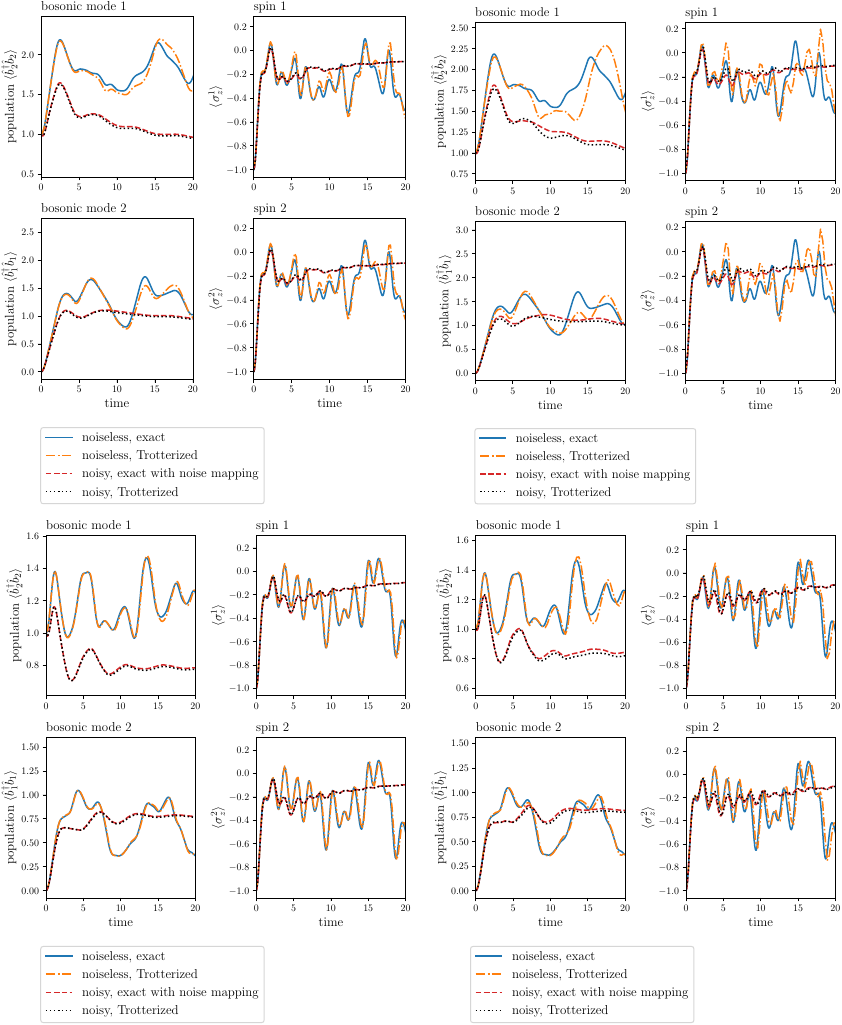}
\caption{Numerical simulations of Trotterized time evolution of the Dicke model as defined in Fig.~\ref{fig:noise_mapping},
for different number of included energy levels~$d$ and different Trotter time steps~$\tau$.
(top left)~$d=8, \tau=0.2$,
(top right)~$d=8, \tau=0.3$,
(bottom left)~$d=4, \tau=0.2$,
(bottom right)~$d=4, \tau=0.3$.
We see that the Trotter error (i.e., the difference between the exact and the Trotterized solutions)
increases with~$\tau$ but decreases significantly when lowering~$d$.
The latter result implies that most of the Trotter error in the simulation presented
in Fig.~\ref{fig:noise_mapping}, corresponding here to result~(top left),
arises from dynamics at high boson numbers.
}
\label{fig:noise_mapping_comparison}
\end{figure*}

Next, in Fig.~\ref{fig:noise_mapping_comparison}, we study the origin of the Trotter error in the simulation shown in Fig.~\ref{fig:noise_mapping}.
Here, the Trotter error is the difference between the numerically ``exact'' solution and the ``Trotterized'' solution.
In particular, we look at the size of the error 
for different number of included energy levels~$d$ and different Trotter time steps~$\tau$.
In Fig.~\ref{fig:noise_mapping_comparison}, we observe that the error increases with increasing the time step~$\tau$, as expected.
We also see that the Trotter error decreases strongly with lowering~$d$.
The reduction of Trotter error when reducing~$N_\textrm{bosons}$ is indeed expected,
as derived in Sec.~\ref{sec:Trotter_error_formulas}.
Furthermore, the observation of strong reduction implies that most of the Trotter error in Fig.~\ref{fig:noise_mapping} arises from
time propagating dynamics at high populations of the bosonic modes.
In additional simulations, we also find that the error does not depend on the relative shift of the time discretization,
e.g., if the values of the coupling phases are taken at the midpoints or beginnings of each Trotter time interval.

\subsection{Coherent JC gate errors and their mitigation}
An important difference to conventional all-qubit computation (representation of the bosonic modes by qubits)
is that errors in a device utilizing native bosonic modes are expected to be described by simple models~\cite{Liu2024}.
This makes it more plausible to understand, account for, and mitigate such errors.
Instead, for all-qubit processors, individual errors of qubits representing the bosonic modes
cannot generally be assumed to be described by a simple collective error model,
nor to map any sound physics in the simulated model.
The effect of resonator dissipation was addressed in Sec.~\ref{sec:noise_mapping} and mapped to broadening of simulated bosonic modes.
Below, we shortly discuss the expected form of coherent errors
that can rise due to a finite non-linearity of the used transmon qubits~\cite{Koch2007,Blais2021_RevModPhys}.

We can expect theoretically (at least) three types of coherent errors: (i) ZZ-type interaction between the qubit and the resonator
during the JC gates, (ii) a small shift of resonator frequency during the JC gate, and (iii)
Kerr anharmonicity of the resonator induced by the presence of the coupler and/or the logical qubit~\cite{Kirchmair2013}.

A simple estimate for the ZZ-interaction is obtained by perturbation theory with regards to
the presence of the
first non-computational level of the transmon, usually referred as the level $\vert f\rangle$.
It locates at frequency $2\omega_{eg}-K_q$, where $\omega_{eg}$ is the energy level difference between the ground
and the first excited state and $K_q$ is the anharmonicity.
At qubit-resonator resonance ($\omega_{ge} = \omega_r$),
the shift of the computation-space (split) energy levels are approximately~$\sim n g^2/2K_q= n\chi$,
where $g$ is the qubit-resonator coupling and $n$ is the number of photons in the resonator.
The shift is due to coupling between the levels~$\vert e\rangle$ and~$\vert f\rangle$.
(Here, also a similar shift in the effective resonator frequency may appear.)
We can then expect an effectively applied (error) gate which is of the form
$\tilde U(t_\textrm{gate}\chi) \sim \exp(-\textrm{i}t_\textrm{gate}\chi \sigma_z \hat b^\dagger \hat b/2)$.
The effect of such term is to be compared with the frequency of the gate
Jaynes-Cummings gate itself, which is~$g$.
When implemented using $g$ of few Mhz, and anharmonicity $K_q>100$~MHz,
we can expect this term to stay small.

Important is that coherent error of such form can be taken into account on algorithmic level
and mitigated. In particular, when considering quantum Rabi type models (Sec.~\ref{sec:Decomposing_QR}), with
circuits where every second JC gate is surrounded by X gates,
we effectively changes every second such coherent error to
$\tilde U(t_\textrm{gate}\chi)\rightarrow \tilde U(-t_\textrm{gate}\chi)$.
This then removes the total error in the leading-order of the small parameter $t_\textrm{gate}\chi$.

The other mentioned coherent errors have the form of momentary resonator frequency shift
and possibly constantly present Kerr anharmonicity.
In particular, the Kerr effect introduces an term
of the form $-K \hat b^\dagger\hat b^\dagger\hat b\hat b/2$~\cite{Kirchmair2013}.
We again estimate that in our setup the size of $K\sim\chi^2/4K_q$ is negligible.
Nevertheless, in principle, such error may also be accounted in
in the simulated model and possibly mitigated on the algorithmic level.
It is also clear that momentary shifts in the resonator frequency map to effective bosonic-mode rotation gates,
mapping to Trotterization of finite bosonic-mode frequencies,
which can also be accounted for and removed on the algorithmic level.

\bibliography{references}

\begin{thebibliography}{75}%
\makeatletter
\providecommand \@ifxundefined [1]{%
 \@ifx{#1\undefined}
}%
\providecommand \@ifnum [1]{%
 \ifnum #1\expandafter \@firstoftwo
 \else \expandafter \@secondoftwo
 \fi
}%
\providecommand \@ifx [1]{%
 \ifx #1\expandafter \@firstoftwo
 \else \expandafter \@secondoftwo
 \fi
}%
\providecommand \natexlab [1]{#1}%
\providecommand \enquote  [1]{``#1''}%
\providecommand \bibnamefont  [1]{#1}%
\providecommand \bibfnamefont [1]{#1}%
\providecommand \citenamefont [1]{#1}%
\providecommand \href@noop [0]{\@secondoftwo}%
\providecommand \href [0]{\begingroup \@sanitize@url \@href}%
\providecommand \@href[1]{\@@startlink{#1}\@@href}%
\providecommand \@@href[1]{\endgroup#1\@@endlink}%
\providecommand \@sanitize@url [0]{\catcode `\\12\catcode `\$12\catcode
  `\&12\catcode `\#12\catcode `\^12\catcode `\_12\catcode `\%12\relax}%
\providecommand \@@startlink[1]{}%
\providecommand \@@endlink[0]{}%
\providecommand \url  [0]{\begingroup\@sanitize@url \@url }%
\providecommand \@url [1]{\endgroup\@href {#1}{\urlprefix }}%
\providecommand \urlprefix  [0]{URL }%
\providecommand \Eprint [0]{\href }%
\providecommand \doibase [0]{https://doi.org/}%
\providecommand \selectlanguage [0]{\@gobble}%
\providecommand \bibinfo  [0]{\@secondoftwo}%
\providecommand \bibfield  [0]{\@secondoftwo}%
\providecommand \translation [1]{[#1]}%
\providecommand \BibitemOpen [0]{}%
\providecommand \bibitemStop [0]{}%
\providecommand \bibitemNoStop [0]{.\EOS\space}%
\providecommand \EOS [0]{\spacefactor3000\relax}%
\providecommand \BibitemShut  [1]{\csname bibitem#1\endcsname}%
\let\auto@bib@innerbib\@empty
%</preamble>
\bibitem [{\citenamefont {Fauseweh}(2024)}]{Fauseweh2024}%
  \BibitemOpen
  \bibfield  {author} {\bibinfo {author} {\bibfnamefont {B.}~\bibnamefont
  {Fauseweh}},\ }\bibfield  {title} {\bibinfo {title} {Quantum many-body
  simulations on digital quantum computers: State-of-the-art and future
  challenges},\ }\href {https://doi.org/10.1038/s41467-024-46402-9} {\bibfield
  {journal} {\bibinfo  {journal} {Nature Communications}\ }\textbf {\bibinfo
  {volume} {15}},\ \bibinfo {pages} {2123} (\bibinfo {year}
  {2024})}\BibitemShut {NoStop}%
\bibitem [{\citenamefont {Kippelen}\ and\ \citenamefont
  {Br\'edas}(2009)}]{Kippelen2009}%
  \BibitemOpen
  \bibfield  {author} {\bibinfo {author} {\bibfnamefont {B.}~\bibnamefont
  {Kippelen}}\ and\ \bibinfo {author} {\bibfnamefont {J.~L.}\ \bibnamefont
  {Br\'edas}},\ }\bibfield  {title} {\bibinfo {title} {Organic photovoltaics},\
  }\href {https://doi.org/10.1039/b812502n} {\bibfield  {journal} {\bibinfo
  {journal} {Energy and Environmental Science}\ }\textbf {\bibinfo {volume}
  {2}},\ \bibinfo {pages} {251} (\bibinfo {year} {2009})}\BibitemShut {NoStop}%
\bibitem [{\citenamefont {Ostroverkhova}(2016)}]{Ostroverkhova2016}%
  \BibitemOpen
  \bibfield  {author} {\bibinfo {author} {\bibfnamefont {O.}~\bibnamefont
  {Ostroverkhova}},\ }\bibfield  {title} {\bibinfo {title} {Organic
  optoelectronic materials: Mechanisms and applications},\ }\href
  {https://doi.org/10.1021/acs.chemrev.6b00127} {\bibfield  {journal} {\bibinfo
   {journal} {Chemical Reviews}\ }\textbf {\bibinfo {volume} {116}},\ \bibinfo
  {pages} {13279} (\bibinfo {year} {2016})}\BibitemShut {NoStop}%
\bibitem [{\citenamefont {Ishizaki}\ and\ \citenamefont
  {Fleming}(2012)}]{Ishizaki2012}%
  \BibitemOpen
  \bibfield  {author} {\bibinfo {author} {\bibfnamefont {A.}~\bibnamefont
  {Ishizaki}}\ and\ \bibinfo {author} {\bibfnamefont {G.~R.}\ \bibnamefont
  {Fleming}},\ }\bibfield  {title} {\bibinfo {title} {Quantum coherence in
  photosynthetic light harvesting},\ }\href
  {https://doi.org/10.1146/annurev-conmatphys-020911-125126} {\bibfield
  {journal} {\bibinfo  {journal} {Annu. Rev. Condens. Matter Phys.}\ }\textbf
  {\bibinfo {volume} {3}},\ \bibinfo {pages} {333} (\bibinfo {year}
  {2012})}\BibitemShut {NoStop}%
\bibitem [{\citenamefont {Huelga}\ and\ \citenamefont
  {Plenio}(2013)}]{Huelga_2013}%
  \BibitemOpen
  \bibfield  {author} {\bibinfo {author} {\bibfnamefont {S.}~\bibnamefont
  {Huelga}}\ and\ \bibinfo {author} {\bibfnamefont {M.}~\bibnamefont
  {Plenio}},\ }\bibfield  {title} {\bibinfo {title} {Vibrations, quanta and
  biology},\ }\href {https://doi.org/10.1080/00405000.2013.829687} {\bibfield
  {journal} {\bibinfo  {journal} {Contemporary Physics}\ }\textbf {\bibinfo
  {volume} {54}},\ \bibinfo {pages} {181–207} (\bibinfo {year}
  {2013})}\BibitemShut {NoStop}%
\bibitem [{\citenamefont {Ingold}\ and\ \citenamefont
  {Nazarov}(1992)}]{Ingold1992}%
  \BibitemOpen
  \bibfield  {author} {\bibinfo {author} {\bibfnamefont {G.-L.}\ \bibnamefont
  {Ingold}}\ and\ \bibinfo {author} {\bibfnamefont {Y.~V.}\ \bibnamefont
  {Nazarov}},\ }\href@noop {} {\emph {\bibinfo {title} {Single Charge
  Tunneling: Coulomb Blockade Phenomena in Nanostructures}}},\ edited by\
  \bibinfo {editor} {\bibfnamefont {H.}~\bibnamefont {Grabert}}\ and\ \bibinfo
  {editor} {\bibfnamefont {M.~H.}\ \bibnamefont {Devoret}}\ (\bibinfo
  {publisher} {Plenum, New York},\ \bibinfo {year} {1992})\BibitemShut
  {NoStop}%
\bibitem [{\citenamefont {Schoeller}\ and\ \citenamefont
  {Sch\"on}(1994)}]{SchoellerSchon1994}%
  \BibitemOpen
  \bibfield  {author} {\bibinfo {author} {\bibfnamefont {H.}~\bibnamefont
  {Schoeller}}\ and\ \bibinfo {author} {\bibfnamefont {G.}~\bibnamefont
  {Sch\"on}},\ }\bibfield  {title} {\bibinfo {title} {Mesoscopic quantum
  transport: Resonant tunneling in the presence of a strong coulomb
  interaction},\ }\href {https://doi.org/10.1103/PhysRevB.50.18436} {\bibfield
  {journal} {\bibinfo  {journal} {Phys. Rev. B}\ }\textbf {\bibinfo {volume}
  {50}},\ \bibinfo {pages} {18436} (\bibinfo {year} {1994})}\BibitemShut
  {NoStop}%
\bibitem [{\citenamefont {Lepp\"akangas}\ \emph {et~al.}(2023)\citenamefont
  {Lepp\"akangas}, \citenamefont {Vogt}, \citenamefont {Fratus}, \citenamefont
  {Bark}, \citenamefont {Vaitkus}, \citenamefont {Stadler}, \citenamefont
  {Reiner}, \citenamefont {Zanker},\ and\ \citenamefont
  {Marthaler}}]{Leppakangas2023}%
  \BibitemOpen
  \bibfield  {author} {\bibinfo {author} {\bibfnamefont {J.}~\bibnamefont
  {Lepp\"akangas}}, \bibinfo {author} {\bibfnamefont {N.}~\bibnamefont {Vogt}},
  \bibinfo {author} {\bibfnamefont {K.~R.}\ \bibnamefont {Fratus}}, \bibinfo
  {author} {\bibfnamefont {K.}~\bibnamefont {Bark}}, \bibinfo {author}
  {\bibfnamefont {J.~A.}\ \bibnamefont {Vaitkus}}, \bibinfo {author}
  {\bibfnamefont {P.}~\bibnamefont {Stadler}}, \bibinfo {author} {\bibfnamefont
  {J.-M.}\ \bibnamefont {Reiner}}, \bibinfo {author} {\bibfnamefont
  {S.}~\bibnamefont {Zanker}},\ and\ \bibinfo {author} {\bibfnamefont
  {M.}~\bibnamefont {Marthaler}},\ }\bibfield  {title} {\bibinfo {title}
  {Quantum algorithm for solving open-system dynamics on quantum computers
  using noise},\ }\href {https://doi.org/10.1103/PhysRevA.108.062424}
  {\bibfield  {journal} {\bibinfo  {journal} {Phys. Rev. A}\ }\textbf {\bibinfo
  {volume} {108}},\ \bibinfo {pages} {062424} (\bibinfo {year}
  {2023})}\BibitemShut {NoStop}%
\bibitem [{\citenamefont {Bohm}\ and\ \citenamefont {Pines}(1953)}]{Bohm1953}%
  \BibitemOpen
  \bibfield  {author} {\bibinfo {author} {\bibfnamefont {D.}~\bibnamefont
  {Bohm}}\ and\ \bibinfo {author} {\bibfnamefont {D.}~\bibnamefont {Pines}},\
  }\bibfield  {title} {\bibinfo {title} {A collective description of electron
  interactions: Iii.~coulomb interactions in a degenerate electron gas},\
  }\href {https://doi.org/10.1103/PhysRev.92.609} {\bibfield  {journal}
  {\bibinfo  {journal} {Phys. Rev.}\ }\textbf {\bibinfo {volume} {92}},\
  \bibinfo {pages} {609} (\bibinfo {year} {1953})}\BibitemShut {NoStop}%
\bibitem [{\citenamefont {Fröhlich}(1954)}]{Froehlich1954}%
  \BibitemOpen
  \bibfield  {author} {\bibinfo {author} {\bibfnamefont {H.}~\bibnamefont
  {Fröhlich}},\ }\bibfield  {title} {\bibinfo {title} {Electrons in lattice
  fields},\ }\href {https://doi.org/10.1080/00018735400101213} {\bibfield
  {journal} {\bibinfo  {journal} {Advances in Physics}\ }\textbf {\bibinfo
  {volume} {3}},\ \bibinfo {pages} {325} (\bibinfo {year} {1954})}\BibitemShut
  {NoStop}%
\bibitem [{\citenamefont {Cooper}(1956)}]{Cooper1956}%
  \BibitemOpen
  \bibfield  {author} {\bibinfo {author} {\bibfnamefont {L.~N.}\ \bibnamefont
  {Cooper}},\ }\bibfield  {title} {\bibinfo {title} {Bound electron pairs in a
  degenerate fermi gas},\ }\href {https://doi.org/10.1103/PhysRev.104.1189}
  {\bibfield  {journal} {\bibinfo  {journal} {Phys. Rev.}\ }\textbf {\bibinfo
  {volume} {104}},\ \bibinfo {pages} {1189} (\bibinfo {year}
  {1956})}\BibitemShut {NoStop}%
\bibitem [{\citenamefont {Holstein}(1959)}]{Holstein1959}%
  \BibitemOpen
  \bibfield  {author} {\bibinfo {author} {\bibfnamefont {T.}~\bibnamefont
  {Holstein}},\ }\bibfield  {title} {\bibinfo {title} {Studies of polaron
  motion: Part i. the molecular-crystal model},\ }\href
  {https://doi.org/https://doi.org/10.1016/0003-4916(59)90002-8} {\bibfield
  {journal} {\bibinfo  {journal} {Annals of Physics}\ }\textbf {\bibinfo
  {volume} {8}},\ \bibinfo {pages} {325} (\bibinfo {year} {1959})}\BibitemShut
  {NoStop}%
\bibitem [{\citenamefont {Shnirman}\ \emph {et~al.}(2002)\citenamefont
  {Shnirman}, \citenamefont {Makhlin},\ and\ \citenamefont
  {Sch\"on}}]{Shnirman2002}%
  \BibitemOpen
  \bibfield  {author} {\bibinfo {author} {\bibfnamefont {A.}~\bibnamefont
  {Shnirman}}, \bibinfo {author} {\bibfnamefont {Y.}~\bibnamefont {Makhlin}},\
  and\ \bibinfo {author} {\bibfnamefont {G.}~\bibnamefont {Sch\"on}},\
  }\bibfield  {title} {\bibinfo {title} {Noise and decoherence in quantum
  two-level systems},\ }\href
  {https://doi.org/10.1238/physica.topical.102a00147} {\bibfield  {journal}
  {\bibinfo  {journal} {Physica Scripta}\ }\textbf {\bibinfo {volume} {T102}},\
  \bibinfo {pages} {147} (\bibinfo {year} {2002})}\BibitemShut {NoStop}%
\bibitem [{\citenamefont {Bharti}\ \emph {et~al.}(2022)\citenamefont {Bharti},
  \citenamefont {Cervera-Lierta}, \citenamefont {Kyaw}, \citenamefont {Haug},
  \citenamefont {Alperin-Lea}, \citenamefont {Anand}, \citenamefont {Degroote},
  \citenamefont {Heimonen}, \citenamefont {Kottmann}, \citenamefont {Menke},
  \citenamefont {Mok}, \citenamefont {Sim}, \citenamefont {Kwek},\ and\
  \citenamefont {Aspuru-Guzik}}]{Barthi2022}%
  \BibitemOpen
  \bibfield  {author} {\bibinfo {author} {\bibfnamefont {K.}~\bibnamefont
  {Bharti}}, \bibinfo {author} {\bibfnamefont {A.}~\bibnamefont
  {Cervera-Lierta}}, \bibinfo {author} {\bibfnamefont {T.~H.}\ \bibnamefont
  {Kyaw}}, \bibinfo {author} {\bibfnamefont {T.}~\bibnamefont {Haug}}, \bibinfo
  {author} {\bibfnamefont {S.}~\bibnamefont {Alperin-Lea}}, \bibinfo {author}
  {\bibfnamefont {A.}~\bibnamefont {Anand}}, \bibinfo {author} {\bibfnamefont
  {M.}~\bibnamefont {Degroote}}, \bibinfo {author} {\bibfnamefont
  {H.}~\bibnamefont {Heimonen}}, \bibinfo {author} {\bibfnamefont {J.~S.}\
  \bibnamefont {Kottmann}}, \bibinfo {author} {\bibfnamefont {T.}~\bibnamefont
  {Menke}}, \bibinfo {author} {\bibfnamefont {W.-K.}\ \bibnamefont {Mok}},
  \bibinfo {author} {\bibfnamefont {S.}~\bibnamefont {Sim}}, \bibinfo {author}
  {\bibfnamefont {L.-C.}\ \bibnamefont {Kwek}},\ and\ \bibinfo {author}
  {\bibfnamefont {A.}~\bibnamefont {Aspuru-Guzik}},\ }\bibfield  {title}
  {\bibinfo {title} {Noisy intermediate-scale quantum algorithms},\ }\href
  {https://doi.org/10.1103/RevModPhys.94.015004} {\bibfield  {journal}
  {\bibinfo  {journal} {Rev. Mod. Phys.}\ }\textbf {\bibinfo {volume} {94}},\
  \bibinfo {pages} {015004} (\bibinfo {year} {2022})}\BibitemShut {NoStop}%
\bibitem [{\citenamefont {Sawaya}\ \emph {et~al.}(2020)\citenamefont {Sawaya},
  \citenamefont {Menke}, \citenamefont {Kyaw}, \citenamefont {Johri},
  \citenamefont {Aspuru-Guzik},\ and\ \citenamefont {Guerreschi}}]{Sawaya2020}%
  \BibitemOpen
  \bibfield  {author} {\bibinfo {author} {\bibfnamefont {N.~P.~D.}\
  \bibnamefont {Sawaya}}, \bibinfo {author} {\bibfnamefont {T.}~\bibnamefont
  {Menke}}, \bibinfo {author} {\bibfnamefont {T.~H.}\ \bibnamefont {Kyaw}},
  \bibinfo {author} {\bibfnamefont {S.}~\bibnamefont {Johri}}, \bibinfo
  {author} {\bibfnamefont {A.}~\bibnamefont {Aspuru-Guzik}},\ and\ \bibinfo
  {author} {\bibfnamefont {G.~G.}\ \bibnamefont {Guerreschi}},\ }\bibfield
  {title} {\bibinfo {title} {Resource-efficient digital quantum simulation of
  d-level systems for photonic, vibrational, and spin-s hamiltonians},\ }\href
  {https://doi.org/10.1038/s41534-020-0278-0} {\bibfield  {journal} {\bibinfo
  {journal} {npj Quantum Information}\ }\textbf {\bibinfo {volume} {6}},\
  \bibinfo {pages} {49} (\bibinfo {year} {2020})}\BibitemShut {NoStop}%
\bibitem [{\citenamefont {Mezzacapo}\ \emph {et~al.}(2014)\citenamefont
  {Mezzacapo}, \citenamefont {Las~Heras}, \citenamefont {Pedernales},
  \citenamefont {DiCarlo}, \citenamefont {Solano},\ and\ \citenamefont
  {Lamata}}]{Mezzacapo2014}%
  \BibitemOpen
  \bibfield  {author} {\bibinfo {author} {\bibfnamefont {A.}~\bibnamefont
  {Mezzacapo}}, \bibinfo {author} {\bibfnamefont {U.}~\bibnamefont
  {Las~Heras}}, \bibinfo {author} {\bibfnamefont {J.~S.}\ \bibnamefont
  {Pedernales}}, \bibinfo {author} {\bibfnamefont {L.}~\bibnamefont {DiCarlo}},
  \bibinfo {author} {\bibfnamefont {E.}~\bibnamefont {Solano}},\ and\ \bibinfo
  {author} {\bibfnamefont {L.}~\bibnamefont {Lamata}},\ }\bibfield  {title}
  {\bibinfo {title} {Digital quantum rabi and dicke models in superconducting
  circuits},\ }\href {https://doi.org/10.1038/srep07482} {\bibfield  {journal}
  {\bibinfo  {journal} {Scientific Reports}\ }\textbf {\bibinfo {volume} {4}},\
  \bibinfo {pages} {7482} (\bibinfo {year} {2014})}\BibitemShut {NoStop}%
\bibitem [{\citenamefont {Langford}\ \emph {et~al.}(2017)\citenamefont
  {Langford}, \citenamefont {Sagastizabal}, \citenamefont {Kounalakis},
  \citenamefont {Dickel}, \citenamefont {Bruno}, \citenamefont {Luthi},
  \citenamefont {Thoen}, \citenamefont {Endo},\ and\ \citenamefont
  {DiCarlo}}]{Langford2017}%
  \BibitemOpen
  \bibfield  {author} {\bibinfo {author} {\bibfnamefont {N.~K.}\ \bibnamefont
  {Langford}}, \bibinfo {author} {\bibfnamefont {R.}~\bibnamefont
  {Sagastizabal}}, \bibinfo {author} {\bibfnamefont {M.}~\bibnamefont
  {Kounalakis}}, \bibinfo {author} {\bibfnamefont {C.}~\bibnamefont {Dickel}},
  \bibinfo {author} {\bibfnamefont {A.}~\bibnamefont {Bruno}}, \bibinfo
  {author} {\bibfnamefont {F.}~\bibnamefont {Luthi}}, \bibinfo {author}
  {\bibfnamefont {D.~J.}\ \bibnamefont {Thoen}}, \bibinfo {author}
  {\bibfnamefont {A.}~\bibnamefont {Endo}},\ and\ \bibinfo {author}
  {\bibfnamefont {L.}~\bibnamefont {DiCarlo}},\ }\bibfield  {title} {\bibinfo
  {title} {Experimentally simulating the dynamics of quantum light and matter
  at deep-strong coupling},\ }\href
  {https://doi.org/10.1038/s41467-017-01061-x} {\bibfield  {journal} {\bibinfo
  {journal} {Nat. Commun.}\ }\textbf {\bibinfo {volume} {8}},\ \bibinfo {pages}
  {1715} (\bibinfo {year} {2017})}\BibitemShut {NoStop}%
\bibitem [{\citenamefont {Wang}\ \emph {et~al.}(2023)\citenamefont {Wang},
  \citenamefont {Frattini}, \citenamefont {Chapman}, \citenamefont {Puri},
  \citenamefont {Girvin}, \citenamefont {Devoret},\ and\ \citenamefont
  {Schoelkopf}}]{Wang2023}%
  \BibitemOpen
  \bibfield  {author} {\bibinfo {author} {\bibfnamefont {C.~S.}\ \bibnamefont
  {Wang}}, \bibinfo {author} {\bibfnamefont {N.~E.}\ \bibnamefont {Frattini}},
  \bibinfo {author} {\bibfnamefont {B.~J.}\ \bibnamefont {Chapman}}, \bibinfo
  {author} {\bibfnamefont {S.}~\bibnamefont {Puri}}, \bibinfo {author}
  {\bibfnamefont {S.~M.}\ \bibnamefont {Girvin}}, \bibinfo {author}
  {\bibfnamefont {M.~H.}\ \bibnamefont {Devoret}},\ and\ \bibinfo {author}
  {\bibfnamefont {R.~J.}\ \bibnamefont {Schoelkopf}},\ }\bibfield  {title}
  {\bibinfo {title} {Observation of wave-packet branching through an engineered
  conical intersection},\ }\href {https://doi.org/10.1103/PhysRevX.13.011008}
  {\bibfield  {journal} {\bibinfo  {journal} {Phys. Rev. X}\ }\textbf {\bibinfo
  {volume} {13}},\ \bibinfo {pages} {011008} (\bibinfo {year}
  {2023})}\BibitemShut {NoStop}%
\bibitem [{\citenamefont {Katz}\ and\ \citenamefont {Monroe}(2023)}]{Katz2023}%
  \BibitemOpen
  \bibfield  {author} {\bibinfo {author} {\bibfnamefont {O.}~\bibnamefont
  {Katz}}\ and\ \bibinfo {author} {\bibfnamefont {C.}~\bibnamefont {Monroe}},\
  }\bibfield  {title} {\bibinfo {title} {Programmable quantum simulations of
  bosonic systems with trapped ions},\ }\href
  {https://doi.org/10.1103/physrevlett.131.033604} {\bibfield  {journal}
  {\bibinfo  {journal} {Physical Review Letters}\ }\textbf {\bibinfo {volume}
  {131}},\ \bibinfo {pages} {033604} (\bibinfo {year} {2023})}\BibitemShut
  {NoStop}%
\bibitem [{\citenamefont {Dutta}\ \emph {et~al.}(2024)\citenamefont {Dutta},
  \citenamefont {Cabral}, \citenamefont {Lyu}, \citenamefont {Vu},
  \citenamefont {Wang}, \citenamefont {Allen}, \citenamefont {Dan},
  \citenamefont {Cortiñas}, \citenamefont {Khazaei}, \citenamefont {Schäfer},
  \citenamefont {Albornoz}, \citenamefont {Smart}, \citenamefont {Nie},
  \citenamefont {Devoret}, \citenamefont {Mazziotti}, \citenamefont {Narang},
  \citenamefont {Wang}, \citenamefont {Whitfield}, \citenamefont {Wilson},
  \citenamefont {Hendrickson}, \citenamefont {Lidar}, \citenamefont
  {Pérez-Bernal}, \citenamefont {Santos}, \citenamefont {Kais}, \citenamefont
  {Geva},\ and\ \citenamefont {Batista}}]{Dutta2024}%
  \BibitemOpen
  \bibfield  {author} {\bibinfo {author} {\bibfnamefont {R.}~\bibnamefont
  {Dutta}}, \bibinfo {author} {\bibfnamefont {D.~G.~A.}\ \bibnamefont
  {Cabral}}, \bibinfo {author} {\bibfnamefont {N.}~\bibnamefont {Lyu}},
  \bibinfo {author} {\bibfnamefont {N.~P.}\ \bibnamefont {Vu}}, \bibinfo
  {author} {\bibfnamefont {Y.}~\bibnamefont {Wang}}, \bibinfo {author}
  {\bibfnamefont {B.}~\bibnamefont {Allen}}, \bibinfo {author} {\bibfnamefont
  {X.}~\bibnamefont {Dan}}, \bibinfo {author} {\bibfnamefont {R.~G.}\
  \bibnamefont {Cortiñas}}, \bibinfo {author} {\bibfnamefont {P.}~\bibnamefont
  {Khazaei}}, \bibinfo {author} {\bibfnamefont {M.}~\bibnamefont {Schäfer}},
  \bibinfo {author} {\bibfnamefont {A.~C. C.~d.}\ \bibnamefont {Albornoz}},
  \bibinfo {author} {\bibfnamefont {S.~E.}\ \bibnamefont {Smart}}, \bibinfo
  {author} {\bibfnamefont {S.}~\bibnamefont {Nie}}, \bibinfo {author}
  {\bibfnamefont {M.~H.}\ \bibnamefont {Devoret}}, \bibinfo {author}
  {\bibfnamefont {D.~A.}\ \bibnamefont {Mazziotti}}, \bibinfo {author}
  {\bibfnamefont {P.}~\bibnamefont {Narang}}, \bibinfo {author} {\bibfnamefont
  {C.}~\bibnamefont {Wang}}, \bibinfo {author} {\bibfnamefont {J.~D.}\
  \bibnamefont {Whitfield}}, \bibinfo {author} {\bibfnamefont {A.~K.}\
  \bibnamefont {Wilson}}, \bibinfo {author} {\bibfnamefont {H.~P.}\
  \bibnamefont {Hendrickson}}, \bibinfo {author} {\bibfnamefont {D.~A.}\
  \bibnamefont {Lidar}}, \bibinfo {author} {\bibfnamefont {F.}~\bibnamefont
  {Pérez-Bernal}}, \bibinfo {author} {\bibfnamefont {L.~F.}\ \bibnamefont
  {Santos}}, \bibinfo {author} {\bibfnamefont {S.}~\bibnamefont {Kais}},
  \bibinfo {author} {\bibfnamefont {E.}~\bibnamefont {Geva}},\ and\ \bibinfo
  {author} {\bibfnamefont {V.~S.}\ \bibnamefont {Batista}},\ }\bibfield
  {title} {\bibinfo {title} {Simulating chemistry on bosonic quantum devices},\
  }\href {https://doi.org/10.1021/acs.jctc.4c00544} {\bibfield  {journal}
  {\bibinfo  {journal} {Journal of Chemical Theory and Computation}\ }\textbf
  {\bibinfo {volume} {20}},\ \bibinfo {pages} {6426} (\bibinfo {year}
  {2024})}\BibitemShut {NoStop}%
\bibitem [{\citenamefont {Liu}\ \emph {et~al.}(2024)\citenamefont {Liu},
  \citenamefont {Singh}, \citenamefont {Smith}, \citenamefont {Crane},
  \citenamefont {Martyn}, \citenamefont {Eickbusch}, \citenamefont {Schuckert},
  \citenamefont {Li}, \citenamefont {Sinanan-Singh}, \citenamefont {Soley},
  \citenamefont {Tsunoda}, \citenamefont {Chuang}, \citenamefont {Wiebe},\ and\
  \citenamefont {Girvin}}]{Liu2024}%
  \BibitemOpen
  \bibfield  {author} {\bibinfo {author} {\bibfnamefont {Y.}~\bibnamefont
  {Liu}}, \bibinfo {author} {\bibfnamefont {S.}~\bibnamefont {Singh}}, \bibinfo
  {author} {\bibfnamefont {K.~C.}\ \bibnamefont {Smith}}, \bibinfo {author}
  {\bibfnamefont {E.}~\bibnamefont {Crane}}, \bibinfo {author} {\bibfnamefont
  {J.~M.}\ \bibnamefont {Martyn}}, \bibinfo {author} {\bibfnamefont
  {A.}~\bibnamefont {Eickbusch}}, \bibinfo {author} {\bibfnamefont
  {A.}~\bibnamefont {Schuckert}}, \bibinfo {author} {\bibfnamefont {R.~D.}\
  \bibnamefont {Li}}, \bibinfo {author} {\bibfnamefont {J.}~\bibnamefont
  {Sinanan-Singh}}, \bibinfo {author} {\bibfnamefont {M.~B.}\ \bibnamefont
  {Soley}}, \bibinfo {author} {\bibfnamefont {T.}~\bibnamefont {Tsunoda}},
  \bibinfo {author} {\bibfnamefont {I.~L.}\ \bibnamefont {Chuang}}, \bibinfo
  {author} {\bibfnamefont {N.}~\bibnamefont {Wiebe}},\ and\ \bibinfo {author}
  {\bibfnamefont {S.~M.}\ \bibnamefont {Girvin}},\ }\bibfield  {title}
  {\bibinfo {title} {Hybrid oscillator-qubit quantum processors: Instruction
  set architectures, abstract machine models, and applications},\ }\bibfield
  {journal} {\bibinfo  {journal} {arXiv.2407.10381}\ }\href
  {https://doi.org/10.48550/arXiv.2407.10381} {10.48550/arXiv.2407.10381}
  (\bibinfo {year} {2024})\BibitemShut {NoStop}%
\bibitem [{\citenamefont {Crane}\ \emph {et~al.}(2024)\citenamefont {Crane},
  \citenamefont {Smith}, \citenamefont {Tomesh}, \citenamefont {Eickbusch},
  \citenamefont {Martyn}, \citenamefont {Kühn}, \citenamefont {Funcke},
  \citenamefont {DeMarco}, \citenamefont {Chuang}, \citenamefont {Wiebe},
  \citenamefont {Schuckert},\ and\ \citenamefont {Girvin}}]{Crane2024}%
  \BibitemOpen
  \bibfield  {author} {\bibinfo {author} {\bibfnamefont {E.}~\bibnamefont
  {Crane}}, \bibinfo {author} {\bibfnamefont {K.~C.}\ \bibnamefont {Smith}},
  \bibinfo {author} {\bibfnamefont {T.}~\bibnamefont {Tomesh}}, \bibinfo
  {author} {\bibfnamefont {A.}~\bibnamefont {Eickbusch}}, \bibinfo {author}
  {\bibfnamefont {J.~M.}\ \bibnamefont {Martyn}}, \bibinfo {author}
  {\bibfnamefont {S.}~\bibnamefont {Kühn}}, \bibinfo {author} {\bibfnamefont
  {L.}~\bibnamefont {Funcke}}, \bibinfo {author} {\bibfnamefont {M.~A.}\
  \bibnamefont {DeMarco}}, \bibinfo {author} {\bibfnamefont {I.~L.}\
  \bibnamefont {Chuang}}, \bibinfo {author} {\bibfnamefont {N.}~\bibnamefont
  {Wiebe}}, \bibinfo {author} {\bibfnamefont {A.}~\bibnamefont {Schuckert}},\
  and\ \bibinfo {author} {\bibfnamefont {S.~M.}\ \bibnamefont {Girvin}},\
  }\bibfield  {title} {\bibinfo {title} {Hybrid oscillator-qubit quantum
  processors: Simulating fermions, bosons, and gauge fields},\ }\bibfield
  {journal} {\bibinfo  {journal} {arXiv:2409.03747}\ }\href
  {https://doi.org/10.48550/arXiv.2409.03747} {10.48550/arXiv.2409.03747}
  (\bibinfo {year} {2024})\BibitemShut {NoStop}%
\bibitem [{\citenamefont {Shapiro}\ \emph {et~al.}(2024)\citenamefont
  {Shapiro}, \citenamefont {Weber}, \citenamefont {Bode}, \citenamefont
  {Wilhelm},\ and\ \citenamefont {Bagrets}}]{Shapiro2024}%
  \BibitemOpen
  \bibfield  {author} {\bibinfo {author} {\bibfnamefont {D.~S.}\ \bibnamefont
  {Shapiro}}, \bibinfo {author} {\bibfnamefont {Y.}~\bibnamefont {Weber}},
  \bibinfo {author} {\bibfnamefont {T.}~\bibnamefont {Bode}}, \bibinfo {author}
  {\bibfnamefont {F.~K.}\ \bibnamefont {Wilhelm}},\ and\ \bibinfo {author}
  {\bibfnamefont {D.}~\bibnamefont {Bagrets}},\ }\bibfield  {title} {\bibinfo
  {title} {Quantum simulation of the dicke-ising model via digital-analog
  algorithms},\ }\bibfield  {journal} {\bibinfo  {journal} {arXiv:2412.14285}\
  }\href {https://doi.org/10.48550/arXiv.2412.14285}
  {10.48550/arXiv.2412.14285} (\bibinfo {year} {2024})\BibitemShut {NoStop}%
\bibitem [{\citenamefont {Vu}\ \emph {et~al.}(2025)\citenamefont {Vu},
  \citenamefont {Dong}, \citenamefont {Dan}, \citenamefont {Lyu}, \citenamefont
  {Batista},\ and\ \citenamefont {Liu}}]{Vu2025}%
  \BibitemOpen
  \bibfield  {author} {\bibinfo {author} {\bibfnamefont {N.~P.}\ \bibnamefont
  {Vu}}, \bibinfo {author} {\bibfnamefont {D.}~\bibnamefont {Dong}}, \bibinfo
  {author} {\bibfnamefont {X.}~\bibnamefont {Dan}}, \bibinfo {author}
  {\bibfnamefont {N.}~\bibnamefont {Lyu}}, \bibinfo {author} {\bibfnamefont
  {V.}~\bibnamefont {Batista}},\ and\ \bibinfo {author} {\bibfnamefont
  {Y.}~\bibnamefont {Liu}},\ }\bibfield  {title} {\bibinfo {title} {A
  computational framework for simulations of dissipative non-adiabatic dynamics
  on hybrid oscillator-qubit quantum devices},\ }\bibfield  {journal} {\bibinfo
   {journal} {2502.17820}\ }\href {https://doi.org/10.48550/arXiv.2502.17820}
  {10.48550/arXiv.2502.17820} (\bibinfo {year} {2025})\BibitemShut {NoStop}%
\bibitem [{\citenamefont {Schönauer}\ \emph {et~al.}(2025)\citenamefont
  {Schönauer}, \citenamefont {Schmitteckert}, \citenamefont {Vogt},
  \citenamefont {Reiner}, \citenamefont {Zanker}, \citenamefont {Marthaler},
  \citenamefont {Eckl},\ and\ \citenamefont {Kuehn}}]{EP4487263A1}%
  \BibitemOpen
  \bibfield  {author} {\bibinfo {author} {\bibfnamefont {B.~M.}\ \bibnamefont
  {Schönauer}}, \bibinfo {author} {\bibfnamefont {P.}~\bibnamefont
  {Schmitteckert}}, \bibinfo {author} {\bibfnamefont {N.}~\bibnamefont {Vogt}},
  \bibinfo {author} {\bibfnamefont {J.-M.}\ \bibnamefont {Reiner}}, \bibinfo
  {author} {\bibfnamefont {S.}~\bibnamefont {Zanker}}, \bibinfo {author}
  {\bibfnamefont {M.}~\bibnamefont {Marthaler}}, \bibinfo {author}
  {\bibfnamefont {T.}~\bibnamefont {Eckl}},\ and\ \bibinfo {author}
  {\bibfnamefont {M.}~\bibnamefont {Kuehn}},\ }\href
  {https://worldwide.espacenet.com/patent/search/family/086162336/publication/EP4487263A1}
  {\bibinfo {title} {A quantum computer for performing quantum operations}},\
  \bibinfo {howpublished} {European Patent No. EP4487263A1, European Patent
  Office} (\bibinfo {year} {2025})\BibitemShut {NoStop}%
\bibitem [{\citenamefont {Lloyd}(1996)}]{Lloyd1996}%
  \BibitemOpen
  \bibfield  {author} {\bibinfo {author} {\bibfnamefont {S.}~\bibnamefont
  {Lloyd}},\ }\bibfield  {title} {\bibinfo {title} {Universal quantum
  simulators},\ }\href {https://doi.org/10.1126/science.273.5278.1073}
  {\bibfield  {journal} {\bibinfo  {journal} {Science}\ }\textbf {\bibinfo
  {volume} {273}},\ \bibinfo {pages} {1073} (\bibinfo {year}
  {1996})}\BibitemShut {NoStop}%
\bibitem [{\citenamefont {Tseng}\ \emph {et~al.}(2000)\citenamefont {Tseng},
  \citenamefont {Somaroo}, \citenamefont {Sharf}, \citenamefont {Knill},
  \citenamefont {Laflamme}, \citenamefont {Havel},\ and\ \citenamefont
  {Cory}}]{Tseng2010}%
  \BibitemOpen
  \bibfield  {author} {\bibinfo {author} {\bibfnamefont {C.~H.}\ \bibnamefont
  {Tseng}}, \bibinfo {author} {\bibfnamefont {S.}~\bibnamefont {Somaroo}},
  \bibinfo {author} {\bibfnamefont {Y.}~\bibnamefont {Sharf}}, \bibinfo
  {author} {\bibfnamefont {E.}~\bibnamefont {Knill}}, \bibinfo {author}
  {\bibfnamefont {R.}~\bibnamefont {Laflamme}}, \bibinfo {author}
  {\bibfnamefont {T.~F.}\ \bibnamefont {Havel}},\ and\ \bibinfo {author}
  {\bibfnamefont {D.~G.}\ \bibnamefont {Cory}},\ }\bibfield  {title} {\bibinfo
  {title} {Quantum simulation with natural decoherence},\ }\href
  {https://doi.org/10.1103/PhysRevA.62.032309} {\bibfield  {journal} {\bibinfo
  {journal} {Phys. Rev. A}\ }\textbf {\bibinfo {volume} {62}},\ \bibinfo
  {pages} {032309} (\bibinfo {year} {2000})}\BibitemShut {NoStop}%
\bibitem [{\citenamefont {Rost}\ \emph {et~al.}(2020)\citenamefont {Rost},
  \citenamefont {Jones}, \citenamefont {Vyushkova}, \citenamefont {Ali},
  \citenamefont {Cullip}, \citenamefont {Vyushkov},\ and\ \citenamefont
  {Nabrzyski}}]{Rost2020}%
  \BibitemOpen
  \bibfield  {author} {\bibinfo {author} {\bibfnamefont {B.}~\bibnamefont
  {Rost}}, \bibinfo {author} {\bibfnamefont {B.}~\bibnamefont {Jones}},
  \bibinfo {author} {\bibfnamefont {M.}~\bibnamefont {Vyushkova}}, \bibinfo
  {author} {\bibfnamefont {A.}~\bibnamefont {Ali}}, \bibinfo {author}
  {\bibfnamefont {C.}~\bibnamefont {Cullip}}, \bibinfo {author} {\bibfnamefont
  {A.}~\bibnamefont {Vyushkov}},\ and\ \bibinfo {author} {\bibfnamefont
  {J.}~\bibnamefont {Nabrzyski}},\ }\bibfield  {title} {\bibinfo {title}
  {Simulation of thermal relaxation in spin chemistry systems on a quantum
  computer using inherent qubit decoherence},\ }\bibfield  {journal} {\bibinfo
  {journal} {arXiv:2001.00794}\ }\href
  {https://doi.org/10.48550/arXiv.2001.00794} {10.48550/arXiv.2001.00794}
  (\bibinfo {year} {2020})\BibitemShut {NoStop}%
\bibitem [{\citenamefont {Sun}\ \emph {et~al.}(2021)\citenamefont {Sun},
  \citenamefont {Shih},\ and\ \citenamefont {Cheng}}]{Sun2021}%
  \BibitemOpen
  \bibfield  {author} {\bibinfo {author} {\bibfnamefont {S.}~\bibnamefont
  {Sun}}, \bibinfo {author} {\bibfnamefont {L.-C.}\ \bibnamefont {Shih}},\ and\
  \bibinfo {author} {\bibfnamefont {Y.-C.}\ \bibnamefont {Cheng}},\ }\bibfield
  {title} {\bibinfo {title} {Efficient quantum simulation of open quantum
  system dynamics on noisy quantum computers},\ }\bibfield  {journal} {\bibinfo
   {journal} {arXiv.2106.12882}\ }\href
  {https://doi.org/10.48550/arXiv.2106.12882} {10.48550/arXiv.2106.12882}
  (\bibinfo {year} {2021})\BibitemShut {NoStop}%
\bibitem [{\citenamefont {Guimarães}\ \emph {et~al.}(2023)\citenamefont
  {Guimarães}, \citenamefont {Lim}, \citenamefont {Vasilevskiy}, \citenamefont
  {Huelga},\ and\ \citenamefont {Plenio}}]{Guimaraes2023}%
  \BibitemOpen
  \bibfield  {author} {\bibinfo {author} {\bibfnamefont {J.~D.}\ \bibnamefont
  {Guimarães}}, \bibinfo {author} {\bibfnamefont {J.}~\bibnamefont {Lim}},
  \bibinfo {author} {\bibfnamefont {M.~I.}\ \bibnamefont {Vasilevskiy}},
  \bibinfo {author} {\bibfnamefont {S.~F.}\ \bibnamefont {Huelga}},\ and\
  \bibinfo {author} {\bibfnamefont {M.~B.}\ \bibnamefont {Plenio}},\ }\bibfield
   {title} {\bibinfo {title} {Noise-assisted digital quantum simulation of open
  systems using partial probabilistic error cancellation},\ }\href
  {https://doi.org/10.1103/prxquantum.4.040329} {\bibfield  {journal} {\bibinfo
   {journal} {PRX Quantum}\ }\textbf {\bibinfo {volume} {4}},\ \bibinfo {pages}
  {040329} (\bibinfo {year} {2023})}\BibitemShut {NoStop}%
\bibitem [{\citenamefont {Ma}\ \emph {et~al.}(2024)\citenamefont {Ma},
  \citenamefont {Hanks}, \citenamefont {Gneusheva},\ and\ \citenamefont
  {Kim}}]{Ma2024}%
  \BibitemOpen
  \bibfield  {author} {\bibinfo {author} {\bibfnamefont {Y.}~\bibnamefont
  {Ma}}, \bibinfo {author} {\bibfnamefont {M.}~\bibnamefont {Hanks}}, \bibinfo
  {author} {\bibfnamefont {E.}~\bibnamefont {Gneusheva}},\ and\ \bibinfo
  {author} {\bibfnamefont {M.~S.}\ \bibnamefont {Kim}},\ }\bibfield  {title}
  {\bibinfo {title} {Reshaping quantum device noise via quantum error
  correction},\ }\bibfield  {journal} {\bibinfo  {journal} {arXiv:2411.00751}\
  }\href {https://doi.org/10.48550/arXiv.2411.00751}
  {10.48550/arXiv.2411.00751} (\bibinfo {year} {2024})\BibitemShut {NoStop}%
\bibitem [{\citenamefont {Fratus}\ \emph {et~al.}(2022)\citenamefont {Fratus},
  \citenamefont {Bark}, \citenamefont {Vogt}, \citenamefont {Lepp\"akangas},
  \citenamefont {Zanker}, \citenamefont {Marthaler},\ and\ \citenamefont
  {Reiner}}]{Fratus2022}%
  \BibitemOpen
  \bibfield  {author} {\bibinfo {author} {\bibfnamefont {K.~R.}\ \bibnamefont
  {Fratus}}, \bibinfo {author} {\bibfnamefont {K.}~\bibnamefont {Bark}},
  \bibinfo {author} {\bibfnamefont {N.}~\bibnamefont {Vogt}}, \bibinfo {author}
  {\bibfnamefont {J.}~\bibnamefont {Lepp\"akangas}}, \bibinfo {author}
  {\bibfnamefont {S.}~\bibnamefont {Zanker}}, \bibinfo {author} {\bibfnamefont
  {M.}~\bibnamefont {Marthaler}},\ and\ \bibinfo {author} {\bibfnamefont
  {J.-M.}\ \bibnamefont {Reiner}},\ }\bibfield  {title} {\bibinfo {title}
  {Describing trotterized time evolutions on noisy quantum computers via static
  effective lindbladians},\ }\bibfield  {journal} {\bibinfo  {journal}
  {arXiv.2210.11371}\ }\href {https://doi.org/10.48550/arXiv.2210.11371}
  {10.48550/arXiv.2210.11371} (\bibinfo {year} {2022})\BibitemShut {NoStop}%
\bibitem [{\citenamefont {Renger}\ \emph {et~al.}(2025)\citenamefont {Renger},
  \citenamefont {Verjauw}, \citenamefont {Wurz}, \citenamefont {Hosseinkhani},
  \citenamefont {Ockeloen-Korppi}, \citenamefont {Liu}, \citenamefont {Rath},
  \citenamefont {Thapa}, \citenamefont {Vigneau}, \citenamefont {Wybo},
  \citenamefont {Bergholm}, \citenamefont {Chan}, \citenamefont {Csatári},
  \citenamefont {Dahl}, \citenamefont {Davletkaliyev}, \citenamefont {Giri},
  \citenamefont {Gusenkova}, \citenamefont {Heimonen}, \citenamefont
  {Hiltunen}, \citenamefont {Hsu}, \citenamefont {Hyyppä}, \citenamefont
  {Ikonen}, \citenamefont {Jones}, \citenamefont {Khalid}, \citenamefont {Kim},
  \citenamefont {Koistinen}, \citenamefont {Komlev}, \citenamefont {Kotilahti},
  \citenamefont {Kukushkin}, \citenamefont {Lamprich}, \citenamefont {Landra},
  \citenamefont {Lee}, \citenamefont {Li}, \citenamefont {Liebermann},
  \citenamefont {Majumder}, \citenamefont {Mäntylä}, \citenamefont {Marxer},
  \citenamefont {van~de Griend}, \citenamefont {Milchakov}, \citenamefont
  {Mrożek}, \citenamefont {Nath}, \citenamefont {Orell}, \citenamefont
  {Papič}, \citenamefont {Partanen}, \citenamefont {Plyushch}, \citenamefont
  {Pogorzalek}, \citenamefont {Ritvas}, \citenamefont {Romero}, \citenamefont
  {Sampo}, \citenamefont {Seppälä}, \citenamefont {Selinmaa}, \citenamefont
  {Sundström}, \citenamefont {Takmakov}, \citenamefont {Tarasinski},
  \citenamefont {Tuorila}, \citenamefont {Tyrkkö}, \citenamefont {Välimaa},
  \citenamefont {Wesdorp}, \citenamefont {Yang}, \citenamefont {Yu},
  \citenamefont {Heinsoo}, \citenamefont {Vepsäläinen}, \citenamefont
  {Kindel}, \citenamefont {Ku},\ and\ \citenamefont {Deppe}}]{Renger2025}%
  \BibitemOpen
  \bibfield  {author} {\bibinfo {author} {\bibfnamefont {M.}~\bibnamefont
  {Renger}}, \bibinfo {author} {\bibfnamefont {J.}~\bibnamefont {Verjauw}},
  \bibinfo {author} {\bibfnamefont {N.}~\bibnamefont {Wurz}}, \bibinfo {author}
  {\bibfnamefont {A.}~\bibnamefont {Hosseinkhani}}, \bibinfo {author}
  {\bibfnamefont {C.}~\bibnamefont {Ockeloen-Korppi}}, \bibinfo {author}
  {\bibfnamefont {W.}~\bibnamefont {Liu}}, \bibinfo {author} {\bibfnamefont
  {A.}~\bibnamefont {Rath}}, \bibinfo {author} {\bibfnamefont {M.~J.}\
  \bibnamefont {Thapa}}, \bibinfo {author} {\bibfnamefont {F.}~\bibnamefont
  {Vigneau}}, \bibinfo {author} {\bibfnamefont {E.}~\bibnamefont {Wybo}},
  \bibinfo {author} {\bibfnamefont {V.}~\bibnamefont {Bergholm}}, \bibinfo
  {author} {\bibfnamefont {C.~F.}\ \bibnamefont {Chan}}, \bibinfo {author}
  {\bibfnamefont {B.}~\bibnamefont {Csatári}}, \bibinfo {author}
  {\bibfnamefont {S.}~\bibnamefont {Dahl}}, \bibinfo {author} {\bibfnamefont
  {R.}~\bibnamefont {Davletkaliyev}}, \bibinfo {author} {\bibfnamefont
  {R.}~\bibnamefont {Giri}}, \bibinfo {author} {\bibfnamefont {D.}~\bibnamefont
  {Gusenkova}}, \bibinfo {author} {\bibfnamefont {H.}~\bibnamefont {Heimonen}},
  \bibinfo {author} {\bibfnamefont {T.}~\bibnamefont {Hiltunen}}, \bibinfo
  {author} {\bibfnamefont {H.}~\bibnamefont {Hsu}}, \bibinfo {author}
  {\bibfnamefont {E.}~\bibnamefont {Hyyppä}}, \bibinfo {author} {\bibfnamefont
  {J.}~\bibnamefont {Ikonen}}, \bibinfo {author} {\bibfnamefont
  {T.}~\bibnamefont {Jones}}, \bibinfo {author} {\bibfnamefont
  {S.}~\bibnamefont {Khalid}}, \bibinfo {author} {\bibfnamefont {S.-G.}\
  \bibnamefont {Kim}}, \bibinfo {author} {\bibfnamefont {M.}~\bibnamefont
  {Koistinen}}, \bibinfo {author} {\bibfnamefont {A.}~\bibnamefont {Komlev}},
  \bibinfo {author} {\bibfnamefont {J.}~\bibnamefont {Kotilahti}}, \bibinfo
  {author} {\bibfnamefont {V.}~\bibnamefont {Kukushkin}}, \bibinfo {author}
  {\bibfnamefont {J.}~\bibnamefont {Lamprich}}, \bibinfo {author}
  {\bibfnamefont {A.}~\bibnamefont {Landra}}, \bibinfo {author} {\bibfnamefont
  {L.-H.}\ \bibnamefont {Lee}}, \bibinfo {author} {\bibfnamefont
  {T.}~\bibnamefont {Li}}, \bibinfo {author} {\bibfnamefont {P.}~\bibnamefont
  {Liebermann}}, \bibinfo {author} {\bibfnamefont {S.}~\bibnamefont
  {Majumder}}, \bibinfo {author} {\bibfnamefont {J.}~\bibnamefont {Mäntylä}},
  \bibinfo {author} {\bibfnamefont {F.}~\bibnamefont {Marxer}}, \bibinfo
  {author} {\bibfnamefont {A.~M.}\ \bibnamefont {van~de Griend}}, \bibinfo
  {author} {\bibfnamefont {V.}~\bibnamefont {Milchakov}}, \bibinfo {author}
  {\bibfnamefont {J.}~\bibnamefont {Mrożek}}, \bibinfo {author} {\bibfnamefont
  {J.}~\bibnamefont {Nath}}, \bibinfo {author} {\bibfnamefont {T.}~\bibnamefont
  {Orell}}, \bibinfo {author} {\bibfnamefont {M.}~\bibnamefont {Papič}},
  \bibinfo {author} {\bibfnamefont {M.}~\bibnamefont {Partanen}}, \bibinfo
  {author} {\bibfnamefont {A.}~\bibnamefont {Plyushch}}, \bibinfo {author}
  {\bibfnamefont {S.}~\bibnamefont {Pogorzalek}}, \bibinfo {author}
  {\bibfnamefont {J.}~\bibnamefont {Ritvas}}, \bibinfo {author} {\bibfnamefont
  {P.~F.}\ \bibnamefont {Romero}}, \bibinfo {author} {\bibfnamefont
  {V.}~\bibnamefont {Sampo}}, \bibinfo {author} {\bibfnamefont
  {M.}~\bibnamefont {Seppälä}}, \bibinfo {author} {\bibfnamefont
  {V.}~\bibnamefont {Selinmaa}}, \bibinfo {author} {\bibfnamefont
  {L.}~\bibnamefont {Sundström}}, \bibinfo {author} {\bibfnamefont
  {I.}~\bibnamefont {Takmakov}}, \bibinfo {author} {\bibfnamefont
  {B.}~\bibnamefont {Tarasinski}}, \bibinfo {author} {\bibfnamefont
  {J.}~\bibnamefont {Tuorila}}, \bibinfo {author} {\bibfnamefont
  {O.}~\bibnamefont {Tyrkkö}}, \bibinfo {author} {\bibfnamefont
  {A.}~\bibnamefont {Välimaa}}, \bibinfo {author} {\bibfnamefont
  {J.}~\bibnamefont {Wesdorp}}, \bibinfo {author} {\bibfnamefont
  {P.}~\bibnamefont {Yang}}, \bibinfo {author} {\bibfnamefont {L.}~\bibnamefont
  {Yu}}, \bibinfo {author} {\bibfnamefont {J.}~\bibnamefont {Heinsoo}},
  \bibinfo {author} {\bibfnamefont {A.}~\bibnamefont {Vepsäläinen}}, \bibinfo
  {author} {\bibfnamefont {W.}~\bibnamefont {Kindel}}, \bibinfo {author}
  {\bibfnamefont {H.-S.}\ \bibnamefont {Ku}},\ and\ \bibinfo {author}
  {\bibfnamefont {F.}~\bibnamefont {Deppe}},\ }\href
  {https://arxiv.org/abs/2503.10903} {\bibinfo {title} {A superconducting
  qubit-resonator quantum processor with effective all-to-all connectivity}}
  (\bibinfo {year} {2025}),\ \Eprint {https://arxiv.org/abs/2503.10903}
  {arXiv:2503.10903 [quant-ph]} \BibitemShut {NoStop}%
\bibitem [{\citenamefont {Somoza}\ \emph {et~al.}(2019)\citenamefont {Somoza},
  \citenamefont {Marty}, \citenamefont {Lim}, \citenamefont {Huelga},\ and\
  \citenamefont {Plenio}}]{Somoza2019}%
  \BibitemOpen
  \bibfield  {author} {\bibinfo {author} {\bibfnamefont {A.~D.}\ \bibnamefont
  {Somoza}}, \bibinfo {author} {\bibfnamefont {O.}~\bibnamefont {Marty}},
  \bibinfo {author} {\bibfnamefont {J.}~\bibnamefont {Lim}}, \bibinfo {author}
  {\bibfnamefont {S.~F.}\ \bibnamefont {Huelga}},\ and\ \bibinfo {author}
  {\bibfnamefont {M.~B.}\ \bibnamefont {Plenio}},\ }\bibfield  {title}
  {\bibinfo {title} {Dissipation-assisted matrix product factorization},\
  }\href {https://doi.org/10.1103/PhysRevLett.123.100502} {\bibfield  {journal}
  {\bibinfo  {journal} {Phys. Rev. Lett}\ }\textbf {\bibinfo {volume} {123}},\
  \bibinfo {pages} {100502} (\bibinfo {year} {2019})}\BibitemShut {NoStop}%
\bibitem [{\citenamefont {Tamascelli}\ \emph {et~al.}(2019)\citenamefont
  {Tamascelli}, \citenamefont {Smirne}, \citenamefont {Lim}, \citenamefont
  {Huelga},\ and\ \citenamefont {Plenio}}]{Tamascelli2019}%
  \BibitemOpen
  \bibfield  {author} {\bibinfo {author} {\bibfnamefont {D.}~\bibnamefont
  {Tamascelli}}, \bibinfo {author} {\bibfnamefont {A.}~\bibnamefont {Smirne}},
  \bibinfo {author} {\bibfnamefont {J.}~\bibnamefont {Lim}}, \bibinfo {author}
  {\bibfnamefont {S.~F.}\ \bibnamefont {Huelga}},\ and\ \bibinfo {author}
  {\bibfnamefont {M.~B.}\ \bibnamefont {Plenio}},\ }\bibfield  {title}
  {\bibinfo {title} {Efficient simulation of finite-temperature open quantum
  systems},\ }\href {https://doi.org/10.1103/PhysRevLett.123.090402} {\bibfield
   {journal} {\bibinfo  {journal} {Phys. Rev. Lett.}\ }\textbf {\bibinfo
  {volume} {123}},\ \bibinfo {pages} {090402} (\bibinfo {year}
  {2019})}\BibitemShut {NoStop}%
\bibitem [{\citenamefont {Lambert}\ \emph {et~al.}(2019)\citenamefont
  {Lambert}, \citenamefont {Ahmed}, \citenamefont {Cirio},\ and\ \citenamefont
  {Nori}}]{Lambert2019}%
  \BibitemOpen
  \bibfield  {author} {\bibinfo {author} {\bibfnamefont {N.}~\bibnamefont
  {Lambert}}, \bibinfo {author} {\bibfnamefont {S.}~\bibnamefont {Ahmed}},
  \bibinfo {author} {\bibfnamefont {M.}~\bibnamefont {Cirio}},\ and\ \bibinfo
  {author} {\bibfnamefont {F.}~\bibnamefont {Nori}},\ }\bibfield  {title}
  {\bibinfo {title} {Modelling the ultra-strongly coupled spin-boson model with
  unphysical modes},\ }\href {https://doi.org/10.1038/s41467-019-11656-1}
  {\bibfield  {journal} {\bibinfo  {journal} {Nature Communications}\ }\textbf
  {\bibinfo {volume} {10}},\ \bibinfo {pages} {3721} (\bibinfo {year}
  {2019})}\BibitemShut {NoStop}%
\bibitem [{\citenamefont {Tamascelli}\ \emph {et~al.}(2018)\citenamefont
  {Tamascelli}, \citenamefont {Smirne}, \citenamefont {Huelga},\ and\
  \citenamefont {Plenio}}]{Tamascelli2018}%
  \BibitemOpen
  \bibfield  {author} {\bibinfo {author} {\bibfnamefont {D.}~\bibnamefont
  {Tamascelli}}, \bibinfo {author} {\bibfnamefont {A.}~\bibnamefont {Smirne}},
  \bibinfo {author} {\bibfnamefont {S.~F.}\ \bibnamefont {Huelga}},\ and\
  \bibinfo {author} {\bibfnamefont {M.~B.}\ \bibnamefont {Plenio}},\ }\bibfield
   {title} {\bibinfo {title} {Nonperturbative treatment of non-markovian
  dynamics of open quantum systems},\ }\href
  {https://doi.org/10.1103/PhysRevLett.120.030402} {\bibfield  {journal}
  {\bibinfo  {journal} {Phys. Rev. Lett.}\ }\textbf {\bibinfo {volume} {120}},\
  \bibinfo {pages} {030402} (\bibinfo {year} {2018})}\BibitemShut {NoStop}%
\bibitem [{\citenamefont {Pleasance}\ \emph {et~al.}(2020)\citenamefont
  {Pleasance}, \citenamefont {Garraway},\ and\ \citenamefont
  {Petruccione}}]{Pleasance2020}%
  \BibitemOpen
  \bibfield  {author} {\bibinfo {author} {\bibfnamefont {G.}~\bibnamefont
  {Pleasance}}, \bibinfo {author} {\bibfnamefont {B.~M.}\ \bibnamefont
  {Garraway}},\ and\ \bibinfo {author} {\bibfnamefont {F.}~\bibnamefont
  {Petruccione}},\ }\bibfield  {title} {\bibinfo {title} {Generalized theory of
  pseudomodes for exact descriptions of non-markovian quantum processes},\
  }\href {https://doi.org/10.1103/PhysRevResearch.2.043058} {\bibfield
  {journal} {\bibinfo  {journal} {Phys. Rev. Research}\ }\textbf {\bibinfo
  {volume} {2}},\ \bibinfo {pages} {043058} (\bibinfo {year}
  {2020})}\BibitemShut {NoStop}%
\bibitem [{\citenamefont {Menczel}\ \emph {et~al.}(2024)\citenamefont
  {Menczel}, \citenamefont {Funo}, \citenamefont {Cirio}, \citenamefont
  {Lambert},\ and\ \citenamefont {Nori}}]{Menczel2024}%
  \BibitemOpen
  \bibfield  {author} {\bibinfo {author} {\bibfnamefont {P.}~\bibnamefont
  {Menczel}}, \bibinfo {author} {\bibfnamefont {K.}~\bibnamefont {Funo}},
  \bibinfo {author} {\bibfnamefont {M.}~\bibnamefont {Cirio}}, \bibinfo
  {author} {\bibfnamefont {N.}~\bibnamefont {Lambert}},\ and\ \bibinfo {author}
  {\bibfnamefont {F.}~\bibnamefont {Nori}},\ }\bibfield  {title} {\bibinfo
  {title} {Non-hermitian pseudomodes for strongly coupled open quantum systems:
  Unravelings, correlations, and thermodynamics},\ }\href
  {https://doi.org/10.1103/physrevresearch.6.033237} {\bibfield  {journal}
  {\bibinfo  {journal} {Physical Review Research}\ }\textbf {\bibinfo {volume}
  {6}},\ \bibinfo {pages} {033237} (\bibinfo {year} {2024})}\BibitemShut
  {NoStop}%
\bibitem [{\citenamefont {Shirazi}\ \emph {et~al.}(2024)\citenamefont
  {Shirazi}, \citenamefont {Rybkin}, \citenamefont {Marthaler},\ and\
  \citenamefont {Golubev}}]{Shirazi2024}%
  \BibitemOpen
  \bibfield  {author} {\bibinfo {author} {\bibfnamefont {R.~G.}\ \bibnamefont
  {Shirazi}}, \bibinfo {author} {\bibfnamefont {V.~V.}\ \bibnamefont {Rybkin}},
  \bibinfo {author} {\bibfnamefont {M.}~\bibnamefont {Marthaler}},\ and\
  \bibinfo {author} {\bibfnamefont {D.}~\bibnamefont {Golubev}},\ }\bibfield
  {title} {\bibinfo {title} {Efficient random phase approximation for
  diradicals},\ }\href {https://doi.org/10.1063/5.0227556} {\bibfield
  {journal} {\bibinfo  {journal} {J. Chem. Phys.}\ }\textbf {\bibinfo {volume}
  {161}},\ \bibinfo {pages} {114110} (\bibinfo {year} {2024})}\BibitemShut
  {NoStop}%
\bibitem [{\citenamefont {Campaioli}\ and\ \citenamefont
  {Cole}(2021)}]{Campaioli2021}%
  \BibitemOpen
  \bibfield  {author} {\bibinfo {author} {\bibfnamefont {F.}~\bibnamefont
  {Campaioli}}\ and\ \bibinfo {author} {\bibfnamefont {J.~H.}\ \bibnamefont
  {Cole}},\ }\bibfield  {title} {\bibinfo {title} {Exciton transport in
  amorphous polymers and the role of morphology and thermalisation},\ }\href
  {https://doi.org/10.1088/1367-2630/ac37c7} {\bibfield  {journal} {\bibinfo
  {journal} {New J. Phys.}\ }\textbf {\bibinfo {volume} {23}},\ \bibinfo
  {pages} {113038} (\bibinfo {year} {2021})}\BibitemShut {NoStop}%
\bibitem [{\citenamefont {Binder}\ \emph {et~al.}(2013)\citenamefont {Binder},
  \citenamefont {Wahl}, \citenamefont {Römer},\ and\ \citenamefont
  {Burghardt}}]{Binder2013}%
  \BibitemOpen
  \bibfield  {author} {\bibinfo {author} {\bibfnamefont {R.}~\bibnamefont
  {Binder}}, \bibinfo {author} {\bibfnamefont {J.}~\bibnamefont {Wahl}},
  \bibinfo {author} {\bibfnamefont {S.}~\bibnamefont {Römer}},\ and\ \bibinfo
  {author} {\bibfnamefont {I.}~\bibnamefont {Burghardt}},\ }\bibfield  {title}
  {\bibinfo {title} {Coherent exciton transport driven by torsional dynamics: a
  quantum dynamical study of phenylene-vinylene type conjugated systems},\
  }\href {https://doi.org/10.1039/C3FD20148A} {\bibfield  {journal} {\bibinfo
  {journal} {Faraday Discuss.}\ }\textbf {\bibinfo {volume} {163}},\ \bibinfo
  {pages} {205} (\bibinfo {year} {2013})}\BibitemShut {NoStop}%
\bibitem [{\citenamefont {Grimsmo}\ \emph {et~al.}(2016)\citenamefont
  {Grimsmo}, \citenamefont {Qassemi}, \citenamefont {Reulet},\ and\
  \citenamefont {Blais}}]{Grimsmo2016}%
  \BibitemOpen
  \bibfield  {author} {\bibinfo {author} {\bibfnamefont {A.~L.}\ \bibnamefont
  {Grimsmo}}, \bibinfo {author} {\bibfnamefont {F.}~\bibnamefont {Qassemi}},
  \bibinfo {author} {\bibfnamefont {B.}~\bibnamefont {Reulet}},\ and\ \bibinfo
  {author} {\bibfnamefont {A.}~\bibnamefont {Blais}},\ }\bibfield  {title}
  {\bibinfo {title} {Quantum optics theory of electronic noise in coherent
  conductors},\ }\href {https://doi.org/10.1103/PhysRevLett.116.043602}
  {\bibfield  {journal} {\bibinfo  {journal} {Phys. Rev. Lett.}\ }\textbf
  {\bibinfo {volume} {116}},\ \bibinfo {pages} {043602} (\bibinfo {year}
  {2016})}\BibitemShut {NoStop}%
\bibitem [{\citenamefont {Westig}\ \emph {et~al.}(2017)\citenamefont {Westig},
  \citenamefont {Kubala}, \citenamefont {Parlavecchio}, \citenamefont
  {Mukharsky}, \citenamefont {Altimiras}, \citenamefont {Joyez}, \citenamefont
  {Vion}, \citenamefont {Roche}, \citenamefont {Esteve}, \citenamefont
  {Hofheinz}, \citenamefont {Trif}, \citenamefont {Simon}, \citenamefont
  {Ankerhold},\ and\ \citenamefont {Portier}}]{Westig2017}%
  \BibitemOpen
  \bibfield  {author} {\bibinfo {author} {\bibfnamefont {M.}~\bibnamefont
  {Westig}}, \bibinfo {author} {\bibfnamefont {B.}~\bibnamefont {Kubala}},
  \bibinfo {author} {\bibfnamefont {O.}~\bibnamefont {Parlavecchio}}, \bibinfo
  {author} {\bibfnamefont {Y.}~\bibnamefont {Mukharsky}}, \bibinfo {author}
  {\bibfnamefont {C.}~\bibnamefont {Altimiras}}, \bibinfo {author}
  {\bibfnamefont {P.}~\bibnamefont {Joyez}}, \bibinfo {author} {\bibfnamefont
  {D.}~\bibnamefont {Vion}}, \bibinfo {author} {\bibfnamefont {P.}~\bibnamefont
  {Roche}}, \bibinfo {author} {\bibfnamefont {D.}~\bibnamefont {Esteve}},
  \bibinfo {author} {\bibfnamefont {M.}~\bibnamefont {Hofheinz}}, \bibinfo
  {author} {\bibfnamefont {M.}~\bibnamefont {Trif}}, \bibinfo {author}
  {\bibfnamefont {P.}~\bibnamefont {Simon}}, \bibinfo {author} {\bibfnamefont
  {J.}~\bibnamefont {Ankerhold}},\ and\ \bibinfo {author} {\bibfnamefont
  {F.}~\bibnamefont {Portier}},\ }\bibfield  {title} {\bibinfo {title}
  {Emission of nonclassical radiation by inelastic cooper pair tunneling},\
  }\href {https://doi.org/10.1103/physrevlett.119.137001} {\bibfield  {journal}
  {\bibinfo  {journal} {Physical Review Letters}\ }\textbf {\bibinfo {volume}
  {119}},\ \bibinfo {pages} {137001} (\bibinfo {year} {2017})}\BibitemShut
  {NoStop}%
\bibitem [{\citenamefont {Grimm}\ \emph {et~al.}(2019)\citenamefont {Grimm},
  \citenamefont {Blanchet}, \citenamefont {Albert}, \citenamefont
  {Lepp\"akangas}, \citenamefont {Jebari}, \citenamefont {Hazra}, \citenamefont
  {Gustavo}, \citenamefont {Thomassin}, \citenamefont {Dupont-Ferrier},
  \citenamefont {Portier},\ and\ \citenamefont {Hofheinz}}]{Grimm2019}%
  \BibitemOpen
  \bibfield  {author} {\bibinfo {author} {\bibfnamefont {A.}~\bibnamefont
  {Grimm}}, \bibinfo {author} {\bibfnamefont {F.}~\bibnamefont {Blanchet}},
  \bibinfo {author} {\bibfnamefont {R.}~\bibnamefont {Albert}}, \bibinfo
  {author} {\bibfnamefont {J.}~\bibnamefont {Lepp\"akangas}}, \bibinfo {author}
  {\bibfnamefont {S.}~\bibnamefont {Jebari}}, \bibinfo {author} {\bibfnamefont
  {D.}~\bibnamefont {Hazra}}, \bibinfo {author} {\bibfnamefont
  {F.}~\bibnamefont {Gustavo}}, \bibinfo {author} {\bibfnamefont {J.-L.}\
  \bibnamefont {Thomassin}}, \bibinfo {author} {\bibfnamefont {E.}~\bibnamefont
  {Dupont-Ferrier}}, \bibinfo {author} {\bibfnamefont {F.}~\bibnamefont
  {Portier}},\ and\ \bibinfo {author} {\bibfnamefont {M.}~\bibnamefont
  {Hofheinz}},\ }\bibfield  {title} {\bibinfo {title} {Bright on-demand source
  of antibunched microwave photons based on inelastic cooper pair tunneling},\
  }\href {https://doi.org/10.1103/PhysRevX.9.021016} {\bibfield  {journal}
  {\bibinfo  {journal} {Phys. Rev. X}\ }\textbf {\bibinfo {volume} {9}},\
  \bibinfo {pages} {021016} (\bibinfo {year} {2019})}\BibitemShut {NoStop}%
\bibitem [{\citenamefont {Peugeot}\ \emph {et~al.}(2021)\citenamefont
  {Peugeot}, \citenamefont {M\'enard}, \citenamefont {Dambach}, \citenamefont
  {Westig}, \citenamefont {Kubala}, \citenamefont {Mukharsky}, \citenamefont
  {Altimiras}, \citenamefont {Joyez}, \citenamefont {Vion}, \citenamefont
  {Roche}, \citenamefont {Esteve}, \citenamefont {Milman}, \citenamefont
  {Lepp\"akangas}, \citenamefont {Johansson}, \citenamefont {Hofheinz},
  \citenamefont {Ankerhold},\ and\ \citenamefont {Portier}}]{Peugeot2021}%
  \BibitemOpen
  \bibfield  {author} {\bibinfo {author} {\bibfnamefont {A.}~\bibnamefont
  {Peugeot}}, \bibinfo {author} {\bibfnamefont {G.}~\bibnamefont {M\'enard}},
  \bibinfo {author} {\bibfnamefont {S.}~\bibnamefont {Dambach}}, \bibinfo
  {author} {\bibfnamefont {M.}~\bibnamefont {Westig}}, \bibinfo {author}
  {\bibfnamefont {B.}~\bibnamefont {Kubala}}, \bibinfo {author} {\bibfnamefont
  {Y.}~\bibnamefont {Mukharsky}}, \bibinfo {author} {\bibfnamefont
  {C.}~\bibnamefont {Altimiras}}, \bibinfo {author} {\bibfnamefont
  {P.}~\bibnamefont {Joyez}}, \bibinfo {author} {\bibfnamefont
  {D.}~\bibnamefont {Vion}}, \bibinfo {author} {\bibfnamefont {P.}~\bibnamefont
  {Roche}}, \bibinfo {author} {\bibfnamefont {D.}~\bibnamefont {Esteve}},
  \bibinfo {author} {\bibfnamefont {P.}~\bibnamefont {Milman}}, \bibinfo
  {author} {\bibfnamefont {J.}~\bibnamefont {Lepp\"akangas}}, \bibinfo {author}
  {\bibfnamefont {G.}~\bibnamefont {Johansson}}, \bibinfo {author}
  {\bibfnamefont {M.}~\bibnamefont {Hofheinz}}, \bibinfo {author}
  {\bibfnamefont {J.}~\bibnamefont {Ankerhold}},\ and\ \bibinfo {author}
  {\bibfnamefont {F.}~\bibnamefont {Portier}},\ }\bibfield  {title} {\bibinfo
  {title} {Generating two continuous entangled microwave beams using a
  dc-biased josephson junction},\ }\href
  {https://doi.org/10.1103/PhysRevX.11.031008} {\bibfield  {journal} {\bibinfo
  {journal} {Phys. Rev. X}\ }\textbf {\bibinfo {volume} {11}},\ \bibinfo
  {pages} {031008} (\bibinfo {year} {2021})}\BibitemShut {NoStop}%
\bibitem [{\citenamefont {Walls}\ and\ \citenamefont
  {Milburn}(2008)}]{WallsMilburn}%
  \BibitemOpen
  \bibfield  {author} {\bibinfo {author} {\bibfnamefont {D.~F.}\ \bibnamefont
  {Walls}}\ and\ \bibinfo {author} {\bibfnamefont {G.}~\bibnamefont
  {Milburn}},\ }\href@noop {} {\emph {\bibinfo {title} {Quantum Optics}}}\
  (\bibinfo  {publisher} {Springer, Berlin},\ \bibinfo {year}
  {2008})\BibitemShut {NoStop}%
\bibitem [{\citenamefont {Schwartz}\ \emph {et~al.}(2011)\citenamefont
  {Schwartz}, \citenamefont {Hutchison}, \citenamefont {Genet},\ and\
  \citenamefont {Ebbesen}}]{Schwarz2011}%
  \BibitemOpen
  \bibfield  {author} {\bibinfo {author} {\bibfnamefont {T.}~\bibnamefont
  {Schwartz}}, \bibinfo {author} {\bibfnamefont {J.~A.}\ \bibnamefont
  {Hutchison}}, \bibinfo {author} {\bibfnamefont {C.}~\bibnamefont {Genet}},\
  and\ \bibinfo {author} {\bibfnamefont {T.~W.}\ \bibnamefont {Ebbesen}},\
  }\bibfield  {title} {\bibinfo {title} {Reversible switching of ultrastrong
  light-molecule coupling},\ }\href
  {https://doi.org/10.1103/PhysRevLett.106.196405} {\bibfield  {journal}
  {\bibinfo  {journal} {Phys. Rev. Lett.}\ }\textbf {\bibinfo {volume} {106}},\
  \bibinfo {pages} {196405} (\bibinfo {year} {2011})}\BibitemShut {NoStop}%
\bibitem [{\citenamefont {Forn-D\'{\i}az}\ \emph {et~al.}(2019)\citenamefont
  {Forn-D\'{\i}az}, \citenamefont {Lamata}, \citenamefont {Rico}, \citenamefont
  {Kono},\ and\ \citenamefont {Solano}}]{Forn_Diaz_RevModPhys2019}%
  \BibitemOpen
  \bibfield  {author} {\bibinfo {author} {\bibfnamefont {P.}~\bibnamefont
  {Forn-D\'{\i}az}}, \bibinfo {author} {\bibfnamefont {L.}~\bibnamefont
  {Lamata}}, \bibinfo {author} {\bibfnamefont {E.}~\bibnamefont {Rico}},
  \bibinfo {author} {\bibfnamefont {J.}~\bibnamefont {Kono}},\ and\ \bibinfo
  {author} {\bibfnamefont {E.}~\bibnamefont {Solano}},\ }\bibfield  {title}
  {\bibinfo {title} {Ultrastrong coupling regimes of light-matter
  interaction},\ }\href {https://doi.org/10.1103/RevModPhys.91.025005}
  {\bibfield  {journal} {\bibinfo  {journal} {Rev. Mod. Phys.}\ }\textbf
  {\bibinfo {volume} {91}},\ \bibinfo {pages} {025005} (\bibinfo {year}
  {2019})}\BibitemShut {NoStop}%
\bibitem [{\citenamefont {Qin}\ \emph {et~al.}(2024)\citenamefont {Qin},
  \citenamefont {Kockum}, \citenamefont {Muñoz}, \citenamefont {Miranowicz},\
  and\ \citenamefont {Nori}}]{Qin2024}%
  \BibitemOpen
  \bibfield  {author} {\bibinfo {author} {\bibfnamefont {W.}~\bibnamefont
  {Qin}}, \bibinfo {author} {\bibfnamefont {A.~F.}\ \bibnamefont {Kockum}},
  \bibinfo {author} {\bibfnamefont {C.~S.}\ \bibnamefont {Muñoz}}, \bibinfo
  {author} {\bibfnamefont {A.}~\bibnamefont {Miranowicz}},\ and\ \bibinfo
  {author} {\bibfnamefont {F.}~\bibnamefont {Nori}},\ }\bibfield  {title}
  {\bibinfo {title} {Quantum amplification and simulation of strong and
  ultrastrong coupling of light and matter},\ }\href
  {https://doi.org/10.1016/j.physrep.2024.05.003} {\bibfield  {journal}
  {\bibinfo  {journal} {Physics Reports}\ }\textbf {\bibinfo {volume} {1078}},\
  \bibinfo {pages} {1–59} (\bibinfo {year} {2024})}\BibitemShut {NoStop}%
\bibitem [{\citenamefont {Frisk~Kockum}\ \emph {et~al.}(2019)\citenamefont
  {Frisk~Kockum}, \citenamefont {Miranowicz}, \citenamefont {De~Liberato},
  \citenamefont {Savasta},\ and\ \citenamefont {Nori}}]{FriskKockum2019}%
  \BibitemOpen
  \bibfield  {author} {\bibinfo {author} {\bibfnamefont {A.}~\bibnamefont
  {Frisk~Kockum}}, \bibinfo {author} {\bibfnamefont {A.}~\bibnamefont
  {Miranowicz}}, \bibinfo {author} {\bibfnamefont {S.}~\bibnamefont
  {De~Liberato}}, \bibinfo {author} {\bibfnamefont {S.}~\bibnamefont
  {Savasta}},\ and\ \bibinfo {author} {\bibfnamefont {F.}~\bibnamefont
  {Nori}},\ }\bibfield  {title} {\bibinfo {title} {Ultrastrong coupling between
  light and matter},\ }\href {https://doi.org/10.1038/s42254-018-0006-2}
  {\bibfield  {journal} {\bibinfo  {journal} {Nature Reviews Physics}\ }\textbf
  {\bibinfo {volume} {1}},\ \bibinfo {pages} {19} (\bibinfo {year}
  {2019})}\BibitemShut {NoStop}%
\bibitem [{\citenamefont {Koch}\ \emph {et~al.}(2007)\citenamefont {Koch},
  \citenamefont {Yu}, \citenamefont {Gambetta}, \citenamefont {Houck},
  \citenamefont {Schuster}, \citenamefont {Majer}, \citenamefont {Blais},
  \citenamefont {Devoret}, \citenamefont {Girvin},\ and\ \citenamefont
  {Schoelkopf}}]{Koch2007}%
  \BibitemOpen
  \bibfield  {author} {\bibinfo {author} {\bibfnamefont {J.}~\bibnamefont
  {Koch}}, \bibinfo {author} {\bibfnamefont {T.~M.}\ \bibnamefont {Yu}},
  \bibinfo {author} {\bibfnamefont {J.}~\bibnamefont {Gambetta}}, \bibinfo
  {author} {\bibfnamefont {A.~A.}\ \bibnamefont {Houck}}, \bibinfo {author}
  {\bibfnamefont {D.~I.}\ \bibnamefont {Schuster}}, \bibinfo {author}
  {\bibfnamefont {J.}~\bibnamefont {Majer}}, \bibinfo {author} {\bibfnamefont
  {A.}~\bibnamefont {Blais}}, \bibinfo {author} {\bibfnamefont {M.~H.}\
  \bibnamefont {Devoret}}, \bibinfo {author} {\bibfnamefont {S.~M.}\
  \bibnamefont {Girvin}},\ and\ \bibinfo {author} {\bibfnamefont {R.~J.}\
  \bibnamefont {Schoelkopf}},\ }\bibfield  {title} {\bibinfo {title}
  {Charge-insensitive qubit design derived from the cooper pair box},\ }\href
  {http://link.aps.org/doi/10.1103/PhysRevA.76.042319} {\bibfield  {journal}
  {\bibinfo  {journal} {Phys. Rev. A}\ }\textbf {\bibinfo {volume} {76}},\
  \bibinfo {pages} {042319} (\bibinfo {year} {2007})}\BibitemShut {NoStop}%
\bibitem [{\citenamefont {Blais}\ \emph {et~al.}(2021)\citenamefont {Blais},
  \citenamefont {Grimsmo}, \citenamefont {Girvin},\ and\ \citenamefont
  {Wallraff}}]{Blais2021_RevModPhys}%
  \BibitemOpen
  \bibfield  {author} {\bibinfo {author} {\bibfnamefont {A.}~\bibnamefont
  {Blais}}, \bibinfo {author} {\bibfnamefont {A.~L.}\ \bibnamefont {Grimsmo}},
  \bibinfo {author} {\bibfnamefont {S.~M.}\ \bibnamefont {Girvin}},\ and\
  \bibinfo {author} {\bibfnamefont {A.}~\bibnamefont {Wallraff}},\ }\bibfield
  {title} {\bibinfo {title} {Circuit quantum electrodynamics},\ }\href
  {https://doi.org/10.1103/RevModPhys.93.025005} {\bibfield  {journal}
  {\bibinfo  {journal} {Rev. Mod. Phys.}\ }\textbf {\bibinfo {volume} {93}},\
  \bibinfo {pages} {025005} (\bibinfo {year} {2021})}\BibitemShut {NoStop}%
\bibitem [{\citenamefont {Dunning}\ and\ \citenamefont
  {McKoy}(1967)}]{Dunning1967}%
  \BibitemOpen
  \bibfield  {author} {\bibinfo {author} {\bibfnamefont {T.~H.}\ \bibnamefont
  {Dunning}}\ and\ \bibinfo {author} {\bibfnamefont {V.}~\bibnamefont
  {McKoy}},\ }\bibfield  {title} {\bibinfo {title} {Nonempirical calculations
  on excited states: The ethylene molecule},\ }\href
  {https://doi.org/10.1063/1.1712158} {\bibfield  {journal} {\bibinfo
  {journal} {The Journal of Chemical Physics}\ }\textbf {\bibinfo {volume}
  {47}},\ \bibinfo {pages} {1735} (\bibinfo {year} {1967})}\BibitemShut
  {NoStop}%
\bibitem [{\citenamefont {Childs}\ \emph {et~al.}(2021)\citenamefont {Childs},
  \citenamefont {Su}, \citenamefont {Tran}, \citenamefont {Wiebe},\ and\
  \citenamefont {Zhu}}]{Childs2021}%
  \BibitemOpen
  \bibfield  {author} {\bibinfo {author} {\bibfnamefont {A.~M.}\ \bibnamefont
  {Childs}}, \bibinfo {author} {\bibfnamefont {Y.}~\bibnamefont {Su}}, \bibinfo
  {author} {\bibfnamefont {M.~C.}\ \bibnamefont {Tran}}, \bibinfo {author}
  {\bibfnamefont {N.}~\bibnamefont {Wiebe}},\ and\ \bibinfo {author}
  {\bibfnamefont {S.}~\bibnamefont {Zhu}},\ }\bibfield  {title} {\bibinfo
  {title} {A theory of trotter error},\ }\href
  {https://doi.org/10.1103/PhysRevX.11.011020} {\bibfield  {journal} {\bibinfo
  {journal} {PRX Quantum}\ }\textbf {\bibinfo {volume} {11}},\ \bibinfo {pages}
  {011020} (\bibinfo {year} {2021})}\BibitemShut {NoStop}%
\bibitem [{\citenamefont {Georges}\ \emph {et~al.}(1996)\citenamefont
  {Georges}, \citenamefont {Kotliar}, \citenamefont {Krauth},\ and\
  \citenamefont {Rozenberg}}]{Rozenberg1996}%
  \BibitemOpen
  \bibfield  {author} {\bibinfo {author} {\bibfnamefont {A.}~\bibnamefont
  {Georges}}, \bibinfo {author} {\bibfnamefont {G.}~\bibnamefont {Kotliar}},
  \bibinfo {author} {\bibfnamefont {W.}~\bibnamefont {Krauth}},\ and\ \bibinfo
  {author} {\bibfnamefont {M.~J.}\ \bibnamefont {Rozenberg}},\ }\bibfield
  {title} {\bibinfo {title} {Dynamical mean-field theory of strongly correlated
  fermion systems and the limit of infinite dimensions},\ }\href
  {https://doi.org/10.1103/RevModPhys.68.13} {\bibfield  {journal} {\bibinfo
  {journal} {Rev. Mod. Phys.}\ }\textbf {\bibinfo {volume} {68}},\ \bibinfo
  {pages} {13} (\bibinfo {year} {1996})}\BibitemShut {NoStop}%
\bibitem [{\citenamefont {Wouters}\ \emph {et~al.}(2016)\citenamefont
  {Wouters}, \citenamefont {Jiménez-Hoyos}, \citenamefont {Sun},\ and\
  \citenamefont {Chan}}]{Wouters2016}%
  \BibitemOpen
  \bibfield  {author} {\bibinfo {author} {\bibfnamefont {S.}~\bibnamefont
  {Wouters}}, \bibinfo {author} {\bibfnamefont {C.~A.}\ \bibnamefont
  {Jiménez-Hoyos}}, \bibinfo {author} {\bibfnamefont {Q.}~\bibnamefont
  {Sun}},\ and\ \bibinfo {author} {\bibfnamefont {G.~K.-L.}\ \bibnamefont
  {Chan}},\ }\bibfield  {title} {\bibinfo {title} {A practical guide to density
  matrix embedding theory in quantum chemistry},\ }\href
  {https://doi.org/10.1021/acs.jctc.6b00316} {\bibfield  {journal} {\bibinfo
  {journal} {Journal of Chemical Theory and Computation}\ }\textbf {\bibinfo
  {volume} {12}},\ \bibinfo {pages} {2706–2719} (\bibinfo {year}
  {2016})}\BibitemShut {NoStop}%
\bibitem [{\citenamefont {Sun}\ \emph {et~al.}(2020)\citenamefont {Sun},
  \citenamefont {Ray}, \citenamefont {Cui}, \citenamefont {Stoudenmire},
  \citenamefont {Ferrero},\ and\ \citenamefont {Chan}}]{Sun_2020}%
  \BibitemOpen
  \bibfield  {author} {\bibinfo {author} {\bibfnamefont {C.}~\bibnamefont
  {Sun}}, \bibinfo {author} {\bibfnamefont {U.}~\bibnamefont {Ray}}, \bibinfo
  {author} {\bibfnamefont {Z.-H.}\ \bibnamefont {Cui}}, \bibinfo {author}
  {\bibfnamefont {M.}~\bibnamefont {Stoudenmire}}, \bibinfo {author}
  {\bibfnamefont {M.}~\bibnamefont {Ferrero}},\ and\ \bibinfo {author}
  {\bibfnamefont {G.~K.-L.}\ \bibnamefont {Chan}},\ }\bibfield  {title}
  {\bibinfo {title} {Finite-temperature density matrix embedding theory},\
  }\href {https://doi.org/10.1103/physrevb.101.075131} {\bibfield  {journal}
  {\bibinfo  {journal} {Physical Review B}\ }\textbf {\bibinfo {volume}
  {101}},\ \bibinfo {pages} {075131} (\bibinfo {year} {2020})}\BibitemShut
  {NoStop}%
\bibitem [{\citenamefont {Hofheinz}\ \emph {et~al.}(2009)\citenamefont
  {Hofheinz}, \citenamefont {Wang}, \citenamefont {Ansmann}, \citenamefont
  {Bialczak}, \citenamefont {Lucero}, \citenamefont {Neeley}, \citenamefont
  {O'Connell}, \citenamefont {Sank}, \citenamefont {Wenner}, \citenamefont
  {Martinis},\ and\ \citenamefont {Cleland}}]{Hofheinz2009}%
  \BibitemOpen
  \bibfield  {author} {\bibinfo {author} {\bibfnamefont {M.}~\bibnamefont
  {Hofheinz}}, \bibinfo {author} {\bibfnamefont {H.}~\bibnamefont {Wang}},
  \bibinfo {author} {\bibfnamefont {M.}~\bibnamefont {Ansmann}}, \bibinfo
  {author} {\bibfnamefont {R.~C.}\ \bibnamefont {Bialczak}}, \bibinfo {author}
  {\bibfnamefont {E.}~\bibnamefont {Lucero}}, \bibinfo {author} {\bibfnamefont
  {M.}~\bibnamefont {Neeley}}, \bibinfo {author} {\bibfnamefont {A.~D.}\
  \bibnamefont {O'Connell}}, \bibinfo {author} {\bibfnamefont {D.}~\bibnamefont
  {Sank}}, \bibinfo {author} {\bibfnamefont {J.}~\bibnamefont {Wenner}},
  \bibinfo {author} {\bibfnamefont {J.~M.}\ \bibnamefont {Martinis}},\ and\
  \bibinfo {author} {\bibfnamefont {A.~M.}\ \bibnamefont {Cleland}},\
  }\bibfield  {title} {\bibinfo {title} {Synthesizing arbitrary quantum states
  in a superconducting resonator},\ }\href
  {http://dx.doi.org/10.1038/nature08005} {\bibfield  {journal} {\bibinfo
  {journal} {Nature}\ }\textbf {\bibinfo {volume} {459}},\ \bibinfo {pages}
  {546} (\bibinfo {year} {2009})}\BibitemShut {NoStop}%
\bibitem [{\citenamefont {Vlastakis}\ \emph {et~al.}(2013)\citenamefont
  {Vlastakis}, \citenamefont {Kirchmair}, \citenamefont {Leghtas},
  \citenamefont {Nigg}, \citenamefont {Frunzio}, \citenamefont {Girvin},
  \citenamefont {Mirrahimi}, \citenamefont {Devoret},\ and\ \citenamefont
  {Schoelkopf}}]{Vlastakis2013}%
  \BibitemOpen
  \bibfield  {author} {\bibinfo {author} {\bibfnamefont {B.}~\bibnamefont
  {Vlastakis}}, \bibinfo {author} {\bibfnamefont {G.}~\bibnamefont
  {Kirchmair}}, \bibinfo {author} {\bibfnamefont {Z.}~\bibnamefont {Leghtas}},
  \bibinfo {author} {\bibfnamefont {S.~E.}\ \bibnamefont {Nigg}}, \bibinfo
  {author} {\bibfnamefont {L.}~\bibnamefont {Frunzio}}, \bibinfo {author}
  {\bibfnamefont {S.~M.}\ \bibnamefont {Girvin}}, \bibinfo {author}
  {\bibfnamefont {M.}~\bibnamefont {Mirrahimi}}, \bibinfo {author}
  {\bibfnamefont {M.~H.}\ \bibnamefont {Devoret}},\ and\ \bibinfo {author}
  {\bibfnamefont {R.~J.}\ \bibnamefont {Schoelkopf}},\ }\bibfield  {title}
  {\bibinfo {title} {Deterministically encoding quantum information using
  100-photon schrödinger cat states},\ }\href
  {https://doi.org/10.1126/science.1243289} {\bibfield  {journal} {\bibinfo
  {journal} {Science}\ }\textbf {\bibinfo {volume} {342}},\ \bibinfo {pages}
  {607} (\bibinfo {year} {2013})}\BibitemShut {NoStop}%
\bibitem [{\citenamefont {Kudra}\ \emph {et~al.}(2022)\citenamefont {Kudra},
  \citenamefont {Kervinen}, \citenamefont {Strandberg}, \citenamefont {Ahmed},
  \citenamefont {Scigliuzzo}, \citenamefont {Osman}, \citenamefont {Lozano},
  \citenamefont {Thol\'en}, \citenamefont {Borgani}, \citenamefont {Haviland},
  \citenamefont {Ferrini}, \citenamefont {Bylander}, \citenamefont {Kockum},
  \citenamefont {Quijandr\'ia}, \citenamefont {Delsing},\ and\ \citenamefont
  {Gasparinetti}}]{Kudra2022}%
  \BibitemOpen
  \bibfield  {author} {\bibinfo {author} {\bibfnamefont {M.}~\bibnamefont
  {Kudra}}, \bibinfo {author} {\bibfnamefont {M.}~\bibnamefont {Kervinen}},
  \bibinfo {author} {\bibfnamefont {I.}~\bibnamefont {Strandberg}}, \bibinfo
  {author} {\bibfnamefont {S.}~\bibnamefont {Ahmed}}, \bibinfo {author}
  {\bibfnamefont {M.}~\bibnamefont {Scigliuzzo}}, \bibinfo {author}
  {\bibfnamefont {A.}~\bibnamefont {Osman}}, \bibinfo {author} {\bibfnamefont
  {D.~P.}\ \bibnamefont {Lozano}}, \bibinfo {author} {\bibfnamefont {M.~O.}\
  \bibnamefont {Thol\'en}}, \bibinfo {author} {\bibfnamefont {R.}~\bibnamefont
  {Borgani}}, \bibinfo {author} {\bibfnamefont {D.~B.}\ \bibnamefont
  {Haviland}}, \bibinfo {author} {\bibfnamefont {G.}~\bibnamefont {Ferrini}},
  \bibinfo {author} {\bibfnamefont {J.}~\bibnamefont {Bylander}}, \bibinfo
  {author} {\bibfnamefont {A.~F.}\ \bibnamefont {Kockum}}, \bibinfo {author}
  {\bibfnamefont {F.}~\bibnamefont {Quijandr\'ia}}, \bibinfo {author}
  {\bibfnamefont {P.}~\bibnamefont {Delsing}},\ and\ \bibinfo {author}
  {\bibfnamefont {S.}~\bibnamefont {Gasparinetti}},\ }\bibfield  {title}
  {\bibinfo {title} {Deterministically encoding quantum information using
  100-photon schrödinger cat states},\ }\href
  {https://doi.org/10.1103/PRXQuantum.3.030301} {\bibfield  {journal} {\bibinfo
   {journal} {PRX Quantum}\ }\textbf {\bibinfo {volume} {3}},\ \bibinfo {pages}
  {030301} (\bibinfo {year} {2022})}\BibitemShut {NoStop}%
\bibitem [{\citenamefont {McKay}\ \emph {et~al.}(2017)\citenamefont {McKay},
  \citenamefont {Wood}, \citenamefont {Sheldon}, \citenamefont {Chow},\ and\
  \citenamefont {Gambetta}}]{McKay_2017}%
  \BibitemOpen
  \bibfield  {author} {\bibinfo {author} {\bibfnamefont {D.~C.}\ \bibnamefont
  {McKay}}, \bibinfo {author} {\bibfnamefont {C.~J.}\ \bibnamefont {Wood}},
  \bibinfo {author} {\bibfnamefont {S.}~\bibnamefont {Sheldon}}, \bibinfo
  {author} {\bibfnamefont {J.~M.}\ \bibnamefont {Chow}},\ and\ \bibinfo
  {author} {\bibfnamefont {J.~M.}\ \bibnamefont {Gambetta}},\ }\bibfield
  {title} {\bibinfo {title} {Efficient $z$-gates for quantum computing},\
  }\href {https://doi.org/10.1103/physreva.96.022330} {\bibfield  {journal}
  {\bibinfo  {journal} {Physical Review A}\ }\textbf {\bibinfo {volume} {96}},\
  \bibinfo {pages} {022330} (\bibinfo {year} {2017})}\BibitemShut {NoStop}%
\bibitem [{\citenamefont {Kivlichan}\ \emph {et~al.}(2018)\citenamefont
  {Kivlichan}, \citenamefont {McClean}, \citenamefont {Wiebe}, \citenamefont
  {Gidney}, \citenamefont {Aspuru-Guzik}, \citenamefont {Kin-Lic~Chan},\ and\
  \citenamefont {Babbush}}]{Kivlichan2018}%
  \BibitemOpen
  \bibfield  {author} {\bibinfo {author} {\bibfnamefont {I.~D.}\ \bibnamefont
  {Kivlichan}}, \bibinfo {author} {\bibfnamefont {J.}~\bibnamefont {McClean}},
  \bibinfo {author} {\bibfnamefont {N.}~\bibnamefont {Wiebe}}, \bibinfo
  {author} {\bibfnamefont {C.}~\bibnamefont {Gidney}}, \bibinfo {author}
  {\bibfnamefont {A.}~\bibnamefont {Aspuru-Guzik}}, \bibinfo {author}
  {\bibfnamefont {G.}~\bibnamefont {Kin-Lic~Chan}},\ and\ \bibinfo {author}
  {\bibfnamefont {R.}~\bibnamefont {Babbush}},\ }\bibfield  {title} {\bibinfo
  {title} {Quantum simulation of electronic structure with linear depth and
  connectivity},\ }\href {https://doi.org/10.1103/PhysRevLett.120.110501}
  {\bibfield  {journal} {\bibinfo  {journal} {Phys. Rev. Lett.}\ }\textbf
  {\bibinfo {volume} {120}},\ \bibinfo {pages} {110501} (\bibinfo {year}
  {2018})}\BibitemShut {NoStop}%
\bibitem [{\citenamefont {Hagge}(2020)}]{Hagge2020}%
  \BibitemOpen
  \bibfield  {author} {\bibinfo {author} {\bibfnamefont {T.}~\bibnamefont
  {Hagge}},\ }\bibfield  {title} {\bibinfo {title} {Optimal fermionic swap
  networks for hubbard models},\ }\bibfield  {journal} {\bibinfo  {journal}
  {arXiv:2001.08324}\ }\href {https://doi.org/10.48550/arXiv.2001.08324}
  {10.48550/arXiv.2001.08324} (\bibinfo {year} {2020})\BibitemShut {NoStop}%
\bibitem [{\citenamefont {Fratus}\ \emph {et~al.}(2023)\citenamefont {Fratus},
  \citenamefont {Lepp\"akangas}, \citenamefont {Marthaler},\ and\ \citenamefont
  {Reiner}}]{Fratus2023}%
  \BibitemOpen
  \bibfield  {author} {\bibinfo {author} {\bibfnamefont {K.~R.}\ \bibnamefont
  {Fratus}}, \bibinfo {author} {\bibfnamefont {J.}~\bibnamefont
  {Lepp\"akangas}}, \bibinfo {author} {\bibfnamefont {M.}~\bibnamefont
  {Marthaler}},\ and\ \bibinfo {author} {\bibfnamefont {J.-M.}\ \bibnamefont
  {Reiner}},\ }\bibfield  {title} {\bibinfo {title} {The discrete noise
  approximation in quantum circuits},\ }\bibfield  {journal} {\bibinfo
  {journal} {arXiv.2311.00135}\ }\href
  {https://doi.org/10.48550/arXiv.2311.00135} {10.48550/arXiv.2311.00135}
  (\bibinfo {year} {2023})\BibitemShut {NoStop}%
\bibitem [{\citenamefont {Yan}\ \emph {et~al.}(2018)\citenamefont {Yan},
  \citenamefont {Krantz}, \citenamefont {Sung}, \citenamefont {Kjaergaard},
  \citenamefont {Campbell}, \citenamefont {Orlando}, \citenamefont
  {Gustavsson},\ and\ \citenamefont {Oliver}}]{Yan2018}%
  \BibitemOpen
  \bibfield  {author} {\bibinfo {author} {\bibfnamefont {F.}~\bibnamefont
  {Yan}}, \bibinfo {author} {\bibfnamefont {P.}~\bibnamefont {Krantz}},
  \bibinfo {author} {\bibfnamefont {Y.}~\bibnamefont {Sung}}, \bibinfo {author}
  {\bibfnamefont {M.}~\bibnamefont {Kjaergaard}}, \bibinfo {author}
  {\bibfnamefont {D.~L.}\ \bibnamefont {Campbell}}, \bibinfo {author}
  {\bibfnamefont {T.~P.}\ \bibnamefont {Orlando}}, \bibinfo {author}
  {\bibfnamefont {S.}~\bibnamefont {Gustavsson}},\ and\ \bibinfo {author}
  {\bibfnamefont {W.~D.}\ \bibnamefont {Oliver}},\ }\bibfield  {title}
  {\bibinfo {title} {Tunable coupling scheme for implementing high-fidelity
  two-qubit gates},\ }\href {https://doi.org/10.1103/physrevapplied.10.054062}
  {\bibfield  {journal} {\bibinfo  {journal} {Physical Review Applied}\
  }\textbf {\bibinfo {volume} {10}},\ \bibinfo {pages} {054062} (\bibinfo
  {year} {2018})}\BibitemShut {NoStop}%
\bibitem [{\citenamefont {Marxer}\ \emph {et~al.}(2023)\citenamefont {Marxer},
  \citenamefont {Vepsäläinen}, \citenamefont {Jolin}, \citenamefont
  {Tuorila}, \citenamefont {Landra}, \citenamefont {Ockeloen-Korppi},
  \citenamefont {Liu}, \citenamefont {Ahonen}, \citenamefont {Auer},
  \citenamefont {Belzane}, \citenamefont {Bergholm}, \citenamefont {Chan},
  \citenamefont {Chan}, \citenamefont {Hiltunen}, \citenamefont {Hotari},
  \citenamefont {Hyyppä}, \citenamefont {Ikonen}, \citenamefont {Janzso},
  \citenamefont {Koistinen}, \citenamefont {Kotilahti}, \citenamefont {Li},
  \citenamefont {Luus}, \citenamefont {Papic}, \citenamefont {Partanen},
  \citenamefont {Räbinä}, \citenamefont {Rosti}, \citenamefont {Savytskyi},
  \citenamefont {Seppälä}, \citenamefont {Sevriuk}, \citenamefont {Takala},
  \citenamefont {Tarasinski}, \citenamefont {Thapa}, \citenamefont {Tosto},
  \citenamefont {Vorobeva}, \citenamefont {Yu}, \citenamefont {Tan},
  \citenamefont {Hassel}, \citenamefont {Möttönen},\ and\ \citenamefont
  {Heinsoo}}]{Marxer2023}%
  \BibitemOpen
  \bibfield  {author} {\bibinfo {author} {\bibfnamefont {F.}~\bibnamefont
  {Marxer}}, \bibinfo {author} {\bibfnamefont {A.}~\bibnamefont
  {Vepsäläinen}}, \bibinfo {author} {\bibfnamefont {S.~W.}\ \bibnamefont
  {Jolin}}, \bibinfo {author} {\bibfnamefont {J.}~\bibnamefont {Tuorila}},
  \bibinfo {author} {\bibfnamefont {A.}~\bibnamefont {Landra}}, \bibinfo
  {author} {\bibfnamefont {C.}~\bibnamefont {Ockeloen-Korppi}}, \bibinfo
  {author} {\bibfnamefont {W.}~\bibnamefont {Liu}}, \bibinfo {author}
  {\bibfnamefont {O.}~\bibnamefont {Ahonen}}, \bibinfo {author} {\bibfnamefont
  {A.}~\bibnamefont {Auer}}, \bibinfo {author} {\bibfnamefont {L.}~\bibnamefont
  {Belzane}}, \bibinfo {author} {\bibfnamefont {V.}~\bibnamefont {Bergholm}},
  \bibinfo {author} {\bibfnamefont {C.~F.}\ \bibnamefont {Chan}}, \bibinfo
  {author} {\bibfnamefont {K.~W.}\ \bibnamefont {Chan}}, \bibinfo {author}
  {\bibfnamefont {T.}~\bibnamefont {Hiltunen}}, \bibinfo {author}
  {\bibfnamefont {J.}~\bibnamefont {Hotari}}, \bibinfo {author} {\bibfnamefont
  {E.}~\bibnamefont {Hyyppä}}, \bibinfo {author} {\bibfnamefont
  {J.}~\bibnamefont {Ikonen}}, \bibinfo {author} {\bibfnamefont
  {D.}~\bibnamefont {Janzso}}, \bibinfo {author} {\bibfnamefont
  {M.}~\bibnamefont {Koistinen}}, \bibinfo {author} {\bibfnamefont
  {J.}~\bibnamefont {Kotilahti}}, \bibinfo {author} {\bibfnamefont
  {T.}~\bibnamefont {Li}}, \bibinfo {author} {\bibfnamefont {J.}~\bibnamefont
  {Luus}}, \bibinfo {author} {\bibfnamefont {M.}~\bibnamefont {Papic}},
  \bibinfo {author} {\bibfnamefont {M.}~\bibnamefont {Partanen}}, \bibinfo
  {author} {\bibfnamefont {J.}~\bibnamefont {Räbinä}}, \bibinfo {author}
  {\bibfnamefont {J.}~\bibnamefont {Rosti}}, \bibinfo {author} {\bibfnamefont
  {M.}~\bibnamefont {Savytskyi}}, \bibinfo {author} {\bibfnamefont
  {M.}~\bibnamefont {Seppälä}}, \bibinfo {author} {\bibfnamefont
  {V.}~\bibnamefont {Sevriuk}}, \bibinfo {author} {\bibfnamefont
  {E.}~\bibnamefont {Takala}}, \bibinfo {author} {\bibfnamefont
  {B.}~\bibnamefont {Tarasinski}}, \bibinfo {author} {\bibfnamefont {M.~J.}\
  \bibnamefont {Thapa}}, \bibinfo {author} {\bibfnamefont {F.}~\bibnamefont
  {Tosto}}, \bibinfo {author} {\bibfnamefont {N.}~\bibnamefont {Vorobeva}},
  \bibinfo {author} {\bibfnamefont {L.}~\bibnamefont {Yu}}, \bibinfo {author}
  {\bibfnamefont {K.~Y.}\ \bibnamefont {Tan}}, \bibinfo {author} {\bibfnamefont
  {J.}~\bibnamefont {Hassel}}, \bibinfo {author} {\bibfnamefont
  {M.}~\bibnamefont {Möttönen}},\ and\ \bibinfo {author} {\bibfnamefont
  {J.}~\bibnamefont {Heinsoo}},\ }\bibfield  {title} {\bibinfo {title}
  {Long-distance transmon coupler with cz-gate fidelity above 99.8\%},\ }\href
  {https://doi.org/10.1103/prxquantum.4.010314} {\bibfield  {journal} {\bibinfo
   {journal} {PRX Quantum}\ }\textbf {\bibinfo {volume} {4}},\ \bibinfo {pages}
  {010314} (\bibinfo {year} {2023})}\BibitemShut {NoStop}%
\bibitem [{\citenamefont {Algaba}\ \emph {et~al.}(2022)\citenamefont {Algaba},
  \citenamefont {Ponce-Martinez}, \citenamefont {Munuera-Javaloy},
  \citenamefont {Pina-Canelles}, \citenamefont {Thapa}, \citenamefont
  {Taketani}, \citenamefont {Leib}, \citenamefont {de~Vega}, \citenamefont
  {Casanova},\ and\ \citenamefont {Heimonen}}]{Algaba2022}%
  \BibitemOpen
  \bibfield  {author} {\bibinfo {author} {\bibfnamefont {M.~G.}\ \bibnamefont
  {Algaba}}, \bibinfo {author} {\bibfnamefont {M.}~\bibnamefont
  {Ponce-Martinez}}, \bibinfo {author} {\bibfnamefont {C.}~\bibnamefont
  {Munuera-Javaloy}}, \bibinfo {author} {\bibfnamefont {V.}~\bibnamefont
  {Pina-Canelles}}, \bibinfo {author} {\bibfnamefont {M.~J.}\ \bibnamefont
  {Thapa}}, \bibinfo {author} {\bibfnamefont {B.~G.}\ \bibnamefont {Taketani}},
  \bibinfo {author} {\bibfnamefont {M.}~\bibnamefont {Leib}}, \bibinfo {author}
  {\bibfnamefont {I.}~\bibnamefont {de~Vega}}, \bibinfo {author} {\bibfnamefont
  {J.}~\bibnamefont {Casanova}},\ and\ \bibinfo {author} {\bibfnamefont
  {H.}~\bibnamefont {Heimonen}},\ }\bibfield  {title} {\bibinfo {title}
  {Co-design quantum simulation of nanoscale nmr},\ }\href
  {https://doi.org/10.1103/PhysRevResearch.4.043089} {\bibfield  {journal}
  {\bibinfo  {journal} {Phys. Rev. Res.}\ }\textbf {\bibinfo {volume} {4}},\
  \bibinfo {pages} {043089} (\bibinfo {year} {2022})}\BibitemShut {NoStop}%
\bibitem [{\citenamefont {Arute}\ \emph {et~al.}(2020)\citenamefont {Arute},
  \citenamefont {Arya}, \citenamefont {Babbush}, \citenamefont {Bacon},
  \citenamefont {Bardin}, \citenamefont {Barends}, \citenamefont {Bengtsson},
  \citenamefont {Boixo}, \citenamefont {Broughton}, \citenamefont {Buckley},
  \citenamefont {Buell}, \citenamefont {Burkett}, \citenamefont {Bushnell},
  \citenamefont {Chen}, \citenamefont {Chen}, \citenamefont {Chen},
  \citenamefont {Chiaro}, \citenamefont {Collins}, \citenamefont {Cotton},
  \citenamefont {Courtney}, \citenamefont {Demura}, \citenamefont {Derk},
  \citenamefont {Dunsworth}, \citenamefont {Eppens}, \citenamefont {Eckl},
  \citenamefont {Erickson}, \citenamefont {Farhi}, \citenamefont {Fowler},
  \citenamefont {Foxen}, \citenamefont {Gidney}, \citenamefont {Giustina},
  \citenamefont {Graff}, \citenamefont {Gross}, \citenamefont {Habegger},
  \citenamefont {Harrigan}, \citenamefont {Ho}, \citenamefont {Hong},
  \citenamefont {Huang}, \citenamefont {Huggins}, \citenamefont {Ioffe},
  \citenamefont {Isakov}, \citenamefont {Jeffrey}, \citenamefont {Jiang},
  \citenamefont {Jones}, \citenamefont {Kafri}, \citenamefont {Kechedzhi},
  \citenamefont {Kelly}, \citenamefont {Kim}, \citenamefont {Klimov},
  \citenamefont {Korotkov}, \citenamefont {Kostritsa}, \citenamefont
  {Landhuis}, \citenamefont {Laptev}, \citenamefont {Lindmark}, \citenamefont
  {Lucero}, \citenamefont {Marthaler}, \citenamefont {Martin}, \citenamefont
  {Martinis}, \citenamefont {Marusczyk}, \citenamefont {McArdle}, \citenamefont
  {McClean}, \citenamefont {McCourt}, \citenamefont {McEwen}, \citenamefont
  {Megrant}, \citenamefont {Mejuto-Zaera}, \citenamefont {Mi}, \citenamefont
  {Mohseni}, \citenamefont {Mruczkiewicz}, \citenamefont {Mutus}, \citenamefont
  {Naaman}, \citenamefont {Neeley}, \citenamefont {Neill}, \citenamefont
  {Neven}, \citenamefont {Newman}, \citenamefont {Niu}, \citenamefont
  {O'Brien}, \citenamefont {Ostby}, \citenamefont {Pató}, \citenamefont
  {Petukhov}, \citenamefont {Putterman}, \citenamefont {Quintana},
  \citenamefont {Reiner}, \citenamefont {Roushan}, \citenamefont {Rubin},
  \citenamefont {Sank}, \citenamefont {Satzinger}, \citenamefont {Smelyanskiy},
  \citenamefont {Strain}, \citenamefont {Sung}, \citenamefont {Schmitteckert},
  \citenamefont {Szalay}, \citenamefont {Tubman}, \citenamefont {Vainsencher},
  \citenamefont {White}, \citenamefont {Vogt}, \citenamefont {Yao},
  \citenamefont {Yeh}, \citenamefont {Zalcman},\ and\ \citenamefont
  {Zanker}}]{Arute2020}%
  \BibitemOpen
  \bibfield  {author} {\bibinfo {author} {\bibfnamefont {F.}~\bibnamefont
  {Arute}}, \bibinfo {author} {\bibfnamefont {K.}~\bibnamefont {Arya}},
  \bibinfo {author} {\bibfnamefont {R.}~\bibnamefont {Babbush}}, \bibinfo
  {author} {\bibfnamefont {D.}~\bibnamefont {Bacon}}, \bibinfo {author}
  {\bibfnamefont {J.~C.}\ \bibnamefont {Bardin}}, \bibinfo {author}
  {\bibfnamefont {R.}~\bibnamefont {Barends}}, \bibinfo {author} {\bibfnamefont
  {A.}~\bibnamefont {Bengtsson}}, \bibinfo {author} {\bibfnamefont
  {S.}~\bibnamefont {Boixo}}, \bibinfo {author} {\bibfnamefont
  {M.}~\bibnamefont {Broughton}}, \bibinfo {author} {\bibfnamefont {B.~B.}\
  \bibnamefont {Buckley}}, \bibinfo {author} {\bibfnamefont {D.~A.}\
  \bibnamefont {Buell}}, \bibinfo {author} {\bibfnamefont {B.}~\bibnamefont
  {Burkett}}, \bibinfo {author} {\bibfnamefont {N.}~\bibnamefont {Bushnell}},
  \bibinfo {author} {\bibfnamefont {Y.}~\bibnamefont {Chen}}, \bibinfo {author}
  {\bibfnamefont {Z.}~\bibnamefont {Chen}}, \bibinfo {author} {\bibfnamefont
  {Y.-A.}\ \bibnamefont {Chen}}, \bibinfo {author} {\bibfnamefont
  {B.}~\bibnamefont {Chiaro}}, \bibinfo {author} {\bibfnamefont
  {R.}~\bibnamefont {Collins}}, \bibinfo {author} {\bibfnamefont {S.~J.}\
  \bibnamefont {Cotton}}, \bibinfo {author} {\bibfnamefont {W.}~\bibnamefont
  {Courtney}}, \bibinfo {author} {\bibfnamefont {S.}~\bibnamefont {Demura}},
  \bibinfo {author} {\bibfnamefont {A.}~\bibnamefont {Derk}}, \bibinfo {author}
  {\bibfnamefont {A.}~\bibnamefont {Dunsworth}}, \bibinfo {author}
  {\bibfnamefont {D.}~\bibnamefont {Eppens}}, \bibinfo {author} {\bibfnamefont
  {T.}~\bibnamefont {Eckl}}, \bibinfo {author} {\bibfnamefont {C.}~\bibnamefont
  {Erickson}}, \bibinfo {author} {\bibfnamefont {E.}~\bibnamefont {Farhi}},
  \bibinfo {author} {\bibfnamefont {A.}~\bibnamefont {Fowler}}, \bibinfo
  {author} {\bibfnamefont {B.}~\bibnamefont {Foxen}}, \bibinfo {author}
  {\bibfnamefont {C.}~\bibnamefont {Gidney}}, \bibinfo {author} {\bibfnamefont
  {M.}~\bibnamefont {Giustina}}, \bibinfo {author} {\bibfnamefont
  {R.}~\bibnamefont {Graff}}, \bibinfo {author} {\bibfnamefont {J.~A.}\
  \bibnamefont {Gross}}, \bibinfo {author} {\bibfnamefont {S.}~\bibnamefont
  {Habegger}}, \bibinfo {author} {\bibfnamefont {M.~P.}\ \bibnamefont
  {Harrigan}}, \bibinfo {author} {\bibfnamefont {A.}~\bibnamefont {Ho}},
  \bibinfo {author} {\bibfnamefont {S.}~\bibnamefont {Hong}}, \bibinfo {author}
  {\bibfnamefont {T.}~\bibnamefont {Huang}}, \bibinfo {author} {\bibfnamefont
  {W.}~\bibnamefont {Huggins}}, \bibinfo {author} {\bibfnamefont {L.~B.}\
  \bibnamefont {Ioffe}}, \bibinfo {author} {\bibfnamefont {S.~V.}\ \bibnamefont
  {Isakov}}, \bibinfo {author} {\bibfnamefont {E.}~\bibnamefont {Jeffrey}},
  \bibinfo {author} {\bibfnamefont {Z.}~\bibnamefont {Jiang}}, \bibinfo
  {author} {\bibfnamefont {C.}~\bibnamefont {Jones}}, \bibinfo {author}
  {\bibfnamefont {D.}~\bibnamefont {Kafri}}, \bibinfo {author} {\bibfnamefont
  {K.}~\bibnamefont {Kechedzhi}}, \bibinfo {author} {\bibfnamefont
  {J.}~\bibnamefont {Kelly}}, \bibinfo {author} {\bibfnamefont
  {S.}~\bibnamefont {Kim}}, \bibinfo {author} {\bibfnamefont {P.~V.}\
  \bibnamefont {Klimov}}, \bibinfo {author} {\bibfnamefont {A.~N.}\
  \bibnamefont {Korotkov}}, \bibinfo {author} {\bibfnamefont {F.}~\bibnamefont
  {Kostritsa}}, \bibinfo {author} {\bibfnamefont {D.}~\bibnamefont {Landhuis}},
  \bibinfo {author} {\bibfnamefont {P.}~\bibnamefont {Laptev}}, \bibinfo
  {author} {\bibfnamefont {M.}~\bibnamefont {Lindmark}}, \bibinfo {author}
  {\bibfnamefont {E.}~\bibnamefont {Lucero}}, \bibinfo {author} {\bibfnamefont
  {M.}~\bibnamefont {Marthaler}}, \bibinfo {author} {\bibfnamefont
  {O.}~\bibnamefont {Martin}}, \bibinfo {author} {\bibfnamefont {J.~M.}\
  \bibnamefont {Martinis}}, \bibinfo {author} {\bibfnamefont {A.}~\bibnamefont
  {Marusczyk}}, \bibinfo {author} {\bibfnamefont {S.}~\bibnamefont {McArdle}},
  \bibinfo {author} {\bibfnamefont {J.~R.}\ \bibnamefont {McClean}}, \bibinfo
  {author} {\bibfnamefont {T.}~\bibnamefont {McCourt}}, \bibinfo {author}
  {\bibfnamefont {M.}~\bibnamefont {McEwen}}, \bibinfo {author} {\bibfnamefont
  {A.}~\bibnamefont {Megrant}}, \bibinfo {author} {\bibfnamefont
  {C.}~\bibnamefont {Mejuto-Zaera}}, \bibinfo {author} {\bibfnamefont
  {X.}~\bibnamefont {Mi}}, \bibinfo {author} {\bibfnamefont {M.}~\bibnamefont
  {Mohseni}}, \bibinfo {author} {\bibfnamefont {W.}~\bibnamefont
  {Mruczkiewicz}}, \bibinfo {author} {\bibfnamefont {J.}~\bibnamefont {Mutus}},
  \bibinfo {author} {\bibfnamefont {O.}~\bibnamefont {Naaman}}, \bibinfo
  {author} {\bibfnamefont {M.}~\bibnamefont {Neeley}}, \bibinfo {author}
  {\bibfnamefont {C.}~\bibnamefont {Neill}}, \bibinfo {author} {\bibfnamefont
  {H.}~\bibnamefont {Neven}}, \bibinfo {author} {\bibfnamefont
  {M.}~\bibnamefont {Newman}}, \bibinfo {author} {\bibfnamefont {M.~Y.}\
  \bibnamefont {Niu}}, \bibinfo {author} {\bibfnamefont {T.~E.}\ \bibnamefont
  {O'Brien}}, \bibinfo {author} {\bibfnamefont {E.}~\bibnamefont {Ostby}},
  \bibinfo {author} {\bibfnamefont {B.}~\bibnamefont {Pató}}, \bibinfo
  {author} {\bibfnamefont {A.}~\bibnamefont {Petukhov}}, \bibinfo {author}
  {\bibfnamefont {H.}~\bibnamefont {Putterman}}, \bibinfo {author}
  {\bibfnamefont {C.}~\bibnamefont {Quintana}}, \bibinfo {author}
  {\bibfnamefont {J.-M.}\ \bibnamefont {Reiner}}, \bibinfo {author}
  {\bibfnamefont {P.}~\bibnamefont {Roushan}}, \bibinfo {author} {\bibfnamefont
  {N.~C.}\ \bibnamefont {Rubin}}, \bibinfo {author} {\bibfnamefont
  {D.}~\bibnamefont {Sank}}, \bibinfo {author} {\bibfnamefont {K.~J.}\
  \bibnamefont {Satzinger}}, \bibinfo {author} {\bibfnamefont {V.}~\bibnamefont
  {Smelyanskiy}}, \bibinfo {author} {\bibfnamefont {D.}~\bibnamefont {Strain}},
  \bibinfo {author} {\bibfnamefont {K.~J.}\ \bibnamefont {Sung}}, \bibinfo
  {author} {\bibfnamefont {P.}~\bibnamefont {Schmitteckert}}, \bibinfo {author}
  {\bibfnamefont {M.}~\bibnamefont {Szalay}}, \bibinfo {author} {\bibfnamefont
  {N.~M.}\ \bibnamefont {Tubman}}, \bibinfo {author} {\bibfnamefont
  {A.}~\bibnamefont {Vainsencher}}, \bibinfo {author} {\bibfnamefont
  {T.}~\bibnamefont {White}}, \bibinfo {author} {\bibfnamefont
  {N.}~\bibnamefont {Vogt}}, \bibinfo {author} {\bibfnamefont {Z.~J.}\
  \bibnamefont {Yao}}, \bibinfo {author} {\bibfnamefont {P.}~\bibnamefont
  {Yeh}}, \bibinfo {author} {\bibfnamefont {A.}~\bibnamefont {Zalcman}},\ and\
  \bibinfo {author} {\bibfnamefont {S.}~\bibnamefont {Zanker}},\ }\bibfield
  {title} {\bibinfo {title} {Observation of separated dynamics of charge and
  spin in the fermi-hubbard model},\ }\bibfield  {journal} {\bibinfo  {journal}
  {arXiv.2010.07965}\ }\href {https://doi.org/10.48550/arXiv.2010.07965}
  {10.48550/arXiv.2010.07965} (\bibinfo {year} {2020})\BibitemShut {NoStop}%
\bibitem [{\citenamefont {Huang}\ \emph {et~al.}(2023)\citenamefont {Huang},
  \citenamefont {Wang}, \citenamefont {Wu}, \citenamefont {Ding}, \citenamefont
  {Ye}, \citenamefont {Kong}, \citenamefont {Zhang}, \citenamefont {Ni},
  \citenamefont {Song}, \citenamefont {Shi}, \citenamefont {Zhao},
  \citenamefont {Deng},\ and\ \citenamefont {Chen}}]{Huang2023}%
  \BibitemOpen
  \bibfield  {author} {\bibinfo {author} {\bibfnamefont {C.}~\bibnamefont
  {Huang}}, \bibinfo {author} {\bibfnamefont {T.}~\bibnamefont {Wang}},
  \bibinfo {author} {\bibfnamefont {F.}~\bibnamefont {Wu}}, \bibinfo {author}
  {\bibfnamefont {D.}~\bibnamefont {Ding}}, \bibinfo {author} {\bibfnamefont
  {Q.}~\bibnamefont {Ye}}, \bibinfo {author} {\bibfnamefont {L.}~\bibnamefont
  {Kong}}, \bibinfo {author} {\bibfnamefont {F.}~\bibnamefont {Zhang}},
  \bibinfo {author} {\bibfnamefont {X.}~\bibnamefont {Ni}}, \bibinfo {author}
  {\bibfnamefont {Z.}~\bibnamefont {Song}}, \bibinfo {author} {\bibfnamefont
  {Y.}~\bibnamefont {Shi}}, \bibinfo {author} {\bibfnamefont {H.-H.}\
  \bibnamefont {Zhao}}, \bibinfo {author} {\bibfnamefont {C.}~\bibnamefont
  {Deng}},\ and\ \bibinfo {author} {\bibfnamefont {J.}~\bibnamefont {Chen}},\
  }\bibfield  {title} {\bibinfo {title} {Quantum instruction set design for
  performance},\ }\href {https://doi.org/10.1103/physrevlett.130.070601}
  {\bibfield  {journal} {\bibinfo  {journal} {Physical Review Letters}\
  }\textbf {\bibinfo {volume} {130}},\ \bibinfo {pages} {070601} (\bibinfo
  {year} {2023})}\BibitemShut {NoStop}%
\bibitem [{\citenamefont {Braum\"uller}\ \emph {et~al.}(2017)\citenamefont
  {Braum\"uller}, \citenamefont {Marthaler}, \citenamefont {Schneider},
  \citenamefont {Stehli}, \citenamefont {Rotzinger}, \citenamefont {Weides},\
  and\ \citenamefont {Ustinov}}]{Braumueller2017}%
  \BibitemOpen
  \bibfield  {author} {\bibinfo {author} {\bibfnamefont {J.}~\bibnamefont
  {Braum\"uller}}, \bibinfo {author} {\bibfnamefont {M.}~\bibnamefont
  {Marthaler}}, \bibinfo {author} {\bibfnamefont {A.}~\bibnamefont
  {Schneider}}, \bibinfo {author} {\bibfnamefont {A.}~\bibnamefont {Stehli}},
  \bibinfo {author} {\bibfnamefont {H.}~\bibnamefont {Rotzinger}}, \bibinfo
  {author} {\bibfnamefont {M.}~\bibnamefont {Weides}},\ and\ \bibinfo {author}
  {\bibfnamefont {A.~V.}\ \bibnamefont {Ustinov}},\ }\bibfield  {title}
  {\bibinfo {title} {Analog quantum simulation of the rabi model in the
  ultra-strong coupling regime},\ }\href
  {https://doi.org/10.1038/s41467-017-00894-w} {\bibfield  {journal} {\bibinfo
  {journal} {Nat. Commun.}\ }\textbf {\bibinfo {volume} {8}},\ \bibinfo {pages}
  {779} (\bibinfo {year} {2017})}\BibitemShut {NoStop}%
\bibitem [{IQM()}]{IQM_Pulla}%
  \BibitemOpen
  \href
  {https://www.meetiqm.com/newsroom/blog/how-iqms-pulse-level-access-is-fuelling-value-for-end-users}
  {\bibinfo {title} {Iqm pulla, pulse-level access}}\BibitemShut {NoStop}%
\bibitem [{\citenamefont {Schubert}\ and\ \citenamefont
  {Mendl}(2023)}]{Schubert2023}%
  \BibitemOpen
  \bibfield  {author} {\bibinfo {author} {\bibfnamefont {A.}~\bibnamefont
  {Schubert}}\ and\ \bibinfo {author} {\bibfnamefont {C.~B.}\ \bibnamefont
  {Mendl}},\ }\bibfield  {title} {\bibinfo {title} {Trotter error with
  commutator scaling for the fermi-hubbard model},\ }\href
  {https://doi.org/10.1103/physrevb.108.195105} {\bibfield  {journal} {\bibinfo
   {journal} {Physical Review B}\ }\textbf {\bibinfo {volume} {108}},\ \bibinfo
  {pages} {195105} (\bibinfo {year} {2023})}\BibitemShut {NoStop}%
\bibitem [{\citenamefont {Ikeda}\ \emph {et~al.}(2023)\citenamefont {Ikeda},
  \citenamefont {Abrar}, \citenamefont {Chuang},\ and\ \citenamefont
  {Sugiura}}]{Ikeda2023}%
  \BibitemOpen
  \bibfield  {author} {\bibinfo {author} {\bibfnamefont {T.~N.}\ \bibnamefont
  {Ikeda}}, \bibinfo {author} {\bibfnamefont {A.}~\bibnamefont {Abrar}},
  \bibinfo {author} {\bibfnamefont {I.~L.}\ \bibnamefont {Chuang}},\ and\
  \bibinfo {author} {\bibfnamefont {S.}~\bibnamefont {Sugiura}},\ }\bibfield
  {title} {\bibinfo {title} {Minimum trotterization formulas for a
  time-dependent hamiltonian},\ }\href
  {https://doi.org/10.22331/q-2023-11-06-1168} {\bibfield  {journal} {\bibinfo
  {journal} {Quantum}\ }\textbf {\bibinfo {volume} {7}},\ \bibinfo {pages}
  {1168} (\bibinfo {year} {2023})}\BibitemShut {NoStop}%
\bibitem [{\citenamefont {Kirchmair}\ \emph {et~al.}(2013)\citenamefont
  {Kirchmair}, \citenamefont {Vlastakis}, \citenamefont {Leghtas},
  \citenamefont {Nigg}, \citenamefont {Paik}, \citenamefont {Ginossar},
  \citenamefont {Mirrahimi}, \citenamefont {Frunzio}, \citenamefont {Girvin},\
  and\ \citenamefont {Schoelkopf}}]{Kirchmair2013}%
  \BibitemOpen
  \bibfield  {author} {\bibinfo {author} {\bibfnamefont {G.}~\bibnamefont
  {Kirchmair}}, \bibinfo {author} {\bibfnamefont {B.}~\bibnamefont
  {Vlastakis}}, \bibinfo {author} {\bibfnamefont {Z.}~\bibnamefont {Leghtas}},
  \bibinfo {author} {\bibfnamefont {S.~E.}\ \bibnamefont {Nigg}}, \bibinfo
  {author} {\bibfnamefont {H.}~\bibnamefont {Paik}}, \bibinfo {author}
  {\bibfnamefont {E.}~\bibnamefont {Ginossar}}, \bibinfo {author}
  {\bibfnamefont {M.}~\bibnamefont {Mirrahimi}}, \bibinfo {author}
  {\bibfnamefont {L.}~\bibnamefont {Frunzio}}, \bibinfo {author} {\bibfnamefont
  {S.~M.}\ \bibnamefont {Girvin}},\ and\ \bibinfo {author} {\bibfnamefont
  {R.~J.}\ \bibnamefont {Schoelkopf}},\ }\bibfield  {title} {\bibinfo {title}
  {Observation of quantum state collapse and revival due to the single-photon
  kerr effect},\ }\href {https://doi.org/10.1038/nature11902} {\bibfield
  {journal} {\bibinfo  {journal} {Nature}\ }\textbf {\bibinfo {volume} {495}},\
  \bibinfo {pages} {205} (\bibinfo {year} {2013})}\BibitemShut {NoStop}%
\end{thebibliography}%

\end{document}